\documentclass[structabstract]{aa}  
\usepackage{url,cancel,graphics,latexsym,epsfig,graphicx,amssymb,lscape,color,txfonts,natbib,chngcntr,soul}     
\usepackage[textwidth=3.5cm]{todonotes}
\usepackage{blindtext}

\renewcommand{\d}[0]{{\rm d}}
\newcommand{\e}[0]{{\rm e}}
\renewcommand{\i}[0]{{\rm i}}
\newcommand{\ave}[1]{\langle #1 \rangle}

\newcommand{\Ref}[1]{(\ref{#1})}
\newcommand{\mat}[1]{\tens{#1}}

\newcommand{\msol}[0]{{\rm M}_\odot}

\newcommand{\gabods}{\mbox{GaBoDS}}

\begin{document}
\title{Scale dependence of galaxy biasing investigated by weak
    gravitational lensing: An assessment using semi-analytic galaxies
    and simulated lensing data}

\author{Patrick Simon$^1$ and Stefan Hilbert$^{2,3}$}

\institute{$^1$ Argelander-Institut f\"ur Astronomie, Universit\"at
  Bonn, Auf dem H\"ugel 71, 53121 Bonn, Germany\\
  $^2$ Faculty of Physics, Ludwig-Maximilians University, Scheinerstr. 1, 81679 M\"unchen, Germany\\
  $^3$ Excellence Cluster Universe, Boltzmannstr. 2, 85748 Garching, Germany\\
  \email{psimon@astro.uni-bonn.de}}

\date{Received \today}

\authorrunning{Simon and Hilbert} \titlerunning{Assessment of weak
  lensing for investigating galaxy biasing}

\abstract{Galaxies are biased tracers of the matter density on
  cosmological scales.  For future tests of galaxy models, we refine
  and assess a method to measure galaxy biasing as function of
  physical scale $k$ with weak gravitational lensing. This method
  enables us to reconstruct the galaxy bias factor $b(k)$ as well as
  the galaxy-matter correlation $r(k)$ on spatial scales between
  \mbox{$0.01\,h\,{\rm Mpc^{-1}}\lesssim k\lesssim10\,h\,{\rm
      Mpc^{-1}}$} for redshift-binned lens galaxies below redshift
  \mbox{$z\lesssim0.6$}. In the refinement, we account for an
  intrinsic alignment of source ellipticities, and we correct for the
  magnification bias of the lens galaxies, relevant for the
  galaxy-galaxy lensing signal, to improve the accuracy of the
  reconstructed $r(k)$.  For simulated data, the reconstructions
  achieve an accuracy of \mbox{$3-7\%$} (68\% confidence level) over
  the above $k$-range for a survey area and a typical depth of
  contemporary ground-based surveys.  Realistically the accuracy is,
  however, probably reduced to about \mbox{$10-15\%$}, mainly by
  systematic uncertainties in the assumed intrinsic source alignment,
  the fiducial cosmology, and the redshift distributions of lens and
  source galaxies (in that order). Furthermore, our reconstruction
  technique employs physical templates for $b(k)$ and $r(k)$ that
  elucidate the impact of central galaxies and the halo-occupation
  statistics of satellite galaxies on the scale-dependence of galaxy
  bias, which we discuss in the paper. In a first demonstration, we
  apply this method to previous measurements in the Garching-Bonn-Deep
  Survey and give a physical interpretation of the lens population.}


\maketitle

\section{Introduction}

The standard paradigm of cosmology describes the large-scale
distribution of matter and galaxies in an expanding Universe
\citep[][and references therein]{2003moco.book.....D}.  Strongly
supported by observations, this model assumes a statistically
homogeneous and isotropic Universe with cold dark matter (CDM) as
dominating form of matter. Matter in total has the mean density
\mbox{$\Omega_{\rm m}\approx0.3$} of which ordinary baryonic matter is
just \mbox{$\Omega_{\rm b}\approx0.05$}; as usual densities are in
units of the critical density (or its energy equivalent).  The largest
fraction \mbox{$\Omega_\Lambda\approx0.7$} in the cosmological energy
density is given by a cosmological constant $\Lambda$ or so-called
dark energy, resulting in a flat or approximately flat background
geometry with curvature parameter \mbox{$K=0$} \citep[][and references
therein]{1917SPAW.......142E,2016A&A...594A..13P}.  The exact physical
nature of dark matter is unknown but its presence is consistently
inferred through visible tracers from galactic to cosmological scales
at different epochs in the cosmic history \citep[][for a
review]{2005PhR...405..279B}. In particular the coherent shear of
distant galaxy images (background sources) by the tidal gravitational
field of intervening matter gives direct evidence for the (projected)
density field of dark matter \citep{2004ApJ...604..596C}. The basic
physics of galaxy formation inside dark-matter halos and the galaxy
evolution seems to be identified and reasonably well matched by
observations although various processes, such as star formation and
galaxy-gas feedback, are still not well understood or worked out in
detail \citep{2010gfe..book.....M}. Ultimately, the ability of the
$\Lambda\rm CDM$ model to quantitatively describe the observed
richness of galaxy properties from initial conditions will be a
crucial validation test.

One galaxy property of importance is their spatial distribution.
Galaxies are known to be differently distributed than the matter in
general; they are so-called biased tracers of the matter density field
\citep{1984ApJ...284L...9K}.  The details of the biasing mechanism are
related to galaxy physics
\citep{2017arXiv170703397J,2005Natur.435..629S,2004ApJ...601....1W,2001MNRAS.320..289S,2000MNRAS.311..793B,2000MNRAS.318.1144P}. An
observed galaxy bias for different galaxy types and redshifts
consequently provides input and tests for galaxy models. Additionally,
its measurement is practical for studies that rely on fiducial values
for the biasing of a particular galaxy sample or on the observational
support for a high galaxy-matter correlation on particular spatial
scales
\citep[e.g.][]{2017arXiv170605004V,2013MNRAS.429.3230H,2013A&A...560A..33S,2011ApJ...734...94M,2010Natur.464..256R,2010PhRvD..81f3531B}.
In this context, we investigate the prospects of weak gravitational
lensing to measure the galaxy bias \citep[e.g.][for a
review]{2015RPPh...78h6901K,2008PhR...462...67M,schneider2006gravitational}.

There are clearly various ways to express the statistical relationship
between the galaxy and matter distribution, which both can be seen as
realisations of statistically homogeneous and isotropic random fields
\citep{2016arXiv161109787D}. With focus on second-order statistics we
use the common parameterisation in \citet{1999ApJ...518L..69T}.  This
defines galaxy bias in terms of auto- and cross-correlation power
spectra of the random fields for a given wave number $k$: (i) a bias
factor $b(k)$ for the relative strength between galaxy and matter
clustering; and (ii) a factor $r(k)$ for the galaxy-matter
correlation. The second-order biasing functions can be constrained by
combining galaxy clustering with cosmic-shear information in lensing
surveys
\citep{2016MNRAS.463.3326F,2012MNRAS.426..566C,2012A&A...543A...2S,2003MNRAS.346..994P}.
In applications of these techniques, galaxy biasing is then known to
depend on galaxy type, physical scale, and redshift, thus reflecting
interesting galaxy physics
\citep{2016MNRAS.459.3203C,2016MNRAS.456.3886B,2016MNRAS.462...35P,2016arXiv160908167P,2013MNRAS.433.1146C,2013MNRAS.430.2476S,2012ApJ...750...37J,2007A&A...461..861S,2003MNRAS.346..994P,2002ApJ...577..604H}.

Our interest here is the quality of lensing measurements of galaxy
bias. For this purpose, we focus on the method by
\cite{1998A&A...334....1V} and \citet{1998ApJ...498...43S}, first
applied in \citet{2001ApJ...558L..11H} and
\citet{2002ApJ...577..604H}, where one defines relative aperture
measures of the galaxy number-density and the lensing mass to observe
$b(k)$ and $r(k)$ as projections on the sky, averaged in bands of
radial and transverse direction. The advantage of this method is its
model independence apart from a cosmology-dependent normalisation. As
improvement we define a new procedure to deproject the lensing
measurements of the projected biasing functions, giving direct
estimates of $b(k)$ and $r(k)$ for a selected galaxy population. In
addition, we account for the intrinsic alignment of source galaxies
that are utilised in the lensing analysis
\citep{2015SSRv..193..139K}. To eventually assess the accuracy and
precision of our deprojection technique, we compare the results to the
true biasing functions for various galaxy samples in a simulated,
about $1000\,\rm deg^2$ wide survey, constructed with a semi-analytic
galaxy model by \citet{2015MNRAS.451.2663H}, H15 hereafter, and data
from the Millennium Simulation \citep{2005Natur.435..629S}. To this
end, a large part of this paper deals with the construction of
flexible template models of $b(k)$ and $r(k)$ that we forward-fit to
the relative aperture measures. These templates are based on a
flexible halo-model prescription, which additionally allows us a
physical interpretation of the biasing functions
\citep{2002PhR...372....1C}. Some time is therefore also spent on a
discussion of the scale-dependence of galaxy bias which will be
eminent in future applications of our technique.

The structure of this paper is as follows. In Sect. \ref{sect:data},
we describe the construction of data for a mock lensing survey to
which we apply our deprojection technique. With respect to number
densities of lens and source galaxies on the sky, the mock data are
similar to realistic galaxy samples in the Canada-France-Hawaii
Telescope Lensing Survey, CFHTLenS hereafter
\citep{2012MNRAS.427..146H}.  We increase the simulated survey area,
however, to $\sim1000\rm\deg^2$ in order to assess the quality of our
methodology for state-of-the-art (ground-based) surveys in future
applications. In Sect. \ref{sect:projectedbias}, we revise the
relation of the spatial biasing functions to their projected
counterparts which are observable through the aperture
statistics. This section also adds to the technique of
\cite{2002ApJ...577..604H} as novelty potentially relevant
higher-order corrections in the lensing formalism. It also
incorporates a treatment of the intrinsic alignment of sources into
the aperture statistics. Section \ref{sect:spatialbias} derives our
template models of the spatial biasing functions, applied for a
deprojection; Section \ref{sect:implementation} summarises the
template parameters and explores their impact on the scale dependence
of galaxy bias. The methodological details for the statistical
inference of $b(k)$ and $r(k)$ from noisy measurements are presented
in Sect. \ref{sect:statinference}. We apply this inference technique
to the mock data in the result Sect. \ref{sect:results} and assess its
accuracy, precision, and robustness. As a first demonstration, we
apply our technique to previous measurements in
\citet{2007A&A...461..861S}. We finally discuss our results in
Sect. \ref{sect:discussion}.

\section{Data}
\label{sect:data}

This section details our mock data, that is lens and source
catalogues, to which we apply our deprojection technique in the
following sections. A reader more interested in the method details for
the recovery of galaxy bias with lensing data could proceed to the
next sections.

\renewcommand{\arraystretch}{1.3}
\begin{table}
  \caption{\label{tab:samples} Selection criteria applied to our mock galaxies to emulate stellar-mass samples consistent with SES13 and for the two additional colour-selected samples RED and BLUE. The samples are further subdivided, as in SES13, into the two redshift bins low-$z$ ($\bar{z}\approx0.36$) and high-$z$ ($\bar{z}\approx0.52$) by a emulated selection in photometric redshift $z_{\rm p}$. The redshift distributions of all samples are summarised by Fig. \ref{fig:pofz}. The sample SOURCES is used as background sample for the mock lensing-analysis.}
  \begin{center}
    \begin{tabular}{ll}
      \hline\hline
      Galaxy Sample & Selection\tablefootmark{a}\\
      \hline\\
      SM1 & $0.5\le M_\ast<1$; $i^\prime<22.5$ \\
      SM2 & $1\le M_\ast<2$; $i^\prime<22.5$ \\
      SM3 & $2\le M_\ast<4$; $i^\prime<22.5$ \\
      SM4 & $4\le M_\ast<8$; $i^\prime<22.5$ \\
      SM5 & $8\le M_\ast<16$; $i^\prime<22.5$ \\
      SM6 & $16\le M_\ast<32$; $i^\prime<22.5$ \\
      RED & $u-r>1.93\,z+1.85$; $i^\prime<22.5$;\\
          & $0.5\le M_\ast<32$\\
      BLUE & $u-r\le1.93\,z+1.85$; $i^\prime<22.5$; \\
           & $0.5\le M_\ast<32$\\\\
      SOURCES & $i^\prime\le24.7$; $0.65\le z_{\rm p}<1.2$
    \end{tabular}
    \tablefoot{\tablefoottext{a}{$M_\ast$ refers to the stellar mass
        in units of $10^{10}\,\rm M_\odot$; $i^\prime,u,r$ are
        apparent magnitudes as defined for CFHTLenS
        \citep{2013MNRAS.433.2545E}; $z$ is the (cosmological) galaxy
        redshift; $z_{\rm p}$ is a photometric redshift with
        errors similar to CFHTLenS}}
  \end{center}
\end{table}
\renewcommand{\arraystretch}{1.0}

\begin{figure}
  \begin{center}
    \epsfig{file=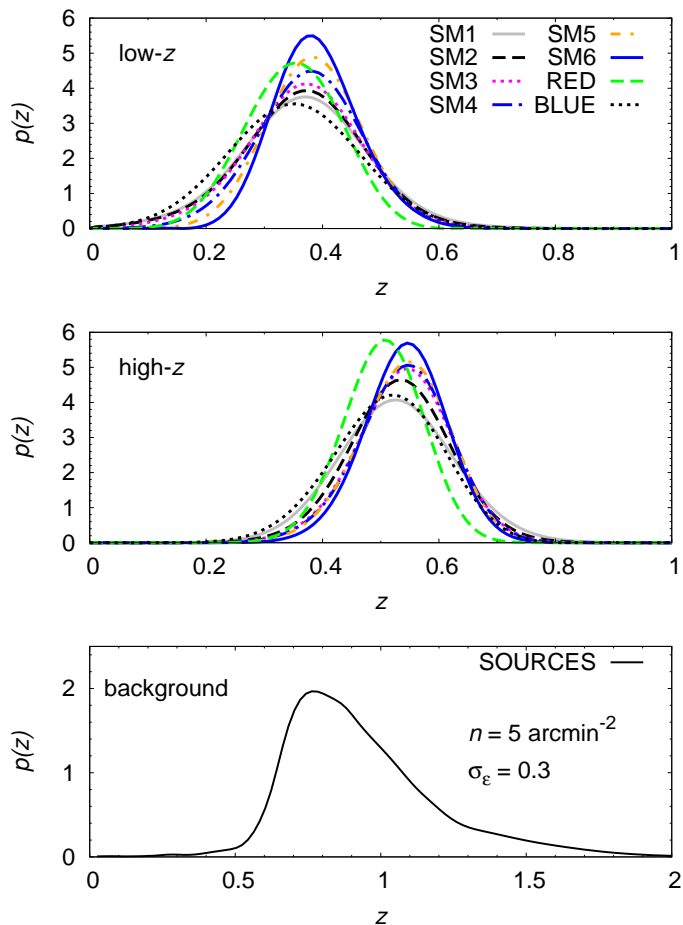,width=125mm,angle=-90}
  \end{center}
  \caption{\label{fig:pofz} Models of the probability densities
    $p_{\rm d}(z)$ of galaxy redshifts in our lens samples SM1 to SM6,
    RED and BLUE (two top panels), and the density $p_{\rm s}(z)$ of
    the source sample (bottom panel).}
\end{figure}

\subsection{Samples of lens galaxies}

Our galaxy samples use a semi-analytic model (SAM) according to H15
which is implemented on the Millennium Simulation
\citep{2006Natur.440.1137S}. These SAMs are the H15 mocks that are
also used in \cite{2016arXiv160808629S}. The Millennium Simulation
(MS) is a $N$-body simulation of the CDM density field inside a
comoving cubic volume of $500\,h^{-1}\,\rm Mpc$ side length, and it
has a spatial resolution of $5\,h^{-1}\,\rm kpc$ sampled by $10^{10}$
mass particles.  The fiducial cosmology of the MS has the density
parameters $\Omega_{\rm m}=0.25=1-\Omega_\Lambda$ and $\Omega_{\rm
  b}=0.045$, $\sigma_8=0.9$ for the normalisation of the linear matter
power-spectrum, a Hubble parameter $H_0=100\,h\,\rm
km\,s^{-1}\,Mpc^{-1}$ with $h=0.73$, and a spectral index for the
primordial matter-power spectrum of $n_{\rm spec}=1.0$. All density
parameters are in units of the critical density \mbox{$\bar{\rho}_{\rm
    crit}=3H_0^2/8\pi\,G_{\rm N}$} where $G_{\rm N}$ denotes Newton's
constant of gravity.  The galaxy mocks are constructed by populating
dark matter halos in the simulation based on the merger history of
halos and in accordance with the SAM details.  We project the
positions of the SAMs inside 64 independent light cones onto a
$4\times4\,\rm deg^2$ piece of sky. The resulting total survey area is
hence $1024\,\rm deg^2$.

We then select galaxies from the mocks to emulate the selection in
redshift and stellar mass in \citet{2013MNRAS.430.2476S}, SES13
henceforth. Details on the emulation process can be found in
\citet{2016arXiv160808629S}. We give only a brief summary here. The
mock-galaxy and source samples are constructed to be compatible with
those in recent lensing studies, dealing with data form the
Canada-France-Hawaii Telescope Survey, CFHTLenS hereafter
\citep{2016arXiv160808629S,2014MNRAS.437.2111V,2013MNRAS.433.2545E,2012MNRAS.427..146H}. Our
selection proceeds in two steps. First, we split the galaxy catalogues
in stellar mass, including emulated measurement errors, and
$i^\prime$-band brightness to produce the stellar-mass samples SM1 to
SM6; the photometry uses the AB-magnitude system. Second, we randomly
discard galaxies in each stellar-mass sample to obtain a redshift
distribution that is comparable to a given target distribution. As
targets, we employ the photometric redshift bins `low-$z$' and
`high-$z$' in SES13 which are the redshift distributions in CFHTLenS
after a cut in photometric redshift $z_{\rm p}$. The low-$z$ bin
applies \mbox{$0.2\le z_{\rm p}<0.44$}, and the high-$z$ bin applies
\mbox{$0.44\le z_{\rm p}<0.6$}. See Fig 5. in SES13 for the different
target distributions.  Our selection criteria for SM1 to SM6 are
listed in Table \ref{tab:samples}. We note here that randomly removing
galaxies at redshift $z$ adds shot noise but does not change the
matter-galaxy correlations and the (shot-noise corrected) galaxy
clustering.

In addition to SM1-6, we define two more samples RED and BLUE based on
the characteristic bimodal distribution of $u-r$ colours (Table
\ref{tab:samples}). Both samples initially consist of all galaxies in
SM1 to SM6 but are then split depending on the $u-r$ colours of
galaxies: the division is at $(u-r)(z)=1.93\,z+1.85$ which varies with
$z$ to account for the reddening with redshift. We crudely found
$(u-r)(z)$ by identifying by eye the mid-points $(u-r)_i$ between the
red and blue mode in $u-r$ histograms of
CFHTLenS\footnote{\url{http://cfhtlens.org}} SM1-6 galaxies in four
photometric-redshift bins with means $\{z_i\}=\{0.25,0.35,0.45,0.55\}$
and width $\Delta z=0.1$ \citep{2012MNRAS.421.2355H}. Then we fit a
straight line to the four empirical data points $\{(z_i,(u-r)_i)\}$
and obtain the above $(u-r)(z)$ as best-fit. For splitting the mocks,
we identify the precise redshifts $z$ in H15 with the photometric
redshifts $z_{\rm p}$ in CFHTLenS which, for the scope of this work,
is a sufficient approximation. Similar to the previous stellar-mass
samples, we combine the redshift posteriors of all CFHTLenS-galaxies
RED or BLUE to define the target distributions for our corresponding
mock samples.

For the following galaxy-bias analysis, we estimate the probability
density function (PDF) $p_{\rm d}(z)$ of each galaxy sample from the
mock catalogues in the foregoing step.  Simply using histograms of the
sample redshifts may seem like a good idea but are, in fact,
problematic because the histograms depend on the adopted binning. This
is especially relevant for the prediction of galaxy clustering which
depends on $p_{\rm d}^2(z)$ (see Eq. \ref{eq:pn}). Instead, we fit for
$p_{\rm d}(z)$ a smooth four-parameter Gram-Charlier series
\begin{equation}
  \label{eq:gramcharlier}
  p_{\rm d}(z|\lambda,\vec{\Theta})=
  \lambda\,
  \e^{-\frac{x^2}{2}}\,
  \left(
    1+\frac{s}{6}H_3(x)+\frac{k}{24}H_4(x)
  \right)
  ~;~
  x=\frac{z-\bar{z}}{\sigma_{\rm z}}
\end{equation}
with the Hermite polynomials $H_3(x)=x^3-3x$ and $H_4(x)=x^4-6x^2+3$
to a mock sample \mbox{$\{z_i:i=1\ldots n\}$} of $n$ galaxy redshifts;
$\lambda$ is a normalisation constant that depends on the parameter
combination \mbox{$\vec{\Theta}=(\bar{z},\sigma_{\rm z},s,k)$} and is
defined by
\begin{equation}
  \int_0^\infty\d z\,p_{\rm d}(z|\lambda,\vec{\Theta})=1\;.
\end{equation} 
For an estimate $\hat{\vec{\Theta}}$ of the parameters $\vec{\Theta}$,
we maximise the log-likelihood
\begin{equation}
  \ln{{\cal L}(\vec{\Theta})}=\sum_{i=1}^n\ln{p_{\rm d}(z_i|\lambda,\vec{\Theta})}
\end{equation}
with respect to $\vec{\Theta}$.  This procedure selects the PDF
$p_{\rm d}(z|\lambda,\hat{\vec{\Theta}})$ that is closest to the
sample distribution of redshifts $z_i$ in a Kullback-Leibler sense
\citep{knight1999mathematical}. The mean $\bar{z}$ and variance
$\sigma_z^2$ in the fit matches that of the redshift distribution in
the mock lens sample. The resulting densities for all our lens samples
are shown in the two top panels of Fig. \ref{fig:pofz}.

\subsection{Shear catalogues}

For mock source catalogues based on the MS data, we construct lensing
data by means of multiple-lens-plane ray-tracing as described in
\cite{2009A&A...499...31H}. The ray-tracing produces the lensing
convergence $\kappa(\vec{\theta}|z_{\rm s})$ and shear distortion
$\gamma(\vec{\theta}|z_{\rm s})$ for $4096^2$ line-of-sight directions
$\vec{\theta}$ on 64 regular angular grids and a sequence of $n_{\rm
  s}=31$ source redshifts $z_{{\rm s},i}$ between $z_{\rm s}=0$ and
$z_{\rm s}=2$; we denote by $\Delta z_i=z_{{\rm s},i+1}-z_{{\rm s},i}$
the difference between neighbouring source redshifts. Each grid covers
a solid angle of $\Omega=4\times4\,\rm deg^2$.  For each grid, we then
compute the average convergence for sources with redshift PDF $p_{\rm
  s}(z)$ by
\begin{equation}
  \kappa(\vec{\theta})=
  \frac{\sum_{i=1}^{n_{\rm s}}p_{\rm s}(z_{{\rm s},i})\,\Delta z_i
    \,\kappa(\vec{\theta}|z_{{\rm s}.i})}
  {\sum_{i=1}^{n_{\rm s}}p_{\rm s}(z_{{\rm s},i})\Delta z_i}\;,
\end{equation}
and the average shear $\gamma(\vec{\theta})$ from the sequence
$\gamma(\vec{\theta}|z_{\rm s})$ accordingly. For $p_{\rm s}(z)$, we
employ the estimated PDF of CFHTLenS sources that is selected through
\mbox{$i^\prime<24.7$} and \mbox{$0.65\le z_{\rm p}<1.2$}, weighted by
their shear-measurement error (SES13); see the bottom panel in
Fig. \ref{fig:pofz}. The mean redshift of sources is
\mbox{$\bar{z}\approx0.93$}. To assign source positions on the sky, we
uniform-randomly pick a sample $\{\vec{\theta}_i:i=1\ldots n\}$ of
positions for each grid; the amount of positions is
$n=\Omega\,\bar{n}_{\rm s}$ for a number density of $\bar{n}_{\rm
  s}=5\,\rm arcmin^{-2}$ sources which roughly equals the effective
number density of sources in SES13.

Depending on the type of our lensing analysis, we assign a source at
$\vec{\theta}_i$ one of the following three values for the simulated
sheared ellipticity $\epsilon_i$: (i)
\mbox{$\epsilon_i=\gamma(\vec{\theta}_i)$} for source without shape
noise; (ii) \mbox{$\epsilon_i=A(\gamma(\vec{\theta}_i),\epsilon_{\rm
    s})$} for noisy sources with shear; and (iii)
\mbox{$\epsilon_i=A\left(g_i,\epsilon_{\rm s}\right)$} for noisy
sources with reduced shear
\mbox{$g_i=\gamma(\vec{\theta}_i)/[1-\kappa(\vec{\theta}_i)]$}. We
define here by \mbox{$A(x,y):=(x+y)\,(1+xy^\ast)^{-1}$} the conformal
mapping of two complex numbers $x$ and $y$, and by $\epsilon_{\rm s}$
a random shape-noise drawn from a bivariate, truncated Gaussian PDF
with zero mean, 1D dispersion $\sigma_\epsilon=0.3$, and an exclusion
of values beyond \mbox{$|\epsilon_{\rm s}|\ge1$}.

\subsection{Power spectra}

We obtain the true spatial galaxy-galaxy, galaxy-matter, and
matter-matter power spectra for all galaxy samples at a given
simulation snapshot with Fast Fourier Transform (FFT) methods.  For a
choice of pair of tracers (i.e. simulation matter particles or
galaxies from different samples) in a snapshot, we compute a series of
raw power spectra by \lq{}chaining the power\rq{} \citep{Smith03}. We
cover the whole simulation volume as well as smaller subvolumes (by a
factor $4^3$ to $256^3$, into which the whole box is folded) by
regular meshes of $512^3$ points (providing a spatial resolution from
$\sim 1 \,h^{-1}\,\text{Mpc}$ for the coarsest mesh to
$\sim5\,h^{-1}\,\text{kpc}$ for the finest mesh). We project the
tracers onto these meshes using clouds-in-cells (CIC) assignment
\citep{HockneyEastwood_book}.

We FFT-transform the meshes, record their raw power spectra, apply a
shot-noise correction (except for cross-spectra), a deconvolution to
correct for the smoothing by the CIC assignment, and an iterative
alias correction \citep[similar to what is described
in][]{2005ApJ...620..559J}. From these power spectra, we discard small
scales beyond half their Nyquist frequency as well as large scales
that are already covered by a coarser mesh, and combine them into a
single power spectrum covering a range of scales from modes $\sim
0.01\,h\,\text{Mpc}^{-1}$ to modes $\sim 100\,h\,\text{Mpc}^{-1}$.

The composite power spectra are then used as input to estimate alias
corrections for the partial power spectra from the individual meshes
with different resolutions, and the process is repeated until
convergence. From the resulting power spectra, we then compute the
true biasing functions, Eq.~\Ref{eq:brdef}, which we compare to our
lensing-based reconstructions in Sect.~\ref{sect:results}.

\section{Projected biasing functions as observed with lensing
  techniques}
\label{sect:projectedbias}

The combination of suitable statistics for galaxy clustering,
galaxy-galaxy lensing, and cosmic-shear correlations on the sky allows
us to infer, without a concrete physical model, the $z$-averaged
spatial biasing-functions $b(k)$ and $r(k)$ as projections $b_{\rm
  2D}(\theta_{\rm ap})$ and $r_{\rm 2D}(\theta_{\rm ap})$ for varying
angular scales $\theta_{\rm ap}$. Later on, we forward-fit templates
of spatial biasing functions to these projected functions to perform a
stable deprojection. We summarise here the relation between $(b(k),
r(k))$ and the observable ratio-statistics $(b_{\rm 2D}(\theta_{\rm
  ap}),r_{\rm 2D}(\theta_{\rm ap}))$.  We include corrections to the
first-order Born approximation for galaxy-galaxy lensing and galaxy
clustering, and corrections for the intrinsic alignment of sources.

\subsection{Spatial biasing functions}

We define galaxy bias in terms of two biasing functions $b(k)$ and
$r(k)$ for a given spatial scale $2\pi\,k^{-1}$ or wave number $k$ in
the following way. Let $\delta(\vec{x})$ in
$\rho(\vec{x})=\overline{\rho}\,[1+\delta(\vec{x})]$ be the density
fluctuations at position $\vec{x}$ of a random density field
$\rho(\vec{x})$, and $\overline{\rho}$ denotes the mean density. A
density field is either the matter density $\rho_{\rm m}(\vec{x})$ or
the galaxy number density $n_{\rm g}(\vec{x})$ with density contrasts
$\delta_{\rm m}(\vec{x})$ and $\delta_{\rm g}(\vec{x})$,
respectively. We determine the fluctuation amplitude for a density
mode $\vec{k}$ by the Fourier transform of $\delta(\vec{x})$,
\begin{equation}
\tilde{\delta}(\vec{k})=
\int\d^3\!x\;\delta(\vec{x})\,\e^{-\i\vec{x}\cdot\vec{k}}\;.
\end{equation}
All information on the two-point correlations of
$\tilde{\delta}(\vec{k})$ is contained in the power spectrum $P(k)$
defined through the second-order correlation function of modes,
\begin{equation}
  \ave{\tilde{\delta}(\vec{k})\tilde{\delta}(\vec{k}^\prime)}
  =(2\pi)^3\delta_{\rm D}(\vec{k}+\vec{k}^\prime)P(k)\;,
\end{equation}
where $k=|\vec{k}|$ is the scalar wave-number and $\delta_{\rm
  D}(\vec{s})$ is the Dirac Delta distribution. Specifically, we
utilise three kinds of power spectra,
\begin{eqnarray}
  \ave{\tilde{\delta}_{\rm m}(\vec{k})\tilde{\delta}_{\rm m}(\vec{k}^\prime)}
  &=&(2\pi)^3\delta_{\rm D}(\vec{k}+\vec{k}^\prime)P_{\rm m}(k)\;;
  \\
  \ave{\tilde{\delta}_{\rm m}(\vec{k})\tilde{\delta}_{\rm g}(\vec{k}^\prime)}
  &=&(2\pi)^3\delta_{\rm D}(\vec{k}+\vec{k}^\prime)P_{\rm gm}(k)\;;
  \\
  \ave{\tilde{\delta}_{\rm g}(\vec{k})\tilde{\delta}_{\rm g}(\vec{k}^\prime)}
  &=&(2\pi)^3\delta_{\rm D}(\vec{k}+\vec{k}^\prime)
  \left(P_{\rm g}(k)+\bar{n}_{\rm g}^{-1}\right)\;,
\end{eqnarray}
namely the matter power spectrum $P_{\rm m}(k)$, the galaxy-matter
cross-power spectrum $P_{\rm gm}(k)$, and the galaxy power-spectrum
$P_{\rm g}(k)$. The latter subtracts the shot-noise $\bar{n}_{\rm
  g}^{-1}$ from the galaxy power spectrum by definition. In contrast
to the smooth matter density, the galaxy number-density is subject to
shot noise because it consists of a finite number of discrete points
that make up the number density field. Traditionally, the definition
of $P_{\rm g}(k)$ assumes a Poisson process for the shot noise in the
definition of $P_{\rm g}(k)$ \citep{peebles80}.

The biasing functions (of the second order) express galaxy bias in
terms of ratios of the foregoing power spectra,
\begin{equation}
 \label{eq:brdef}
  b(k):=
  \sqrt{\frac{P_{\rm g}(k)}{P_{\rm m}(k)}}~;~
  r(k):=
  \frac{P_{\rm gm}(k)}{\sqrt{P_{\rm g}(k)\,P_{\rm m}(k)}}\;.
\end{equation}
Galaxies that sample the matter density by a Poisson process have
\mbox{$b(k)=r(k)=1$} for all scales $k$ and are dubbed `unbiased'; for
\mbox{$b(k)>1$}, we find that galaxies cluster stronger than matter at
scale $k$, and vice versa for \mbox{$b(k)<1$}; a decorrelation of
\mbox{$r(k)\ne1$} indicates either stochastic bias, non-linear bias, a
sampling process that is non-Poisson, or combinations of these cases
\citep{1999ApJ...520...24D,2001MNRAS.321..439G}.

\subsection{Aperture statistics and galaxy-bias normalisation}
\label{sect:biassmoothing}

The projected biasing functions $b(k)$ and $r(k)$ are observable by
taking ratios of the (co-)variances of the aperture mass and aperture
number count of galaxies
\citep{1998A&A...334....1V,1998ApJ...498...43S}. To see this, let
$\kappa_{\rm g}(\vec{\theta})=N_{\rm
  g}(\vec{\theta})/\overline{N}_{\rm g}-1$ be the density contrast of
the number density of galaxies $N_{\rm g}(\vec{\theta})$ on the sky in
the direction $\vec{\theta}$, and $\overline{N}_{\rm g}=\ave{N_{\rm
    g}(\vec{\theta})}$ is their mean number density. We define the
aperture number count of $N_{\rm g}(\vec{\theta})$ for an angular
scale $\theta_{\rm ap}$ at position $\vec{\theta}$ by
\begin{equation}
  \label{eq:apcount}
  {\cal N}(\theta_{\rm ap};\vec{\theta})=
  \int\d^2\theta^\prime\;U(|\vec{\theta}^\prime|;\theta_{\rm ap})\,
  \kappa_{\rm g}(\vec{\theta}^\prime+\vec{\theta})\;,
\end{equation}
where 
\begin{equation}
  \label{eq:apfilter}
  U(\theta;\theta_{\rm ap})=
  \frac{1}{\theta_{\rm ap}^2}\,u(\theta\,\theta_{\rm ap}^{-1})~;~
  u(x)=\frac{9}{\pi}\,(1-x^2)\,\left(\frac{1}{3}-x^2\right)\,{\rm H}(1-x)
\end{equation}
is the aperture filter of the density field, and ${\rm H}(x)$ is the
Heaviside step function of our polynomial filter profile $u(x)$. The
aperture filter is compensated, that is $\int_0^\infty\d
x\;x\,u(x)=0$. Similarly for the (average) lensing convergence
$\kappa(\vec{\theta})$ of sources in direction $\vec{\theta}$, the
aperture mass is given by
\begin{equation}
  \label{eq:apmass}
  M_{\rm ap}(\theta_{\rm ap};\vec{\theta})=
  \int\d^2\theta^\prime\;U(|\vec{\theta}^\prime|;\theta_{\rm ap})\,
  \kappa(\vec{\theta}^\prime+\vec{\theta})\;.
\end{equation}
The aperture statistics consider the variances $\ave{{\cal
    N}^2}(\theta_{\rm ap})$ and $\ave{M_{\rm ap}^2}(\theta_{\rm ap})$
of ${\cal N}(\theta_{\rm ap};\vec{\theta})$ and $M_{\rm
  ap}(\theta_{\rm ap};\vec{\theta})$, respectively, across the sky as
well as their co-variance $\ave{{\cal N}M_{\rm ap}}(\theta_{\rm ap})$
at zero lag.

From these observable aperture statistics, we obtain the galaxy-bias
factor $b_{\rm 2D}(\theta_{\rm ap})$ and correlation factor $r_{\rm
  2D}(\theta_{\rm ap})$ through the ratios
\begin{eqnarray}
  \label{eq:b2dobs}
  b_{\rm 2D}(\theta_{\rm ap})&=&
  \sqrt{
    \frac{\ave{{\cal N}^2}(\theta_{\rm ap})}
    {\ave{M_{\rm ap}^2}(\theta_{\rm ap})}}\times f_{\rm b}(\theta_{\rm ap})\;,\\
  \label{eq:r2dobs}
  r_{\rm 2D}(\theta_{\rm ap})&=&
  \frac{\ave{{\cal N}M_{\rm ap}}(\theta_{\rm ap})}
  {\sqrt{\ave{{\cal N}^2}(\theta_{\rm ap})\,\ave{M_{\rm ap}^2}(\theta_{\rm
        ap})}}
  \times f_{\rm r}(\theta_{\rm ap})\;,
\end{eqnarray}
where 
\begin{eqnarray}
  \label{eq:calfb}
  f_{\rm b}(\theta_{\rm ap})&:=&
  \sqrt{\frac{\ave{M^2_{\rm ap}}_{\rm th}(\theta_{\rm ap})}{\ave{{\cal
          N}^2}_{\rm th}(\theta_{\rm ap};1)}}\;,
  \\
  \label{eq:calfr}
  f_{\rm r}(\theta_{\rm ap})&:=&
  \frac
  {\sqrt{\ave{M^2_{\rm ap}}_{\rm th}(\theta_{\rm ap})\,\ave{{\cal
          N}^2}_{\rm th}(\theta_{\rm ap};1)}}
  {\ave{{\cal N}M_{\rm ap}}_{\rm th}(\theta_{\rm ap};1)}
\end{eqnarray}
normalise the statistics according to a fiducial cosmology, that means
the aperture statistics with subscript `th' as in $\ave{M^2_{\rm
    ap}}_{\rm th}(\theta_j)$ denote the expected (co-)variance for a
fiducial model. The normalisation is chosen such that we have
\mbox{$b_{\rm 2D}(\theta_{\rm ap})=r_{\rm 2D}(\theta_{\rm ap})=1$} for
unbiased galaxies given the distributions of lenses and sources with
distance $\chi$ as in the survey, hence the `$(\theta_{\rm ap};1)$' in
the arguments of the normalisation.  The normalisation functions
$f_{\rm f}$ and $f_{\rm b}$ are typically weakly varying with angular
scale $\theta_{\rm ap}$ \citep{2002ApJ...577..604H}. In addition, they
depend weakly on the fiducial matter power spectrum \mbox{$P_{\rm
    m}(k;z)$}; they are even invariant with respect to an amplitude
change \mbox{$P_{\rm m}(k;z)\mapsto\upsilon\,P_{\rm m}(k;z)$} with
some number \mbox{$\upsilon>0$}. We explore the dependence on the
fiducial cosmology quantitatively in Sect. \ref{sect:calbias}.

For this study, we assume that the distance distribution of lenses is
sufficiently narrow, which means that the bias evolution in the lens
sample is negligible.  We therefore skip the argument $\chi$ in
$b(k;\chi)$ and $r(k;\chi)$, and we use a $b(k)$ and $r(k)$
independent of $\chi$ for average biasing functions instead.

The relation between $(b(k),r(k))$ and
$(b_{\rm 2D}(\theta_{\rm ap}),r_{\rm 2D}(\theta_{\rm ap}))$ is
discussed in the following. Let $p_{\rm d}(\chi)\,\d\chi$ and
$p_{\rm s}(\chi)\,\d\chi$ be the probability to find a lens or source
galaxy, respectively, at comoving distance $[\chi,\chi+\d\chi)$. The
matter power spectrum at distance $\chi$ shall be $P_{\rm m}(k;\chi)$,
and \mbox{$k_\ell^\chi:=(\ell+0.5)/f_K(\chi)$} is a shorthand for the
transverse spatial wave-number $k$ at distance $\chi$ that corresponds
to the angular wave-number $\ell$. The function $f_K(\chi)$ denotes
the comoving angular-diameter distance in the given fiducial
cosmological model. The additive constant 0.5 in $k_\ell^\chi$ applies
a correction to the standard Limber approximation on the flat sky
which gives more accurate results for large angular scales
\citep{2017arXiv170205301K,2008PhRvD..78l3506L}.  According to theory,
the aperture statistics are then
\begin{eqnarray}
  \label{eq:n2b}
  \ave{{\cal N}^2}_{\rm th}(\theta_{\rm ap};b)&=&
  2\pi\int\limits_0^\infty\d\ell\,\ell\,P_{\rm n}(\ell;b)\,\left[I(\ell\theta_{\rm
      ap})\right]^2\;,\\
    \label{eq:nmapbr}
    \ave{{\cal N}M_{\rm ap}}_{\rm th}(\theta_{\rm ap};b,r)&=&
  2\pi\int\limits_0^\infty\d\ell\,\ell\,P_{{\rm n}\kappa}(\ell;b,r)\,\left[I(\ell\theta_{\rm
    ap})\right]^2\;,\\
  \label{eq:mapsq}
    \ave{M_{\rm ap}^2}_{\rm th}(\theta_{\rm ap})&=&
  2\pi\int\limits_0^\infty\d\ell\,\ell\,P_\kappa(\ell)\,\left[I(\ell\theta_{\rm
    ap})\right]^2\;,
\end{eqnarray}
with the angular band-pass filter
\begin{equation}
  I(x):=\int\limits_0^\infty\d s\;s\,u(s)\,{\rm J}_0(s\,x)
  =\frac{12}{\pi}\frac{{\rm J}_4(x)}{x^2}\;,
\end{equation}
the angular power spectrum of the galaxy clustering
\begin{equation}
  \label{eq:pn}
  P_{\rm n}(\ell;b)=
  \int\limits_0^{\chi_{\rm h}}\frac{\d\chi\;p_{\rm d}^2(\chi)}{f_K^2(\chi)}\,
  b^2(k_\ell^\chi)\,P_{\rm m}\left(k_\ell^\chi;\chi\right)\;,
\end{equation}
the galaxy-convergence cross-power
\begin{multline}
  \label{eq:pnkappa}
  P_{{\rm n}\kappa}(\ell;b,r)=\\
  \frac{3H_0^2\,\Omega_{\rm m}}{2c^2}
  \int\limits_0^{\chi_{\rm h}}
  \frac{\d\chi\;p_{\rm d}(\chi)\,g_{\rm s}(\chi)}
  {a(\chi)\,f_K(\chi)}\,
  b(k_\ell^\chi)\,r(k_\ell^\chi)\,P_{\rm m}\left(k_\ell^\chi;\chi\right)\;,
\end{multline}
and the convergence power-spectrum
\begin{equation}
  \label{eq:pkappa}
  P_\kappa(\ell)=
  \frac{9H_0^4\,\Omega_{\rm m}^2}{4c^4}
  \int\limits_0^{\chi_{\rm h}}
  \frac{\d\chi\,\;g_{\rm s}^2(\chi)}{a^2(\chi)}\,
  P_{\rm m}\left(k_\ell^\chi;\chi\right)\;,
\end{equation}
all in the Born and Limber approximation. In the integrals, we use the
lensing kernel
\begin{equation}
  g_{\rm s}(\chi)=
  \int\limits_\chi^{\chi_{\rm h}}\d\chi^\prime\;p_{\rm s}(\chi^\prime)\,\frac{f_K(\chi^\prime-\chi)}{f_K(\chi^\prime)}\;,
\end{equation}
the scale factor $a(\chi)$ at distance $\chi$, the maximum distance
$\chi_{\rm h}$ of a source, and the $n$th-order Bessel function ${\rm
  J}_n(x)$ of the first kind. By $c$ we denote the vacuum speed of
light. The power spectra and aperture statistics depend on specific
biasing functions as indicated by the $b$ and $r$ in the arguments.
For given biasing functions $b(k)$ and $r(k)$, we obtain the
normalised galaxy bias inside apertures therefore through
\begin{eqnarray}
  \label{eq:b2d}
  b_{\rm 2D}(\theta_{\rm ap};b)&=&
  \sqrt{
    \frac{\ave{{\cal N}^2}_{\rm th}(\theta_{\rm ap};b)}
    {\ave{{\cal N}^2}_{\rm th}(\theta_{\rm ap};1)}}\;,\\
  \label{eq:r2d}
  r_{\rm 2D}(\theta_{\rm ap};b,r)&=&
  \frac{1}{b_{\rm 2D}(\theta_{\rm ap};b)}
  \,
    \frac{\ave{{\cal N}M_{\rm ap}}_{\rm th}(\theta_{\rm ap};b,r)}
    {\ave{{\cal N}M_{\rm ap}}_{\rm th}(\theta_{\rm ap};1)}\;,
\end{eqnarray}
which can be compared to measurements of the Eqs. \Ref{eq:b2dobs} and
\Ref{eq:r2dobs}.

\subsection{Intrinsic alignment of sources}
\label{sect:IIandGI}

Recent studies of cosmic shear find evidence for an alignment of
intrinsic source-ellipticities that contribute to the
shear-correlation functions
\citep{2017MNRAS.465.1454H,2016PhRvD..94b2001A,2015PhR...558....1T,2013MNRAS.432.2433H,2011A&A...527A..26J,2006MNRAS.367..611M}.
These contributions produce systematic errors in the reconstruction of
$b(k)$ and $r(k)$ if not included in their normalisation $f_{\rm b}$
and $f_{\rm r}$. Relevant are `II'-correlations between intrinsic
shapes of sources in $\ave{M^2_{\rm ap}}$ and `GI'-correlations
between shear and intrinsic shapes in both $\ave{{\cal N}M_{\rm ap}}$
and $\ave{M^2_{\rm ap}}$. The GI term in $\ave{{\cal N}M_{\rm ap}}$
can be suppressed by minimising the redshift overlap between lenses
and sources. Likewise, the II term is suppressed by a broad redshift
distribution of sources which, however, increases the GI
amplitude. The amplitudes of II and GI also vary with galaxy type and
luminosity of the sources \citep{2011A&A...527A..26J}.

An intrinsic alignment (IA) of sources has an impact on the ratio
statistics $b_{\rm 2D}(\theta_{\rm ap})$ and $r_{\rm 2D}(\theta_{\rm
  ap})$, Eqs. \Ref{eq:b2dobs} and \Ref{eq:r2dobs}, mainly through
$\ave{M^2_{\rm ap}}(\theta_{\rm ap})$ if we separate sources and
lenses in redshift. The impact can be mitigated by using an
appropriate model for $\ave{M^2_{\rm ap}}_{\rm th}(\theta_{\rm ap})$
and $\ave{{\cal N}M_{\rm ap}}_{\rm th}(\theta_{\rm ap})$ in the
normalisation of the measurements. For this study, we do not include
II or GI correlations in our synthetic mock data but, instead, predict
the amplitude of potential systematic errors when ignoring the
intrinsic alignment for future applications in
Sect. \ref{sect:calbias}.

For a reasonable prediction of the GI and II contributions to
$\ave{M^2_{\rm ap}}_{\rm th}(\theta_{\rm ap})$, we use the recent
non-evolution model utilised in \cite{2017MNRAS.465.1454H}. This model
is implemented by using
\begin{equation}
  \label{eq:giii}
  P_\kappa^\prime(\ell)=
  P_\kappa(\ell)+P_\kappa^{\rm II}(\ell)+P_\kappa^{\rm GI}(\ell)
\end{equation}
instead of \Ref{eq:pkappa} in Eq. \Ref{eq:mapsq}. The new II and GI
terms are given by
\begin{eqnarray}
  P_\kappa^{\rm II}(\ell)&=&
  \int_0^{\chi_{\rm h}}\frac{\d\chi\;p^2_{\rm s}(\chi)}{f^2_K(\chi)}
  \,F_{\rm ia}^2(\chi)\,P_{\rm m}(k^\chi_\ell;\chi)\;;
  \\
  P_\kappa^{\rm GI}(\ell)&=&
  \frac{3H_0^2\,\Omega_{\rm m}}{c^2}\,
  \int_0^{\chi_{\rm h}}\frac{\d\chi\;p_{\rm s}(\chi)\,g_{\rm s}(\chi)}
  {a(\chi)\,f_K(\chi)}
  \,F_{\rm ia}(\chi)\,P_{\rm m}(k^\chi_\ell;\chi)\;,
\end{eqnarray}
where
\begin{multline}
  F_{\rm ia}(\chi):=
  -A_{\rm ia}\,{\rm C}_1\,\bar{\rho}_{\rm crit}\,\frac{\Omega_{\rm m}}{D_+(\chi)}
  \\
  \approx-2.4\times10^{-2}
  \left(\frac{A_{\rm ia}}{3.0}\right)\,
  \left(\frac{\Omega_{\rm m}}{0.3}\right)\,
  \left(\frac{D_+(\chi)}{0.5}\right)^{-1}
\end{multline}
controls the correlation amplitude in the so-called `non-linear
linear' model; see \citet{2004PhRvD..70f3526H},
\citet{2007NJPh....9..444B}, or \cite{2011A&A...527A..26J} for
details. The factor $A_{\rm ia}$ scales the amplitude; it broadly
falls within $A_{\rm ia}\in[-3,3]$ for recent cosmic-shear surveys and
is consistent with \mbox{$A_{\rm ia}\approx2$} for sources in the
Kilo-Degree Survey
\citep{2017arXiv170706627J,2017MNRAS.465.1454H,2013MNRAS.432.2433H}.
For the normalisation of $F_{\rm ia}(\chi)$, we use ${\rm
  C}_1=5\times10^{-14}\,h^{-2}\,{\rm M}_\odot^{-1}\,{\rm Mpc}^3$, and
the linear structure-growth factor $D_+(\chi)$, normalised to unity
for \mbox{$\chi=0$} \citep{peebles80}. By comparing $P_\kappa^{\rm
  II}(\ell)$ and $P_{\rm n}(\ell)$ in Eq. \Ref{eq:pn} we see that II
contributions are essentially the clustering of sources on the sky
(times a small factor). Likewise, $P_\kappa^{\rm GI}(\ell)$ is
essentially the correlation between source positions and their shear
on the sky, cf. Eq. \Ref{eq:pnkappa}. In this IA model, we assume a
scale-independent galaxy bias for sources in the IA modelling since
$F_{\rm ia}(\chi)$ does not depend on $k$.

\begin{figure}
  \begin{center}
    \epsfig{file=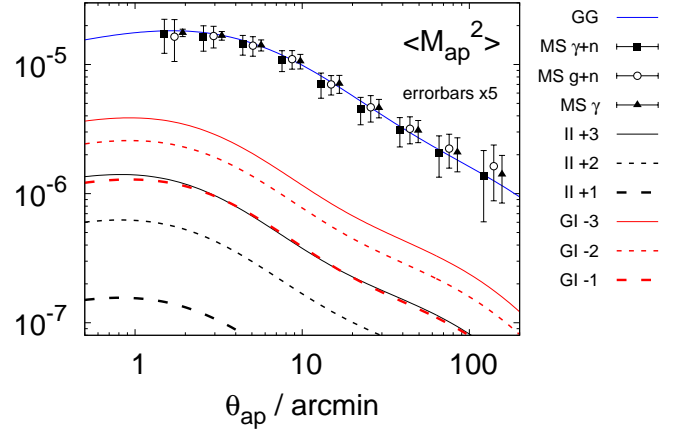,width=65mm,angle=-90}
  \end{center}
  \caption{\label{fig:GIandII} Levels of GI and II contributions to
    $\ave{M^2_{\rm ap}}$ for different values of $A_{\rm ia}$ (red and
    black lines labelled `II $\pm A_{\rm ia}$' and `GI $\pm A_{\rm
      ia}$').  The line `GG' is the theoretical $\ave{M^2_{\rm ap}}$
    without GI and II terms; the data points are measurements on the
    mocks for sources with shear and shape noise (MS $\gamma$+n),
    reduced shear and shape noise (MS $g$+n), and shear without shape
    noise (MS $\gamma$). The error bars indicate jackknife errors
    inflated by a factor of five for clarity (Appendix
    \ref{sect:estimators}).}
\end{figure}

\begin{figure}
  \begin{center}
    \epsfig{file=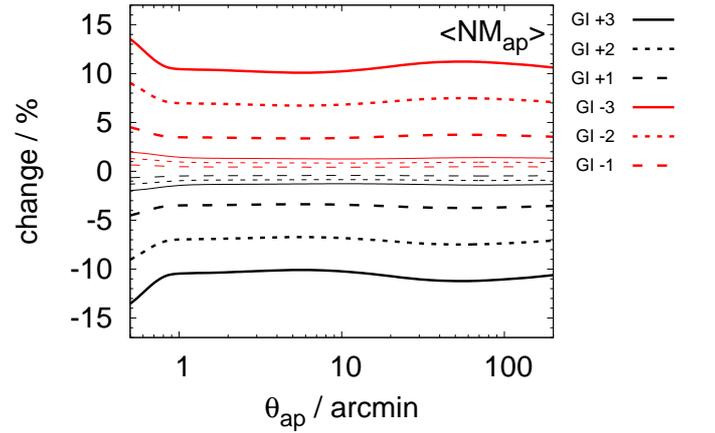,width=65mm,angle=-90}
  \end{center}
  \caption{\label{fig:nmapGI} Relative change of $\ave{{\cal N}M_{\rm
        ap}}$ for present GI correlations with different amplitudes
    $A_{\rm ia}$ as indicated by `GI $\pm A_{\rm ia}$'. The figure uses
    SM4 as fiducial lens-sample; the results for other samples are
    similar. The thin lines within $\pm2\%$ are for the low-$z$
    sample, and the thick lines are for the high-$z$ sample.}
\end{figure}

In Fig. \ref{fig:GIandII}, we plot the predicted levels of II and GI
terms in the observed $\ave{M_{\rm ap}^2}$ for varying values of
$A_{\rm ia}$ as black and red lines for our MS cosmology and the
$p_{\rm s}(z)$ in our mock survey. The corresponding value of $A_{\rm
  ia}$ is shown as number in the figure key. We use negative values of
$A_{\rm ia}$ for GI to produce positive correlations for the plot; the
corresponding predictions for $-A_{\rm ia}$ have the same amplitude as
those for $A_{\rm ia}$ but with opposite sign. II terms, on the other
hand, are invariant with respect to a sign flip of $A_{\rm ia}$. All
curves in the plot use a matter power spectrum $P_{\rm m}(k;\chi)$
computed with \texttt{Halofit} \citep{Smith03} and the update in
\citet{2012ApJ...761..152T}. For comparison, we plot as blue line GG
the theoretical $\ave{M^2_{\rm ap}}$ without GI and II terms. For
\mbox{$|A_{\rm ia}|\approx3$}, GI terms can reach levels up to 10 to
20 per cent of the shear-shear correlation signal for $\theta_{\rm
  ap}\gtrsim1^\prime$, whereas II terms are typically below 10 per
cent. GI and II terms partly cancel each other for \mbox{$A_{\rm
    ia}>0$} so that the contamination is worse for negative $A_{\rm
  ia}$.

\begin{figure*}
  \begin{center}
    \epsfig{file=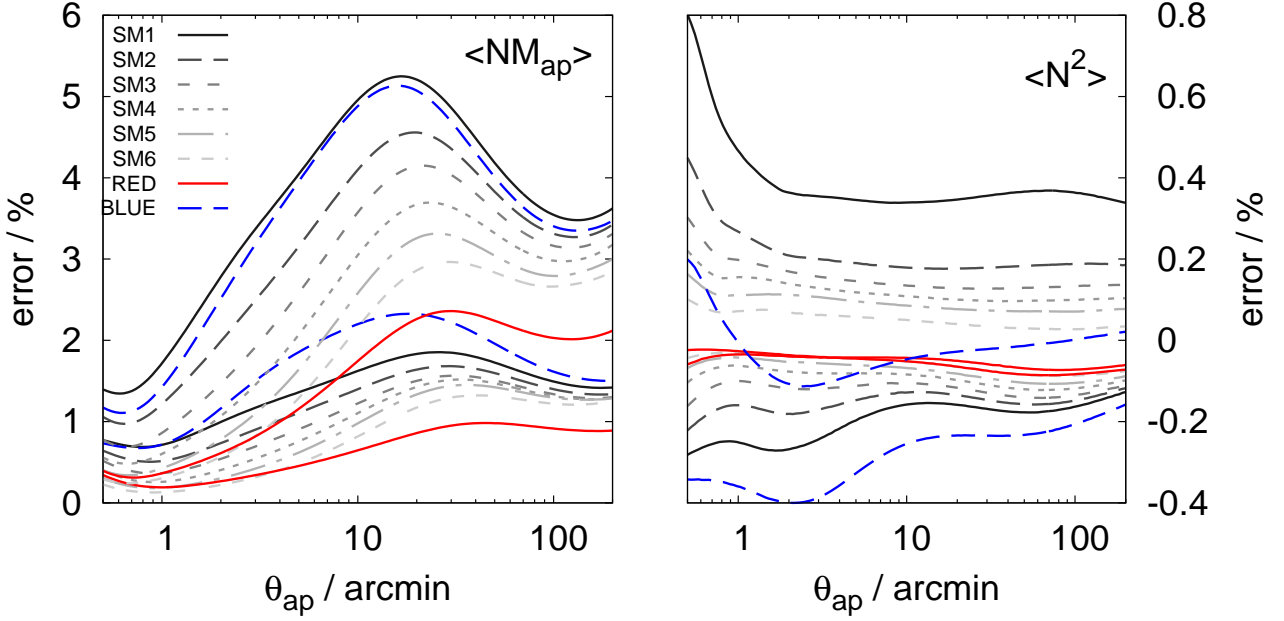,width=90mm,angle=-90}
    \vspace{-0.5cm}
  \end{center}
  \caption{\label{fig:higherorder} Relative errors in the aperture
    statistics due to magnification bias of the lenses. \emph{Left}:
    Errors for $\ave{{\cal N}M_{\rm ap}}$ where different line styles
    distinguish the galaxy samples. Larger errors for the same sample
    correspond to the high-$z$ bin, smaller errors to
    low-$z$. \emph{Right}: Percentage errors for $\ave{{\cal N}^2}$
    where larger errors for the same line style are the high-$z$
    bias.}
\end{figure*}

Moreover, we quantify the GI term in $\ave{{\cal N}M_{\rm ap}}$ by
using in \Ref{eq:nmapbr} the modified power spectrum
\begin{equation}
P_{{\rm n}\kappa}^\prime(\ell;b,r)=
P_{{\rm n}\kappa}(\ell;b,r)+P^{\rm GI}_{{\rm n}\kappa}(\ell;b,r)
\end{equation}
with
\begin{equation}
  P^{\rm GI}_{{\rm n}\kappa}(\ell;b,r)=
  \int_0^{\chi_{\rm h}}\frac{\d\chi\;p_{\rm s}(\chi)\,p_{\rm d}(\chi)}{f^2_K(\chi)}\,
  b(k_\ell^\chi)\,r(k_\ell^\chi)\,F_{\rm ia}(\chi)\,P_{\rm m}(k_\ell^\chi;\chi)\;.
\end{equation}
This is the model in \cite{2017arXiv170706627J}, see their Equation
(11), with an additional term $r(k_\ell^\chi)$ that accounts for a
decorrelation of the lens galaxies. This GI model is essentially the
relative clustering between lenses and unbiased sources on the sky and
therefore vanishes in the absence of an overlap between the lens and
source distributions, which means that \mbox{$\int\d\chi\;p_{\rm
    s}(\chi)\,p_{\rm d}(\chi)=0$}. In Fig. \ref{fig:nmapGI}, we
quantify the relative change in $\ave{{\cal N}M_{\rm ap}}$ owing to
the GI term for different values of $A_{\rm ia}$. Since the change is
very similar for all galaxy samples in the same redshift bin, we plot
only the results for SM4. The overlap between sources and lenses is
only around 4 per cent for low-$z$ samples and, therefore, the change
stays within 2 per cent for all angular scales considered here
(SES13). On the other hand, for high-$z$ samples where we have roughly
14 per cent overlap between the distributions, the change can amount
to almost 10 per cent for \mbox{$A_{\rm ia}\approx\pm2$} and could
have a significant impact on the normalisation.

\subsection{Higher-order corrections}
\label{sect:higherorder}

Corrections to the (first-order) Born approximation or for the
magnification of the lenses cannot always be neglected as done in
Eq. \Ref{eq:pnkappa}
\citep[e.g.,][]{2008PhRvD..78l3517Z,2009A&A...499...31H,2009PhDThesisHartlap}.
This uncorrected equation over-predicts the power spectrum $P_{\rm
  n\kappa}(\ell)$ by up to 10\% depending on the galaxy selection and
the mean redshift of the lens sample; the effect is smaller in a
flux-limited survey but also more elaborate to predict as it depends
on the luminosity function of the
lenses. \citet[][]{2009A&A...499...31H} tests this for the tangential
shear around lenses by comparing \Ref{eq:pnkappa} to the
full-ray-tracing results in the MS data which account for
contributions from lens-lens couplings and the magnification of the
angular number density of lenses.

For a volume-limited lens sample, \citet{2009PhDThesisHartlap}, H09
hereafter, derives the second-order correction (in our notation)
\begin{equation}
P_{{\rm
    n}\kappa}^{(2)}(\ell)=
-\frac{9H_0^4\,\Omega_{\rm m}^2}{2c^4}
\int_0^{\chi_{\rm h}}\d\chi\;
\frac{g_{\rm s}(\chi)\,g_{\rm d}(\chi)}{a^2(\chi)}\,
P_{\rm m}(k^\chi_\ell;\chi)\;,
\end{equation}
where
\begin{equation}
  g_{\rm d}(\chi)=
  \int_\chi^{\chi_{\rm h}}\d\chi^\prime\;p_{\rm d}(\chi^\prime)\,
  \frac{f_K(\chi^\prime-\chi)}{f_K(\chi^\prime)}\;,
\end{equation}
for a more accurate power spectrum $P^\prime_{{\rm
    n}\kappa}(\ell)=P_{{\rm n}\kappa}(\ell)+P_{{\rm
    n}\kappa}^{(2)}(\ell)$ that correctly describes the correlations
in the MS.  Physically, this correction accounts for the magnification
of the projected number density of lens galaxies by matter in the
foreground. We find that the thereby corrected $\ave{{\cal N}M_{\rm
    ap}}_{\rm th}(\theta_{\rm ap};b,r)$ can be different to the
uncorrected aperture statistic by up to a few per cent, see left-hand
panel in Fig. \ref{fig:higherorder}. This directly affects the
normalisation of $r_{\rm 2D}(\theta_{\rm ap})$: the measured,
normalised correlation $r_{\rm 2D}(\theta_{\rm ap})$ would be
systematically low. We obtain Fig. \ref{fig:higherorder} by comparing
the uncorrected to the corrected $\ave{{\cal N}M_{\rm ap}}_{\rm th}$
for each of our lens-galaxy samples. In accordance with H09, we find
that the systematic error is not negligible for some lens sample, and
we therefore include this correction by employing $P^\prime_{{\rm
    n}\kappa}(\ell)$ instead of $P_{{\rm n}\kappa}(\ell)$ in the
normalisation $f_{\rm r}(\theta_{\rm ap})$ and in the prediction
$r_{\rm 2D}(\theta_{\rm ap};b,r)$. This improves the accuracy of the
lensing reconstruction of $r(k)$ by up to a few per cent, most notably
the sample blue high-$z$, especially around \mbox{$k\approx1\,h\,\rm
  Mpc^{-1}$} which corresponds to \mbox{$\theta_{\rm
    ap}\approx10^\prime$}.

Additional second-order terms for $P_{{\rm n}\kappa}(\ell)$ arise due
to a flux limit of the survey (Equations 3.129 and 3.130 in H09), but
they require a detailed model of the luminosity function for the
lenses. We ignore these contributions here because our mock lens
samples, selected in redshift bins of $\Delta z\approx0.2$ and for
stellar masses greater than $5\times10^9\,{\rm M}_\odot$, are
approximately volume limited because of the lower limit of stellar
masses and the redshift binning \citep[see Sect. 4.1 in][which use our
lens samples]{2017arXiv171009902S}.

Similarly, by 
\begin{multline}
  P^{(2)}_{\rm n}(\ell;b,r)=
 \frac{9H_0^4\,\Omega_{\rm m}^2}{c^4}
  \int_0^{\chi_{\rm h}}\d\chi\;
  \frac{g_{\rm d}^2(\chi)}{a^2(\chi)}\,
  P_{\rm m}(k^\chi_\ell;\chi)
  \\
  -\frac{6H_0^2\,\Omega_{\rm m}}{c^2}
  \int_0^{\chi_{\rm h}}\d\chi\;
  \frac{p_{\rm d}(\chi)\,g_{\rm d}(\chi)}{a(\chi)\,f_K(\chi)}\,
  b(k^\chi_\ell)\,r(k^\chi_\ell)\,
  P_{\rm m}(k^\chi_\ell;\chi)
\end{multline}
H09 gives a second-order correction for $P_{\rm n}(\ell;b)$ in
addition to more corrections for flux-limited surveys (Equations
3.140-3.143). We include $P_{\rm n}^{(2)}(\ell;b,r)$ by using $P_{\rm
  n}^\prime(\ell;b,r)=P_{\rm n}(\ell;b)+P_{\rm n}^{(2)}(\ell;b,r)$
instead of Eq. \Ref{eq:pn} in the following for $f_{\rm b}(\theta_{\rm
  ap})$ and $b_{\rm 2D}(\theta_{\rm ap};b,r)$, although this
correction is typically below half a per cent here; see the right-hand
panel in Fig. \ref{fig:higherorder}.

\section{Model templates of biasing functions}
\label{sect:spatialbias}

Apart from the galaxy-bias normalisation, the ratio statistics $b_{\rm
  2D}$ and $r_{\rm 2D}$ are model-free observables of the spatial
biasing functions, averaged for the radial distribution of lenses. The
deprojection of the ratio statistics into (an average) $b(k)$ and
$r(k)$ is not straightforward due to the radial and transverse
smoothing in the projection. Therefore, for a deprojection we
construct a parametric family of templates that we forward-fit to the
ratio statistics. In principle, this family could be any generic
function but we find that physical templates that can be extrapolated
to scales unconstrained by the observations result in a more stable
deprojection. To this end, we pick a template prescription that is
motivated by the halo-model approach but with more freedom that is
commonly devised \citep[][for a review]{2002PhR...372....1C}. Notably,
we derive explicit expressions for $b(k)$ and $r(k)$ in a halo-model
framework.

\subsection{Separation of small and large scales}

Before we outline the details of our version of a halo model, used to
construct model templates, we point out that any halo model splits the
power spectra $P_{\rm m}(k)$, $P_{\rm gm}(k)$, and $P_{\rm g}(k)$ into
one- and two halo terms,
\begin{equation}
  P(k)=P^{\rm 1h}(k)+P^{\rm 2h}(k)\;.
\end{equation}
The one-halo term $P^{\rm 1h}(k)$ dominates at small scales,
quantifying the correlations between density fluctuations within the
same halo, whereas the two-halo term $P^{\rm 2h}(k)$ dominates the
power spectrum at large scales where correlations between fluctuations
in different halos and the clustering of halos become dominant.

We exploit this split to distinguish between galaxy bias on small
scales (one-halo terms) and galaxy bias on large scales (two-halo
terms), namely
\begin{equation}
  b^{\rm 1h}(k):=
  \sqrt{\frac{P_{\rm g}^{\rm 1h}(k)}{P_{\rm m}^{\rm 1h}(k)}}
  ~;~
  b^{\rm 2h}(k):=
  \sqrt{\frac{P_{\rm g}^{\rm 2h}(k)}{P_{\rm m}^{\rm 2h}(k)}}
\end{equation}
and
\begin{equation}
  \label{eq:r12h}
  r^{\rm 1h}(k):=
  \frac{P_{\rm gm}^{\rm 1h}(k)}{\sqrt{P_{\rm g}^{\rm 1h}(k)\,P_{\rm m}^{\rm 1h}(k)}}
  ~;~
  r^{\rm 2h}(k):=
  \frac{P_{\rm gm}^{\rm 2h}(k)}{\sqrt{P_{\rm g}^{\rm 2h}(k)\,P_{\rm m}^{\rm 2h}(k)}}\;,
\end{equation}
and we derive approximations for both regimes separately. We will find
that the two-halo biasing functions are essentially constants, and the
one-halo biasing functions are only determined by the relation between
matter and galaxy density inside halos.

\begin{figure}
  \begin{center}
    \epsfig{file=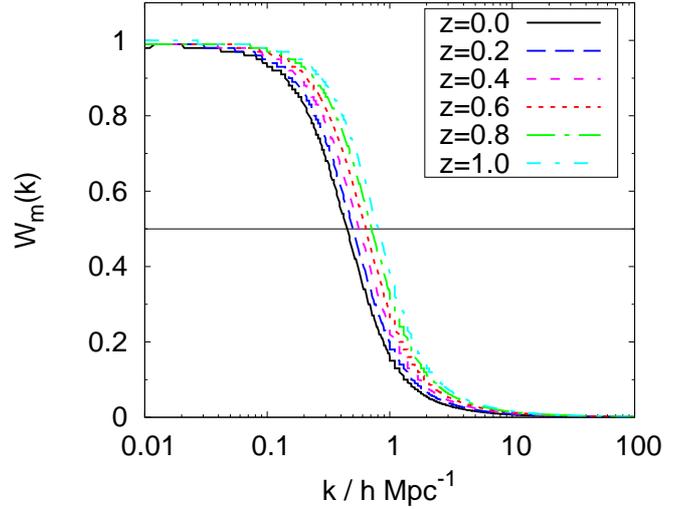,width=75mm,angle=-90}
  \end{center}
  \caption{\label{fig:wm} The weight $W_{\rm m}(k)$ of the two-halo
    term in the matter-power spectrum for varying redshifts $z$.}
\end{figure}

To patch together both approximations of the biasing functions in the
one-halo and two-halo regime, we then do the following.  Based on
Eq. \Ref{eq:brdef}, the function $b^2(k)$ is a weighted mean of
$b^{\rm 1h}(k)$ and $b^{\rm 2h}(k)$:
\begin{eqnarray}
  \nonumber
  b^2(k)&=&
  \frac{P^{\rm 1h}_{\rm g}(k)+P^{\rm 2h}_{\rm g}(k)}
  {P_{\rm m}(k)}
  \\
  \nonumber&=&
  \frac{P^{\rm 1h}_{\rm m}(k)\,[b^{\rm 1h}(k)]^2}
  {P_{\rm m}(k)}
  +
  \frac{P^{\rm 2h}_{\rm m}(k)\,[b^{\rm 2h}(k)]^2  }
  {P_{\rm m}(k)}
  \\
  \label{eq:bkgeneral} 
  &=&
  \Big(1-W_{\rm m}(k)\Big)\,[b^{\rm 1h}(k)]^2+W_{\rm m}(k)\,[b^{\rm
    2h}(k)]^2\;,
\end{eqnarray}
where the weight
\begin{equation}
  \label{eq:Wm}
  W_{\rm m}(k):=
  \frac{P^{\rm 2h}_{\rm m}(k)}{P_{\rm m}(k)}
\end{equation}
is the amplitude of the two-halo matter power spectrum relative to the
total matter power spectrum. Deep in the one-halo regime we have
\mbox{$W_{\rm m}(k)\approx0$} but \mbox{$W_{\rm m}(k)\approx1$} in the
two-halo regime.  Since the two-halo biasing is approximately
constant, the scale-dependence of galaxy bias is mainly a result of
the galaxy physics inside halos and the shape of $W_{\rm m}(k)$.

Once the weight $W_{\rm m}(k)$ is determined for a fiducial cosmology,
it does not rely on galaxy physics, we can use it for any model of
$b^{\rm 1h}(k)$ and $b^{\rm 2h}(k)$. In principle, the weight $W_{\rm
  m}(k)$ could be accurately measured from a cosmological simulation
by correlating only the matter density from different halos for
$P^{\rm 2h}_{\rm m}(k)$ which is then normalised by the full power
spectrum $P_{\rm m}(k)$ in the simulation.  We, however, determine
$W_{\rm m}(k)$ by computing the one-halo and two-halo term of $P_{\rm
  m}(k)$ with the setup of \citet{2009MNRAS.398..807S}. Our results
for $W_{\rm m}(k)$ at different redshifts are plotted in
Fig. \ref{fig:wm}. There we find that the transition between the
one-halo and two-halo regime, \mbox{$W_{\rm m}\sim0.5$}, is at
\mbox{$k\sim0.3\,h\,\rm Mpc^{-1}$} for \mbox{$z=0$}, whereas the
transition point moves to \mbox{$k\sim1\,h\,\rm Mpc^{-1}$} for
\mbox{$z\sim1$}.

Similar to $b(k)$, we can expand the correlation function $r(k)$ in
terms of its one-halo and two-halo biasing functions. To this end, let
\begin{equation}
  W_{\rm  g}(k):=
  \frac{P^{\rm 2h}_{\rm g}(k)}{P_{\rm g}(k)}
  =
  \left(\frac{b^{\rm 2h}(k)}{b(k)}\right)^2\,W_{\rm m}(k)
\end{equation}
be a weight by analogy to $W_{\rm m}(k)$. For unbiased galaxies, that
is \mbox{$b^{\rm 2h}(k)=b(k)=1$}, we simply have \mbox{$W_{\rm
    g}(k)=W_{\rm m}(k)$}. Using the definition of $r(k)$ in
Eq. \Ref{eq:brdef} and Eq. \Ref{eq:r12h}, we generally find
\begin{multline}
  \label{eq:rkgeneral}
  r(k)=\\
  \sqrt{(1-W_{\rm m}(k))(1-W_{\rm g}(k))}\,r^{\rm 1h}(k)+
  \sqrt{W_{\rm m}(k)W_{\rm g}(k)}\,
  r^{\rm 2h}(k)\;.
\end{multline}

\subsection{Halo-model definitions}

For approximations of the biasing functions in the one- and two-halo
regime, we apply the formalism in \citet{2000MNRAS.318..203S} and
briefly summarise it here. All halo-related quantities depend on
redshift. In the fits with the model later on, we use for this the
mean redshift of the lens galaxies.

We shall denote by $n(m)\,\d m$ the (comoving) number density of halos
within the halo-mass range \mbox{$[m,m+\d m)$}; \mbox{$\ave{N|m}$} is
the mean number of galaxies inside a halo of mass $m$;
\mbox{$\ave{N(N-1)|m}$} is the mean number of galaxy pairs inside a
halo of mass $m$. Let $u(r,m)$ be the radial profile of the matter
density inside a halo or the galaxy density-profile. Also let
\begin{equation}
  \tilde{u}(k,m)= 
  \frac{\int_0^\infty\d s\;sk^{-1}\,u(s,m)
    \sin{(ks)}} {\int_0^\infty\d s\;s^2\,u(s,m)}
\end{equation}
be its normalised Fourier transform. Owing to this normalisation,
profiles obey \mbox{$\tilde{u}(k,m)=1$} at \mbox{$k=0$}. To assert a
well-defined normalisation of halos, we truncate them at their virial
radius $r_{\rm vir}$, which we define by the over-density $\Omega_{\rm
  m}\,\bar{\rho}_{\rm crit}\,\Delta_{\rm vir}(z)$ within the distance
$r_{\rm vir}$ from the halo centre and by $\Delta_{\rm vir}(z)$ as in
\cite{2001MNRAS.321..559B}.  Furthermore, the mean matter and galaxy
number density (comoving) are
\begin{equation}
  \label{eq:rhong}
  \bar{\rho}_{\rm m}=\int\d m\;n(m)\,m
  ~;~
  \overline{n}_{\rm g}=\int\d m\;n(m)\,\ave{N|m}\;.
\end{equation}
The one-halo terms of the galaxy power spectrum $P_{\rm g}(k)$, the
matter power-spectrum $P_{\rm m}(k)$, and the galaxy-matter
cross-power spectrum $P_{\rm gm}(k)$ are
\begin{eqnarray}
  \label{eq:pg}
  P^{\rm 1h}_{\rm g}(k)&=&
  \int_0^\infty\frac{\d m\;n(m)}{\overline{n}_{\rm g}^2}\,
  \tilde{u}^{2p}_{\rm g}(k,m)\,
  \ave{N(N-1)|m}\;;\\
  \label{eq:pm}
  P^{\rm 1h}_{\rm m}(k)&=&
  \int_0^\infty\frac{\d m\;n(m)\,m^2}{\bar{\rho}_{\rm m}^2}\,
  \tilde{u}^2_{\rm m}(k,m)\;;\\
  \label{eq:pgm}
  P^{\rm 1h}_{\rm gm}(k)&=&
  \int_0^\infty\frac{\d m\;n(m)\,m}{\bar{\rho}_{\rm m}\overline{n}_{\rm g}}\,
  \tilde{u}_{\rm m}(k,m)\,\tilde{u}^q_{\rm g}(k,m)\,\ave{N|m}\;.
\end{eqnarray}
In these equations, the exponents $p$ and $q$ are modifiers of the
statistics for central galaxies which are accounted for in the
following simplistic way: Central galaxies are by definition at the
halo centre $r=0$; one galaxy inside a halo is always a central
galaxy; their impact on galaxy power spectra is assumed to be only
significant for halos that contain few galaxies.  Depending on whether
a halo contains few galaxies or not, the factors $(p,q)$ switch on or
off a statistics dominated by central galaxies through
\begin{eqnarray}
  \label{eq:pq}
  p&=&\left\{
    \begin{array}{ll}
      1 & ,\,{\rm for~} \ave{N(N-1)|m}>1\;\\
      1/2& ,\,{\rm otherwise}
    \end{array}
  \right.\;;
  \\
  \nonumber
  q&=&\left\{
    \begin{array}{ll}
      1 & ,\,{\rm for~} \ave{N|m}>1\;\\
      0 & ,\,{\rm otherwise}
    \end{array}
  \right.\;.
\end{eqnarray}
We note that $p$ and $q$ are functions of the halo mass $m$. Later in
Sect. \ref{sect:centrals}, we consider also more general models where
there can be a fraction of halos that contains only satellite
galaxies. We achieve this by mixing \Ref{eq:pg}-\Ref{eq:pgm} with
power spectra in a pure-satellite scenario, this means a scenario
where always $p\equiv q\equiv1$.

We now turn to the two-halo terms in this halo model. We approximate
the clustering power of centres of halos with mass $m$ by $b_{\rm
  h}^2(m)\,P_{\rm lin}(k)$, where $P_{\rm lin}(k)$ denotes the linear
matter power spectrum, and $b_{\rm h}(m)$ is the halo bias-factor on
linear scales; the clustering of halos is thus linear and
deterministic in this description. Likewise, this model approximates
the cross-correlation power-spectrum of halos with the masses $m_1$
and $m_2$ by \mbox{$b_{\rm h}(m_1)\,b_{\rm h}(m_2)\,P_{\rm
    lin}(k)$}. The resulting two-halo terms are then
\begin{eqnarray}
  \label{eq:p2halog}
  P^{\rm 2h}_{\rm g}(k)&=&
  \frac{P_{\rm lin}(k)}{\overline{n}_{\rm g}^2}
  \left(\int_0^\infty\d m\;n(m)\ave{N|m}\,b_{\rm h}(m)\tilde{u}_{\rm g}(k,m)\right)^2\;;\\  
  \label{eq:p2halom}
  P^{\rm 2h}_{\rm m}(k)&=&
  \frac{P_{\rm lin}(k)}{\overline{\rho}_{\rm m}^2}
  \left(
   \int_0^\infty\d m\;n(m)m\,b_{\rm h}(m)
    \tilde{u}_{\rm m}(k,m)
  \right)^2\;;\\
  \nonumber
  P^{\rm 2h}_{\rm gm}(k)&=&
  \frac{P_{\rm lin}(k)}{\overline{n}_{\rm g}\,\overline{\rho}_{\rm m}}
  \int_0^\infty\d m\;n(m)\ave{N|m}\,b_{\rm h}(m)\tilde{u}_{\rm g}(k,m)\\
  &&\label{eq:p2halomg}
  \times\int_0^\infty\d m\;n(m)m\,b_{\rm h}(m)\tilde{u}_{\rm m}(k,m)\;.
\end{eqnarray}
The two-halo terms ignore power from central galaxies because it is
negligible in the two-halo regime.

\subsection{A toy model for the small-scale galaxy bias}
\label{sect:toymodel}

We first consider an insightful toy model of $b(k)$ and $r(k)$ at
small scales. In this model, both the matter and the galaxy
distribution shall be completely dominated by halos of mass $m_0$,
such that we find an effective halo-mass function
\mbox{$n(m)\propto\delta_{\rm D}(m-m_0)$}; its normalisation is
irrelevant for the galaxy bias. In addition, the halos of the toy
model shall not cluster so that the two-halo terms of the power
spectra vanish entirely.  The toy model has practical relevance in
what follows later because the one-halo biasing functions that we
derive afterwards are weighted averages of toy models with different
$m_0$. For this reason, most of the features can already be understood
here, albeit not all, and it already elucidates biasing functions on
small scales.

Let us define the variance
$\sigma_N^2(m_0)=\ave{N^2|m_0}-\ave{N|m_0}^2$ of the halo-occupation
distribution (HOD) in excess of a Poisson variance $\ave{N|m_0}$ by
\begin{equation}
  \label{eq:dsigma}
  \Delta\sigma^2_N(m_0)=\sigma^2_N(m_0)-\ave{N|m_0}\;.
\end{equation}
If the model galaxies obey Poisson statistic they have
\mbox{$\Delta\sigma_N^2(m_0)=0$}.  We can now write the mean number of
galaxy pairs as
\begin{multline}
  \ave{N(N-1)|m_0}=\ave{N^2|m_0}-\ave{N|m_0}\\
  =\ave{N|m_0}^2
  \left(1+\frac{\Delta\sigma_N^2(m_0)}{\ave{N|m_0}^2}\right)\;.
\end{multline}
By using the Eqs. \Ref{eq:pg}--\Ref{eq:pgm} with
$n(m)\propto\delta_{\rm D}(m-m_0)$, the correlation factor reads
\begin{multline}
  \label{eq:rktm}
  R(k,m_0)=\\
    \frac{\tilde{u}^{q-p}_{\rm g}(k,m_0)\,\ave{N|m_0}}{\sqrt{\ave{N(N-1)|m_0}}}
    =
    \tilde{u}^{q-p}_{\rm g}(k,m_0)
    \left(1+\frac{\Delta\sigma_N^2(m_0)}{\ave{N|m_0}^2}\right)^{-1/2}\;,
\end{multline}
and the bias factor is
\begin{multline}
  \label{eq:bktm}
  B(k,m_0)=\\
  \frac{\tilde{u}^p_{\rm g}(k,m_0)\,\sqrt{\ave{N(N-1)|m_0}}}
  {\tilde{u}_{\rm m}(k,m_0)\,\ave{N|m_0}}
  =
  \frac{\tilde{u}^q_{\rm g}(k,m_0)}{\tilde{u}_{\rm
      m}(k,m_0)}
  \frac{1}{R(k,m_0)}\;.
\end{multline}
To avoid ambiguities in the following, we use capital letters for the
biasing functions in the toy model.

We dub galaxies `faithful tracers' of the matter density if they have
both (i) \mbox{$\tilde{u}_{\rm g}(k,m)=\tilde{u}_{\rm m}(k,m)$} and
(ii) no central galaxies (\mbox{$p=q=1$}). Halos with relatively small
numbers of galaxies, that is
\mbox{$\ave{N|m_0},\ave{N(N-1)|m_0}\lesssim1$}, are called
`low-occupancy halos' in the following.  This toy model then
illustrates the following points.
\begin{itemize}
\item Owing to galaxy discreteness, faithful tracers are biased if
  they not obey Poisson statistics.  Namely, for a sub-Poisson
  variance, \mbox{$\Delta\sigma_N^2(m_0)<0$}, they produce opposite
  trends \mbox{$R(k,m_0)>1$} and \mbox{$B(k,m_0)<1$} with $k$, and
  vice versa for a super-Poisson sampling, but generally we find the
  relation $R(k,m_0)\times B(k,m_0)=1$.
\item Nevertheless faithful tracers obey
  \mbox{$B(k,m_0),R(k,m_0)\approx1$} if the excess variance becomes
  negligible, that is if
  \mbox{$\Delta\sigma_N^2(m_0)\ll\ave{N|m_0}^2$}. The discreteness of
  galaxies therefore becomes only relevant in low-occupancy halos.
\item A value of \mbox{$R(k,m_0)>1$} occurs once central galaxies are
  present (\mbox{$p,q<1$}).  As a central galaxy is \emph{always}
  placed at the centre, central galaxies produce a non-Poisson
  sampling of the profile $u_{\rm m}(r,m_0)$. In contrast to faithful
  galaxies with a non-Poisson HOD, we then find agreeing trends with
  scale $k$ for $R(k,m_0)$ and $B(k,m_0)$ if \mbox{$\Delta\sigma_{\rm
      N}^2(m_0)=0$}.  Again, this effect is strong only in
  low-occupancy halos.
\item The biasing functions in the toy model are only scale-dependent
  if galaxies are not faithful tracers. The bias function $B(k,m_0)$
  varies with $k$ if either \mbox{$\tilde{u}_{\rm
      m}(k,m_0)\ne\tilde{u}_{\rm g}(k,m_0)$} or for central galaxies
  (\mbox{$p\ne1$}). The correlation function $R(k,m_0)$ is
  scale-dependent only for central galaxies, that is \mbox{$p-q\ne0$},
  which then obeys \mbox{$R(k,m_0)\propto\tilde{u}_{\rm
      g}^{-1/2}(k,m_0)$}.  Variations with $k$ become small for both
  functions, however, if \mbox{$\tilde{u}_{\rm
      m}(k,m_0),\tilde{u}_{\rm g}(k,m_0)\approx1$}, which is on scales
  larger than the size $r_{\rm vir}$ of a halo.
\end{itemize}

We stress again that a counter-intuitive \mbox{$r(k)>1$} is a result
of the definition of $P_{\rm g}(k)$ relative to Poisson shot-noise and
the actual presence of non-Poisson galaxy noise.  One may wonder here
if \mbox{$r>1$} is also allowed for biasing parameters defined in
terms of spatial correlations rather than the power spectra. That this
is indeed the case is shown in Appendix \ref{ap:realspacecorr} for
completeness.

\subsection{Galaxy biasing at small scales}

Compared to the foregoing toy model, no single halo mass scale
dominates the galaxy bias at any wave number $k$ for realistic
galaxies. Nevertheless, we can express the realistic biasing functions
$b^{\rm 1h}(k)$ and $r^{\rm 1h}(k)$ in the one-halo regime as weighted
averages of the toy model $B(k,m)$ and $R(k,m)$ with modifications.

To this end, we introduce by 
\begin{equation}
  \label{eq:flm}
  b(m)= 
  \frac{\ave{N|m}}{m}
  \frac{\overline{\rho}_{\rm m}}{\overline{n}_{\rm g}}
\end{equation}
the `mean biasing function' which is the mean number of halo galaxies
$\ave{N|m}$ per halo mass $m$ in units of the cosmic average
${\overline{n}_{\rm g}}/{\overline{\rho}_{\rm m}}$
\citep{2012MNRAS.426..566C}. If galaxy numbers linearly scale with
halo mass, that means \mbox{$\ave{N|m}\propto m$}, we find a mean
biasing function of \mbox{$b(m)=1$} while halos masses devoid of
galaxies have \mbox{$b(m)=0$}. For convenience, we make use of
\mbox{$\ave{N|m}\propto m\,b(m)$} instead of $\ave{N|m}$ in the
following equations because we typically find \mbox{$\ave{N|m}\propto
  m^\beta$} with \mbox{$\beta\approx1$}: $b(m)$ is therefore usually
not too different from unity.

Using the Eqs. \Ref{eq:pg} and \Ref{eq:pm} we then find
\begin{equation}
  \label{eq:bk}
  [b^{\rm 1h}(k)]^2=
  \frac{P^{\rm 1h}_{\rm g}(k)}{P^{\rm 1h}_{\rm m}(k)}
  =
  \int_0^\infty\d m\;w_{20}^{\rm 1h}(k,m)\,
  b^2(m)\,B^2(k,m)
\end{equation}
with $w_{20}^{\rm 1h}(k,m)$ being one case in a family of (one-halo)
weights,
\begin{equation}
 w_{ij}^{\rm 1h}(k,m):=
  \frac{n(m)\,m^2\,\tilde{u}_{\rm m}^i(k,m)\,[b(m)\,\tilde{u}_{\rm g}(k,m)]^j}
  {\int_0^\infty\!\d m\;n(m)\,m^2\,\tilde{u}_{\rm m}^i(k,m)\,[b(m)\,\tilde{u}_{\rm g}(k,m)]^j}\;.
\end{equation}
This family and the following weights $w(k,m)$ are normalised, which
means that \mbox{$\int\d m\,w(k,m)=1$}. The introduction of these
weight functions underlines that the biasing functions are essentially
weighted averages across the halo-mass spectrum as, for example,
$[b^{\rm 1h}(k)]^2$ which is the weighted average of
\mbox{$b^2(m)\,B^2(k,m)$}.

The effect of $w_{20}^{\rm 1h}(k,m)$ is to down-weight large halo
masses in the bias function because $w_{20}^{\rm
  1h}(k,m_1)/w_{20}^{\rm 1h}(k,m_2)\propto\tilde{u}_{\rm
  m}^2(k,m_1)/\tilde{u}_{\rm m}^2(k,m_2)$ decreases with $m_1$ for a
fixed $m_2<m_1$ and $k$. Additionally, the relative weight of a halo
with mass $m$ decreases towards larger $k$ because $\tilde{u}_{\rm
  m}(k,m)$ tends to decrease with $k$. As a result, at a given scale
$k$ only halos below a typical mass essentially contribute to the
biasing functions \citep{2000MNRAS.318..203S}.

We move on to the correlation factor $r^{\rm 1h}(k)$ in the one-halo
regime. Using the Eqs. \Ref{eq:pg}--\Ref{eq:pgm} and the relations
\begin{eqnarray}
  \ave{N(N-1)|m}&=&R^{-2}(k,m)\,\tilde{u}_{\rm g}^{2q-2p}(k,m)\,\ave{N|m}\;;\\
  \tilde{\rho}_{\rm m}(k,m)&=&m\,\tilde{u}_{\rm m}(k,m)\;;\\
  \tilde{\rho}_{\rm g}(k,m)&=&\ave{N|m}\,\tilde{u}_{\rm g}(k,m)
\end{eqnarray}
we write  
\begin{equation}
  \label{eq:rk}
  r^{\rm 1h}(k)=
  \frac{P^{\rm 1h}_{\rm gm}(k)}{\sqrt{P^{\rm
        1h}_{\rm g}(k)P^{\rm 1h}_{\rm m}(k)}}
  =:\zeta_{\rm sat}(k)\,\zeta_{\rm cen}(k)\,\zeta_{\Delta\sigma}(k)
\end{equation}
as product of the three separate factors
\begin{eqnarray}
  \label{eq:rkpois}
  \zeta_{\rm sat}(k)\!\!\!\!&:=&
  \!\!\!\!
    \frac{\int_0^\infty\d m\;n(m)\,
      \tilde{\rho}_{\rm m}(k,m)\,\tilde{\rho}_{\rm
        g}(k,m)}
    {
      \left(\int_0^\infty\d m\;n(m)\,\tilde{\rho}^2_{\rm g}(k,m)\,
      \int_0^\infty\d m\;n(m)\,\tilde{\rho}^2_{\rm m}(k,m)
      \right)^{1/2}
    }\;;
    \\
    \zeta_{\rm cen}(k)\!\!\!\!&:=&
    \!\!\!\!\int_0^\infty\d m\;w_{11}^{\rm 1h}(k,m)\,\tilde{u}^{q-1}_{\rm
      g}(k,m)\;;
    \\
    \label{eq:rkul}
    \zeta_{\Delta\sigma}(k)\!\!\!\!&:=&\!\!\!\!
    \left(\int_0^\infty\d m\;w_{02}^{\rm 1h}(k,m)\,
      \tilde{u}_{\rm g}^{2q-2}(k,m)\,R^{-2}(k,m)\right)^{-1/2}
\end{eqnarray}
with the following meaning.
\begin{itemize} 
\item The first factor $\zeta_{\rm sat}(k)$ quantifies, at spatial
  scale $k$, the correlation between the radial profiles of the matter
  density $\tilde{\rho}_{\rm m}(k,m)$ and the (average) number density
  of satellite galaxies $\tilde{\rho}_{\rm g}(k,m)$ across the halo
  mass-spectrum $n(m)$. As upper bound we always have
  \mbox{$|\zeta_{\rm sat}(k)|\le1$} because of the Cauchy-Schwarz
  inequality when applied to the nominator of
  Eq. \Ref{eq:rkpois}. Thus $\zeta_{\rm sat}(k)$ probably reflects
  best what we intuitively understand by a correlation factor between
  galaxies and matter densities inside a halo. Since it only involves
  the average satellite profile, the satellite shot-noise owing to a
  HOD variance is irrelevant at this point. The next two factors can
  be seen as corrections to $\zeta_{\rm sat}(k)$ owing to central
  galaxies or a non-Poisson HOD variance.
\item The second factor $\zeta_{\rm cen}(k)$ is only relevant in the
  sense of \mbox{$\zeta_{\rm cen}(k)\ne1$} through low-occupancy halos
  with central galaxies ($q\ne1$). It has the lower limit $\zeta_{\rm
    cen}(k)\ge1$ because of \mbox{$\tilde{u}_{\rm g}(k,m)\le1$} and
  hence \mbox{$\tilde{u}_{\rm g}^{q-1}(k,m)\ge1$}. This correction
  factor can therefore at most increase the correlation $r^{1\rm
    h}(k)$.
\item The third factor $\zeta_{\Delta\sigma}(k)$ is the only one that
  is sensitive to an excess variance \mbox{$\Delta\sigma_{\rm
      N}^2(m)\ne0$} of the HOD, namely through $R(k,m)$.  In the
  absence of central galaxies, that means for \mbox{$p\equiv
    q\equiv1$}, $\zeta^2_{\Delta\sigma}(k)$ is the (weighted) harmonic
  mean of $R^2(k,m)$, or the harmonic mean of the reduced
  $[\tilde{u}_{\rm g}(k,m)\,R(k,m)]^2\le R^2(k,m)$ otherwise.
\end{itemize}

As sanity check, we note the recovery of the toy model by setting
\mbox{$n(m)\propto\delta_{\rm D}(m-m_0)$} in \Ref{eq:bk} and
\Ref{eq:rk}. In contrast to the toy model, the templates $b^{\rm
  1h}(k)$ and $r^{\rm 1h}(k)$ can be scale-dependent even if $B(k,m)$
and $R(k,m)$ are constants. This scale dependence can be produced by a
varying $w_{20}^{\rm 1h}(k,m)$ or $\zeta_{\rm sat}(k)$.

\subsection{Galaxy biasing at large scales}
\label{sect:largescalebias}

From the two-halo terms \Ref{eq:p2halog}--\Ref{eq:p2halomg}, we can
immediately derive the two-halo biasing functions. The bias factor is
\begin{equation}
  b^{\rm 2h}(k)
  =
  \sqrt{\frac{P^{\rm 2h}_{\rm g}(k)}{P^{\rm 2h}_{\rm m}(k)}}
  =
  \frac{\int_0^{\infty}\d m\;w_{10}^{\rm 2h}(m)\,b_{\rm h}(m)\,\tilde{u}_{\rm g}(k,m)}
  {\int_0^{\infty}\d m\;w_{01}^{\rm 2h}(m)\,b_{\rm h}(m)\,\tilde{u}_{\rm m}(k,m)}\;,
\end{equation}
where we have introduced into the integrals the normalised (two-halo)
weights
\begin{equation}
  w_{ij}^{\rm 2h}(m):=
  \frac{n(m)\,[m\,b(m)]^i\,m^j}{\int_0^\infty\d
    m\;n(m)\,[m\,b(m)]^i\,m^j}
  \;.
\end{equation}
We additionally approximate \mbox{$\tilde{u}(k,m)\approx1$} for the
two-halo regime. This is a reasonable approximation because virialised
structures are typically not larger than \mbox{$\sim10\,h^{-1}\,\rm
  Mpc$} and hence exhibit \mbox{$\tilde{u}(k,m)\approx1$} for
\mbox{$k\ll0.5\,h\,\rm Mpc^{-1}$}.  Therefore, we find an essentially
constant bias function at large scales,
\begin{equation}
  \label{eq:lsbias}
  b^{\rm 2h}(k)\approx 
  \int_0^{\infty}\d m\;w_{10}^{\rm 2h}(m)\,b_{\rm h}(m)=:b_{\rm ls}\;.
\end{equation}
We have used here \mbox{$\int\d m\;w_{01}^{\rm 2h}(m)\,b_{\rm
    h}(m)=1$} which follows from the constraint \mbox{$P_{\rm m}(k)\to
  P_{\rm lin}(k)$} for \mbox{$k\to0$} and the Eq. \Ref{eq:p2halom}. To
have more template flexibility, we leave $b_{\rm ls}$ as free
parameter and devise the Eq. \Ref{eq:lsbias} only if no large-scale
information is available by observations.

The two-halo correlation-function at large scales is exactly
\begin{equation}
  \label{eq:lscorr}
  r^{\rm 2h}(k)=
  \frac{P^{\rm 2h}_{\rm gm}(k)}{\sqrt{P^{\rm
    2h}_{\rm g}(k)P^{\rm 2h}_{\rm m}(k)}}=1=r_{\rm ls}
\end{equation}
due to \mbox{$P_{\rm h}(k;m_1,m_2)=b_{\rm h}(m_1)\,b_{\rm
    h}(m_2)\,P_{\rm m}(k)$} for the assumed halo
clustering. Evidently, the large-scale matter-galaxy correlation is
fixed to \mbox{$r_{\rm ls}=1$}. The correlation is necessarily high
because the model galaxies are always inside halos so that galaxies
closely follow the matter distribution at large scales. 

We note that \mbox{$r_{\rm ls}\ne1$} is physically conceivable
although it is usually excluded in halo models
\citep{1998ApJ...500L..79T}.  To test for an actually high correlation
\mbox{$r_{\rm ls}=1$} in real data, we may use $r_{\rm ls}$ as free
parameter in the templates.

\subsection{Fraction of central galaxies}
\label{sect:centrals}

Up to here, we assumed either one central galaxy for every halo that
hosts galaxies or pure samples of satellite galaxies, meaning
\mbox{$p\equiv q\equiv1$}.  In reality where we select sub-populations
of galaxies, not every sub-sample automatically provides a central
galaxy in every halo; a central galaxy could belong to another galaxy
population, for instance.  For more template flexibility, we thus
assume that only a fraction $f_{\rm cen}$ of \emph{halos} can have
central galaxies from the selected galaxy population; the other
fraction $1-f_{\rm cen}$ of halos has either only satellites or
central galaxies from another population. Both halo fractions
nevertheless shall contain $\ave{N|m}$ halo galaxies on
average. Importantly, $f_{\rm cen}$ shall be independent of halo
mass. This is not a strong restriction because the impact of central
galaxies becomes only relevant for low-occupancy halos whose mass
scale $m$ is confined by \mbox{$\ave{N|m}\lesssim1$} anyway.

The extra freedom of \mbox{$f_{\rm cen}\ne1$} in the templates
modifies the foregoing power spectra. On the one hand, the two-halo
power spectra are unaffected because they do not depend on either $p$
or $q$. On the other hand for the one-halo regime, we now find the
linear combination
\begin{eqnarray}
  \label{eq:pgfcen}
  P_{\rm g}^{\rm 1h}(k)&=&f_{\rm cen}\,P_{\rm g}^{\rm
    cen}(k)+(1-f_{\rm cen})\,P_{\rm
    g}^{\rm sat}(k)\;,\\
  \label{eq:pgmfcen}
  P_{\rm gm}^{\rm 1h}(k)&=&f_{\rm cen}\,P_{\rm gm}^{\rm
    cen}(k)+(1-f_{\rm cen})\,P_{\rm
    gm}^{\rm sat}(k)
\end{eqnarray}
because halos with (or without) central galaxies contribute with
probability $f_{\rm cen}$ (or $1-f_{\rm cen}$) to the one-halo
term. In the equations, the $P^{\rm cen}(k)$ denote the one-halo power
spectra of halos with central galaxies, and the $P^{\rm sat}(k)$
denote spectra of halos with only satellites. Both cases are covered
in the foregoing formalism for appropriate values of $p,q$:
Satellite-only halos with superscript `sat' are obtained by using
\mbox{$p\equiv q\equiv1$}; halos with central galaxies, superscript
`cen', use the usual mass-dependent expressions \Ref{eq:pq}.

As result, we can determine the bias factor for the mixture scenario
with \Ref{eq:pgfcen} by
\begin{equation}
  \label{eq:bkcen}
  [b^{\rm 1h}(k)]^2=
  f_{\rm cen}\,[b_{\rm cen}(k)]^2
  +
  (1-f_{\rm cen})\,[b_{\rm sat}(k)]^2\;.
\end{equation}
Here $b_{\rm cen}(k)$ denotes Eq. \Ref{eq:bk} in the central-galaxy
scenario, whereas $b_{\rm sat}(k)$ denotes the satellite-only scenario
of this equation.  Similarly for the correlation, $r^{\rm 1h}(k)$ we
obtain with \Ref{eq:pgfcen} and \Ref{eq:pgmfcen}
\begin{eqnarray}
  \label{eq:rkcen} 
  \lefteqn{r^{\rm 1h}(k)=}\\
  \nonumber
  &&
  f_{\rm cen}\,\frac{P_{\rm gm}^{\rm cen}(k)}
  {\sqrt{P^{\rm 1h}_{\rm g}(k)\,P^{\rm 1h}_{\rm m}(k)}}
  +
  (1-f_{\rm cen})\,\frac{P_{\rm gm}^{\rm sat}(k)}
  {\sqrt{P^{\rm 1h}_{\rm g}(k)\,P^{\rm 1h}_{\rm m}(k)}}
  \\
  \nonumber
  &&=
  f_{\rm cen}\,\sqrt{\frac{P^{\rm cen}_{\rm g}(k)}
    {P^{\rm 1h}_{\rm g}(k)}}r_{\rm cen}(k)+
  (1-f_{\rm cen})
  \sqrt{\frac{P^{\rm sat}_{\rm g}(k)}
    {P^{\rm 1h}_{\rm g}(k)}}r_{\rm sat}(k)
  \\
  \nonumber
  &&=
  \frac{f_{\rm cen}\,r_{\rm cen}(k)}
  {\sqrt{f_{\rm cen}+(1-f_{\rm cen})\,\left(\frac{b_{\rm
            sat}(k)}{b_{\rm cen}(k)}\right)^2}}+
  \frac{(1-f_{\rm cen})\,r_{\rm sat}(k)}
  {\sqrt{1-f_{\rm cen}+f_{\rm cen}\left(\frac{b_{\rm
            cen}(k)}{b_{\rm sat}(k)}\right)^2}}\;,
\end{eqnarray}
because 
\begin{equation}
  P_{\rm g}^{\rm 1h}(k)=
  f_{\rm cen}\,[b_{\rm cen}(k)]^2\,P^{\rm 1h}_{\rm m}(k)+
  (1-f_{\rm cen})\,[b_{\rm sat}(k)]^2\,P^{\rm 1h}_{\rm m}(k)\;.
\end{equation}
The function $r_{\rm cen}(k)$ denotes Eq. \Ref{eq:rk} in the
central-galaxy scenario, and $r_{\rm sat}(k)$ is the satellite-only
scenario.

\section{Parameters of model templates and physical discussion}
\label{sect:implementation}

In this section, we summarise the concrete implementation of our
templates, and we discuss their parameter dependence for a physical
discussion on the scale-dependent galaxy bias.

\subsection{Normalised excess-variance}

For a practical implementation of our templates, we find it useful to
replace $\Delta\sigma^2_N(m)$ in Eq. \Ref{eq:dsigma} by the
`normalised excess-variance'
\begin{equation}
  V(m)=
  \frac{\Delta\sigma^2_N(m)}{\ave{N|m}}
  =\frac{\sigma^2_{\rm N}(m)}{\ave{N|m}}-1\;,
\end{equation}
which typically has a small dynamic range with values between minus
and plus unity. To see this, we discuss its upper and lower limits in
the following.

First, the normalised excess-variance has a lower limit because the
average number of galaxy pairs is always positive,
\begin{equation}
  \ave{N(N-1)|m}=\ave{N|m}^2\,\left(1+\frac{V(m)}{\ave{N|m}}\right)\ge0\,,
\end{equation}
which imposes \mbox{$V(m)\ge-\ave{N|m}$}. As additional constraint we
have a positive variance
\begin{equation}
\sigma_{\rm N}^2(m)=\ave{N|m}\,\Big(V(m)+1\Big)\,\ge0
\end{equation}
or $V(m)\ge-1$ so that we use
\begin{equation}
  V(m)\ge\max{\Big\{-1,-\ave{N|m}\Big\}}
\end{equation}
for a valid set of template parameters. 

Second for the upper limit of $V(m)$, we imagine that there is a
maximum $N_{\rm max}(m)$ for the amount of halo galaxies (of the
selected population) inside a halo of mass $m$. A maximum $N_{\rm
  max}(m)$ makes physically sense because we cannot squeeze an
arbitrary number of galaxies into a halo. Nevertheless, their amount
\mbox{$0\le N(m) \le N_{\rm max}(m)$} shall be random with PDF
$P(N|m)$. Of this PDF we already know that its mean is
$\ave{N|m}$. For its the maximum possible variance $\sigma^2_{\rm
  max}(m)$, we note that $\sigma_{\rm N}^2(m)$ cannot be larger than
that for halos with a bimodal distribution of only two allowed galaxy
numbers \mbox{$N(m)\in\{0,N_{\rm max}(m)\}$} that shall occur with
probability $1-\lambda$ and $\lambda$, respectively. The mean of this
bimodal PDF is $\ave{N|m}=\lambda\,N_{\rm max}(m)$, and its variance
$\sigma_{\rm max}^2(m)= \ave{N^2|m}-\ave{N|m}^2$ consequently
satisfies
\begin{multline}
  \sigma_{\rm max}(m)=
  \\
  \sqrt{N_{\rm max}^2(m)\,\lambda-N_{\rm max}^2(m)\,\lambda^2}=
  \ave{N|m}^{1/2}\,\sqrt{N_{\rm max}(m)-\ave{N|m}}\;,
\end{multline}
which is the upper limit for any $P(N|m)$.  Together with the lower
bound of $V(m)$, we thus arrive at
\begin{equation}
  \max{\Big\{-1,-\ave{N|m}\Big\}}\le V(m)\le N_{\rm max}(m)-1-\ave{N|m}\;.
\end{equation}
This means: halos that are (on average) filled close to the limit,
that is \mbox{$\ave{N|m}\approx N_{\rm max}(m)\ge1$}, have a HOD
variance that is sub-Poisson, close to \mbox{$V(m)=-1$}. This should
be especially the case for halos with \mbox{$\ave{N|m}\approx1$}. On
the other hand, halos with \mbox{$N_{\rm max}(m)\approx1$} and low
occupancy, \mbox{$\ave{N|m}\ll1$}, necessarily obey Poisson statistics
or are close to that, which means that \mbox{$V(m)\approx 0$}. On the
other extreme end, spacious halos well below the fill limit,
\mbox{$N_{\max}(m)\gg1$} and \mbox{$N_{\rm max}(m)\gg\ave{N|m}$}, have
sufficient headroom to allow for a super-Poisson variance which means
that \mbox{$V(m)>0$}. In the following, we adopt the upper limit
\mbox{$V(m)\le+1$} meaning that we a-priori do not allow the HOD
variance to become larger than twice the Poisson variance.

\subsection{Implementation}
\label{sect:modimplement}

\renewcommand{\arraystretch}{1.1}
\begin{table}
  \caption{\label{tab:model} List of free template parameters}
\begin{center}
  \begin{tabular}{llr}
    \hline\hline
    Param & Description & Dim\\
    \hline\\
    $b_{\rm ls}$  & large-scale bias factor & 1\\
    $r_{\rm ls}$  & large-scale correlation factor & (1)\\
    $b(m)$      & mean biasing function (interp.) & 22\\
    $V(m)$      & normalised excess-variance (interp.) & 22\\
    $m_{\rm piv}$  & $\ave{N|m_{\rm piv}}=1$; pivotal halo mass 
    & 1\\
    $f_{\rm cen}$ & halo fraction open for central galaxies & 1\\
    \hline\\
    & & $\Sigma=47\,(48)$
  \end{tabular}
  \tablefoot{The parameters $b(m)$ and $V(m)$ cover the mass range
    $10^4-10^{16}\,h^{-1}\,\msol$. The numbers ``(1)'' in brackets
    indicate optional degrees of freedom of the template. See text for
    more details.}
\end{center}
\end{table}
\renewcommand{\arraystretch}{1.0}

Generally the functions $V(m)$ and $b(m)$ are continuous functions of
the halo mass $m$. We apply, however, an interpolation with $20$
interpolation points on a equidistant logarithmic $m$-scale for these
functions, spanning the range $10^8\,h^{-1}\,\msol$ to
$10^{16}\,h^{-1}\,\msol$; between adjacent sampling points we
interpolate linearly on the log-scale; we set \mbox{$b(m)=V(m)=0$}
outside the interpolation range. Additionally, we find in numerical
experiments with unbiased galaxies that the halo mass-scale has to be
lowered to $10^4\,h^{-1}\,\msol$ to obtain correct descriptions of the
bias. We therefore include two more interpolation points at $10^4$ and
$10^6\,h^{-1}\,\msol$ to extend the mass scale to very low halo
masses. For the large-scale bias, we set $r_{\rm ls}\equiv1$ but leave
$b_{\rm ls}$ as free parameter.

\begin{figure*}
  \begin{center}
    \epsfig{file=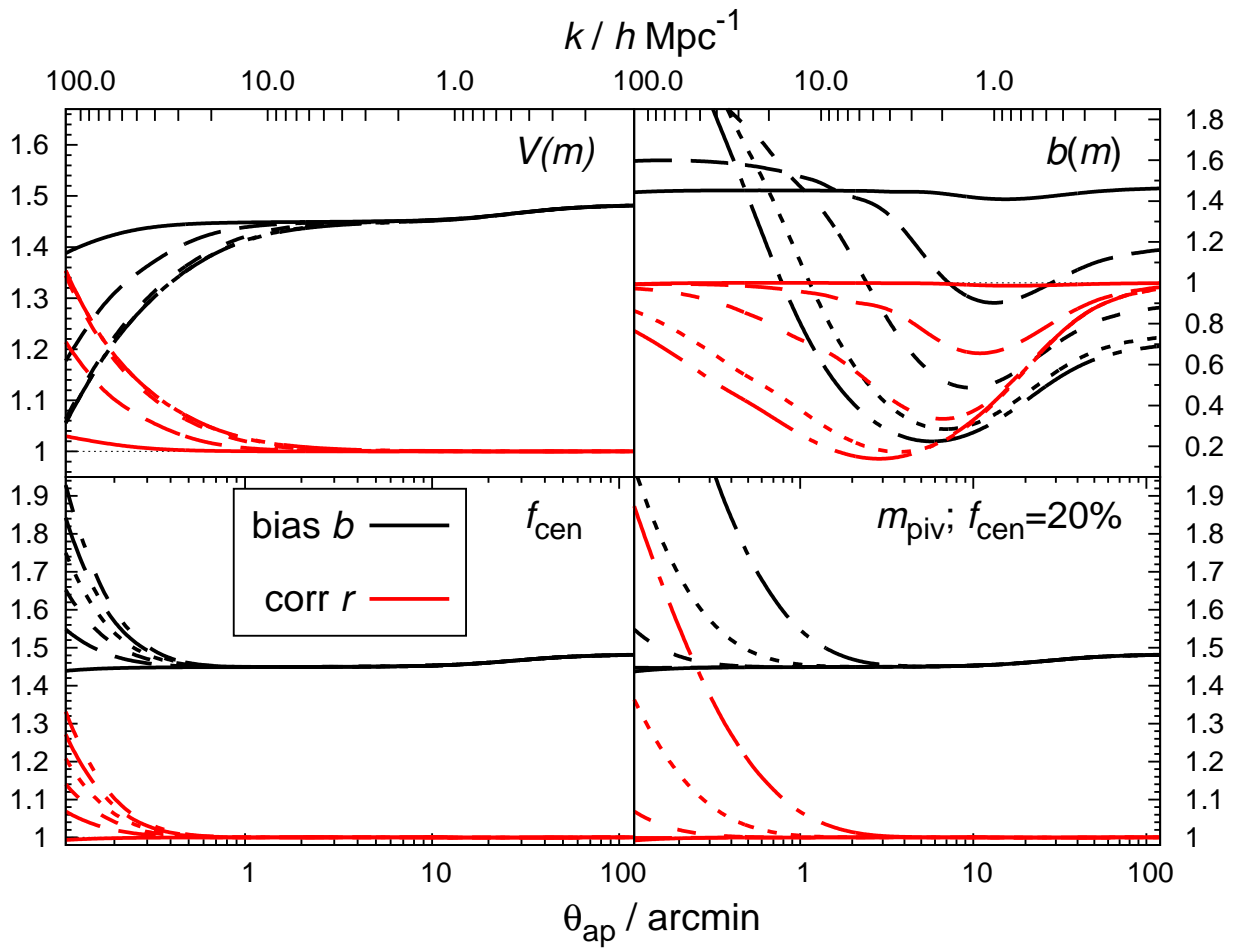,width=125mm,angle=-90}
  \end{center}
  \caption{\label{fig:models} Family of templates $b(k)$ (black lines)
    and $r(k)$ (red lines) for the range of wave numbers $k$ in the
    top axis; the left-hand $y$-axis applies to the panels in the
    first column, the right-hand axis to the second column. The
    aperture scale $\theta_{\rm ap}=4.25/(k\,f_K(z_{\rm d}))$ (bottom
    axis) crudely traces the projected $b_{\rm 2D}(\theta_{\rm ap})$
    and $r_{\rm 2D}(\theta_{\rm ap})$ for lens galaxies at $z_{\rm
      d}=0.3$. Each panel varies only one template parameter.  See
    text for more details.}
\end{figure*}

To predict the number density $\bar{n}_{\rm g}$ of galaxies,
Eq. \Ref{eq:rhong}, and to determine $(p,q)$ for a given mass $m$ we
have to obtain $\ave{N|m}$ from $b(m)$. For this purpose, we introduce
another parameter $m_{\rm piv}$ which is the pivotal mass of
low-occupancy halos, defined by \mbox{$\ave{N|m_{\rm piv}}=1$} such
that
\begin{equation}
  \label{eq:nofm}
  \ave{N|m}=
  \frac{m}{m_{\rm piv}}\,
  \frac{b(m)}{b(m_{\rm piv})}\;.
\end{equation}
The (comoving) number density of galaxies $\bar{n}_{\rm g}$ is then
given by
\begin{equation}
  \label{eq:nbar}
  \frac{\bar{n}_{\rm g}}{\overline{\rho}_{\rm m}}=
  \frac{\int_0^\infty\d m\;n(m)\,\ave{N|m}}{\overline{\rho}_{\rm m}}=
  \int_0^\infty\frac{\d m\;w_{01}^{\rm 2h}(m)}{m_{\rm piv}}\,
  \frac{b(m)}{b(m_{\rm piv})}\;,
\end{equation}
for which we use $\bar{\rho}_{\rm m}=\Omega_{\rm m}\,\bar{\rho}_{\rm
  crit}$. With this parameterisation, the normalisation of $b(m)$ is
irrelevant in all equations of our bias templates. Nevertheless,
$b(m)$ can be shown to obey
\begin{equation}
  \int_0^\infty\!\!\d m\;n(m)\,m\,b(m)=
  \overline{\rho}_{\rm m}\Longleftrightarrow
  \int_0^\infty\!\!\d m\;w_{01}^{\rm 2h}(m)\,b(m)=1
\end{equation}
which follows from the Eqs. \Ref{eq:flm} and \Ref{eq:rhong}. When
plotting $b(m)$, we make sure that it is normalised correspondingly.

Furthermore for the templates, we assume that satellite galaxies
always trace the halo matter density so that \mbox{$\tilde{u}_{\rm
    g}(k,m)\equiv\tilde{u}_{\rm m}(k,m)$}. This assumption could be
relaxed in a future model extension. For the matter density profile
$\tilde{u}_{\rm m}(k,m)$, we assume a NFW profile
\citep{1996ApJ...462..563N} with a mass concentration as in
\citet{2000MNRAS.318..203S} and a halo mass spectrum $n(m)$ according
to \citet{1999MNRAS.308..119S}.  For the average biasing functions
$b(k)$ and $r(k)$, we evaluate $n(m)$, $b_{\rm h}(m)$, and
$\tilde{u}_{\rm m}(k,m)$ at the mean redshift of the lens galaxies. As
model for $P_{\rm m}(k;\chi)$ in Sect. \ref{sect:biassmoothing} we
employ the publicly available code
\texttt{nicaea}\footnote{\url{http://www.cosmostat.org/software/nicaea/}}
version 2.5 \citep{2009A&A...497..677K} that provides an
implementation of \texttt{Halofit} with the recent update by
\citet{2012ApJ...761..152T} and the matter transfer-function in
\cite{1998ApJ...496..605E} for baryonic oscillations.

We list all free parameters of the templates in Table
\ref{tab:model}. Their total number is 47 by default. In a future
application, we may also consider $r_{\rm ls}$ a free parameter to
test, for instance, the validity of \mbox{$r_{\rm ls}=1$}. If no
large-scale information on the aperture statistics is available, we
predict $b_{\rm ls}$ from Eq. \Ref{eq:lsbias}, reducing the degrees of
freedom in the model by one.

To obtain the biasing functions $b(k)$ and $r(k)$ from the set of
parameters we proceed as follows. We first compute the one-halo terms
\Ref{eq:bk} and \Ref{eq:rk} for two separate scenarios: with and
without central galaxies. Both scenarios are then mixed according to
the Eqs. \Ref{eq:bkcen} and \Ref{eq:rkcen} for the given value of
$f_{\rm cen}$.  Finally, we patch together the one- and two-halo
biasing functions according to Eqs. \Ref{eq:rkgeneral} and
\Ref{eq:bkgeneral} with a weight $W_{\rm m}(k)$ for the fiducial
cosmology.

\subsection{Physical discussion}
\label{sect:modeldiscussion}

Fig. \ref{fig:models} is a showcase of conceivable biasing functions
and their relation to the underlying galaxy physics which we compute
in the aforementioned way. The wave number $k$ is plotted on the top
axis, whereas the bottom axis is defined by $\theta_{\rm
  ap}=4.25/(k\,f_K(z_{\rm d}))$ for a lens redshift of $z_{\rm
  d}=0.3$, which is essentially a simplistic prediction for $b_{\rm
  2D}(\theta_{\rm ap})$ and $r_{\rm 2D}(\theta_{\rm ap})$ as observed
by the lensing technique in Sect. \ref{sect:projectedbias}.  For the
discussion here, we concentrate on the spatial biasing functions.

We plot both $b(k)$ and $r(k)$ inside each panel. The black lines show
a family of $b(k)$ that we obtain by varying one template parameter at
a time in a fiducial model; the red lines are families of $r(k)$.  The
varied parameter is indicated in the top right corner of each panel.
We assume a large-scale bias $b_{\rm ls}$ according to
Eq. \Ref{eq:lsbias} with the theoretical halo bias $b_{\rm h}(m)$ in
\cite{2005ApJ...631...41T}. The fiducial model has: (i) no central
galaxies, \mbox{$f_{\rm cen}=0$}; (ii) a constant $b(m)>0$ for
$m\in[10^9,10^{15}]\,h^{-1}\,\msol$ but vanishing everywhere else;
(iii) a Poisson HOD-variance, \mbox{$V(m)=0$}, for all halo masses;
and (iv) a pivotal mass of \mbox{$m_{\rm
    piv}=10^{11}\,h^{-1}\,\msol$}.  This setup results in a
large-scale bias factor of \mbox{$b_{\rm ls}=1.48$}.  The details of
the panels are as follows.
\begin{itemize}
\item The bottom left panel varies $f_{\rm cen}$ between zero and
  100\% in steps of 20\% (bottom to top lines). Affected by a change
  of $f_{\rm cen}$ are only the small scales $k\gtrsim10\,h^{-1}\,\rm
  Mpc$ (or $\theta_{\rm ap}\lesssim1\,\rm arcmin$) that are strongly
  influenced by low-mass, low-occupancy halos.
\item The bottom right panel increases $m_{\rm piv}$ from
  $10^9\,h^{-1}\,\msol$ (bottom line) to $10^{13}\,h^{-1}\,\msol$ (top
  line) in steps of one dex. An impact on the bias functions is only
  visible if we have either a non-Poisson HOD variance or central
  galaxies. We hence set \mbox{$f_{\rm cen}=20\%$} compared to the
  fiducial model. A greater value of $m_{\rm piv}$ shifts the mass
  scale of low-occupancy halos to larger masses and thus their impact
  on the bias functions to larger scales.
\item In the top left panel, we adopt a sub-Poisson model of
  \mbox{$V(m)=\max{\{-0.5,-\ave{N|m}\}}$} for halos with \mbox{$m\le
    m_{\rm v}$}. We step up the mass scale $m_{\rm v}$ from
  \mbox{$10^{10}\,h^{-1}\,\msol$} (bottom line for $r$; top line for
  $b$) to \mbox{$10^{14}\,h^{-1}\,\msol$} (top line for $r$; bottom
  line for $b$) in one dex steps. Similar to the toy model in
  Sect. \ref{sect:toymodel}, a sub-Poisson variance produces opposite
  trends for $b$ and $r$: if $b$ goes up, $r$ goes down, and vice
  versa. The effect is prominent at small scales where low-occupancy
  halos significantly contribute to the bias functions. Conversely to
  what is shown here, these trends in $b$ and $r$ change signs if we
  adopt a super-Poisson variance instead of a sub-Poisson variance for
  \mbox{$m\le m_{\rm v}$}, which means that \mbox{$V(m)>0$}.
\item The top right panel varies the mean biasing function $b(m)$. To
  achieve this we consider a mass-cutoff scale $m_{\rm f}$ beyond
  which halos not harbour any galaxies, that means \mbox{$b(m)=0$}. We
  reduce this cutoff from \mbox{$m_{\rm f}=10^{15}\,h^{-1}\,\msol$}
  down to \mbox{$10^{11}\,h^{-1}\,\msol$} by one dex in each step (top
  to bottom line). This gradually excludes galaxies from high-mass
  halos on the mass scale. Broadly speaking, we remove galaxies from
  massive clusters first, then groups, and retain only field galaxies
  in the end. In the same way as for a non-Poisson HOD or present
  central galaxies this gives rise to a strong scale-dependence in the
  bias functions but now clearly visible on all scales. Despite its
  complex scale-dependence, the correlation factor stays always
  \mbox{$r(k)\le1$} because of the Poisson HOD variance and the
  absence of central galaxies in the default model.
\end{itemize}

This behaviour of the biasing functions is qualitatively similar to
what is seen in the related analytic model by
\cite{2012MNRAS.426..566C}, where deviations from either faithful
galaxies, a Poissonian HOD, or a constant mean biasing function
\mbox{$b(m)\equiv1$} are also necessary for biased galaxies. Moreover,
the scale-dependence that is induced by central galaxies or a
non-Poisson HOD variance is there, as for our templates, restricted to
small scales in the one-halo (low-occupancy halo) regime, typically
below a few $h^{-1}\,\rm Mpc$. However, their model has a different
purpose than our templates and is therefore less flexible. To make
useful predictions of biasing functions for luminosity-selected
galaxies they assume (apart from different technicalities as to the
treatment of centrals and satellites) that: the mean galaxy number
$\ave{N|m}$ is strongly confined by realistic conditional
luminosity-functions ($b(m)$ is not free); their `Poisson function'
\mbox{$\beta(m):=V(m)/\ave{N|m}+1$} is a constant ($V(m)$ is not
free); the large-scale biasing factor $b_{\rm ls}$ is determined by
$b(m)$. Especially, the freedom of $b(m)$ facilitates our templates
with the flexibility to vary over a large range of scales (top right
panel in Fig. \ref{fig:models}), which may be required for galaxies
with a complex selection function.

\section{Practical inference of biasing functions}
\label{sect:statinference}

In this section, we construct a methodology to statistically infer the
biasing functions $b(k)$ and $r(k)$ from noisy observations of the
lensing aperture statistics $\ave{{\cal N}^2}$,
$\ave{{\cal N}M_{\rm ap}}$, and $\ave{M^2_{\rm ap}}$. The general idea
is to utilise the model templates in Sect. \ref{sect:implementation}
and to constrain the space of their parameters by the likelihood of
the observed ratio statistics $b_{\rm 2D}(\theta_{\rm ap})$ and
$r_{\rm 2D}(\theta_{\rm ap})$. The posterior distribution of templates
will constitute the posterior of the deprojected biasing functions.

To estimate the aperture statistics from lens and source catalogues we
employ standard techniques that we summarise in Appendix
\ref{sect:estimators} for a practical reference. We shall assume that
we have measurements of the aperture statistics and their joint error
covariance in the following, based on estimates of lens-lens,
lens-shear, and shear-shear correlation functions between 1.4 arcsec
to 280 arcmin and 64 jackknife samples. The aperture statistics are
computed for nine radii $\theta_{\rm ap}$ between 1.8 arcmin and 140
arcmin.

\subsection{Statistical analysis}
\label{sect:statanalysis}

In our statistical analysis, we fit for a set of $n_{\rm d}$ aperture
radii $\theta_i$ a model of the aperture statistics $b_{\rm
  2D}(\theta_i;b)$ and $r_{\rm 2D}(\theta_i;b,r)$, Eqs. \Ref{eq:b2d}
and \Ref{eq:r2d}, to the measurement of the ratio statistics $b_{\rm
  2D}(\theta_i)$ and $r_{\rm 2D}(\theta_i)$, Eqs. \Ref{eq:b2dobs} and
\Ref{eq:r2dobs}. Ratios of the noisy aperture statistics result in a
skewed error distribution for $b_{\rm 2D}$ and $r_{\rm 2D}$ which we
account for in a non-Gaussian model likelihood that assumes Gaussian
errors for the aperture moments $\ave{{\cal N}^2}$, $\ave{{\cal
    N}M_{\rm ap}}$, $\ave{M_{\rm ap}^2}$ themselves (and positive
values for the variances).

With regard to the validity of a (truncated) Gaussian model for the
aperture moments, at least for current cosmic-shear studies this is
known to be a sufficiently accurate approximation
\citep[e.g.][]{2017MNRAS.465.1454H,2013MNRAS.430.2200K}. Nevertheless,
our statistical tests in Appendix \ref{sect:nongauss} find evidence
for a non-Gaussian statistics in our mock data, especially for the
variance $\ave{M^2_{\rm ap}}$ on scales of one degree or larger. This
may bias the reconstruction of $b(k)$ and $r(k)$ which will eventually
be contained in our assessment of systematic errors later on.

To motivate our model likelihood for $b_{\rm 2D}(\theta_i)$ and
$r_{\rm 2D}(\theta_i)$, let us first consider a simpler case where
$\hat{x}=x+\delta x$ and $\hat{y}=y+\delta y$ are measurements of two
numbers $x$ and $y$, respectively, with a bivariate PDF
$p_\delta(\delta x,\delta y)$ for the noise in the measurement. Our
aim shall be to constrain the ratio \mbox{$R=\sqrt{y/x}$}. The
posterior PDF $p(R|\hat{x},\hat{y})$ of $R$ given $\hat{x}$ and
$\hat{y}$ can be written as the marginal PDF 
\begin{eqnarray}
  p(R|\hat{x},\hat{y})&=&
  \int\d x\;p(R,x|\hat{x},\hat{y})\\
  &\propto&
  \int\d x\;{\cal L}(\hat{x},\hat{y}|R,x)\,p(x)\, p(R)\\
  &=& p(R)\, \int\d
  x\;p_\delta\Big(\hat{x}-x,\hat{y}-R^2\,x\Big)\,p(x)\,,
\end{eqnarray}
where ${\cal L}(\hat{x},\hat{y}|R,x)$ shall be the likelihood of
$(\hat{x},\hat{y})$ given a value pair $(x,R)$, and the product
$p(x)\,p(R)$ is the joint prior of $(x,R)$ \citep[see][for a
introduction to Bayesian statistics]{gelman2003bayesian}. We see that
the integral in the last line,
\begin{equation}
  \label{eq:illustration}
  {\cal L}(\hat{x},\hat{y}|R):=
  \int\d x\;p_\delta\Big(\hat{x}-x,\hat{y}-R^2\,x\Big)\,p(x)\;,
\end{equation}
has to be the likelihood of $(\hat{x},\hat{y})$ for a given ratio
$R$. We are thus essentially fitting a two-parameter model $(R,x)$ to
\mbox{$\hat{y}=R^2\,x$} and \mbox{$\hat{x}=x$} followed by a
marginalising over $x$. Coming back to our statistical analysis of the
aperture statistics, $y$ and $x$ would be here $f^2_{\rm
  b}\,\ave{{\cal N}^2}$ and $\ave{M^2_{\rm ap}}$, for example, and
\mbox{$R=f_{\rm b}\,\ave{{\cal N}^2}^{1/2}\,\ave{M^2_{\rm
      ap}}^{-1/2}$} is the (projected) bias factor $b_{\rm 2D}$. For
our full analysis, however, we have to jointly constrain $b_{\rm 2D}$
and the correlation factor $r_{\rm 2D}$ for a set of aperture radii
$\theta_i$ in a more general approach.

To implement a general approach involving $n_{\rm d}$ aperture radii
and both the bias and correlation factors for all radii
simultaneously, we combine the measurements of aperture moments inside
the data vector with the (observed) elements
\begin{equation}
  d_j=\left\{
    \begin{array}{ll}
      \ave{{\cal N}^2(\theta_j)}\;, & 1\le j\le n_{\rm d}\\
      \ave{{\cal N}M_{\rm ap}(\theta_{j-n_{\rm d}})} \;, &n_{\rm
        d}<j\le2n_{\rm d}\\
      \ave{M^2_{\rm ap}(\theta_{j-2n_{\rm d}})} \;, &2n_{\rm
        d}<j\le3n_{\rm d}
    \end{array}
  \right.\;,
\end{equation}
and we fit this vector by the parameters
$\vec{m}(\vec{\Theta},\vec{x})$ with template parameters
$\vec{\Theta}$ (Table \ref{tab:model}) and (theoretical) vector
elements
\begin{equation}
  m_j(\vec{\Theta},\vec{x})
 =\left\{
   \begin{array}{ll}
     \frac{\ave{{\cal N}^2}_{\rm th}(\theta_j;b)\,x_j}{\ave{M^2_{\rm ap}}_{\rm th}(\theta_j)}\;, & 1\le
     j\le n_{\rm d}\\\\
     \frac{\ave{{\cal N}M_{\rm ap}}_{\rm th}(\theta_{j-n_{\rm
           d}};b,r)\,x_{j-n_{\rm d}}}{\ave{M^2_{\rm ap}}_{\rm th}(\theta_{j-n_{\rm d}})} \;,
   & n_{\rm
     d}<j\le2n_{\rm d}\\\\
     x_{j-2n_{\rm d}}\;, & 2n_{\rm
       d}<j\le3n_{\rm d}
   \end{array}
 \right.
\end{equation}
using a PDF $p_\delta(\delta\vec{d})$ that accounts for the correlated
noise $\delta\vec{d}$ in the aperture statistics. The details of this
PDF are given below. We note that the explicit normalisation $f_{\rm
  b}(\theta_i)$ and $f_{\rm r}(\theta_i)$ disappears here because both
the theoretical and observed ratio-statistics $\{(b_{\rm
  2D}(\theta_i),r_{\rm 2D}(\theta_i))\}$ are normalised exactly the
same way. However, the normalisation is indirectly present through the
ratio of theoretical aperture moments in
$\vec{m}(\vec{\Theta},\vec{x})$ so that a wrong normalisation will
introduce a bias in the reconstruction. Similar to the previous
illustration, we integrate over the nuisance parameter
$x_i=\ave{M^2_{\rm ap}(\theta_i)}$ to obtain the marginal likelihood
\begin{equation}
  \label{eq:likelihood}
  {\cal L}(\vec{d}|\vec{\Theta})=
  \int\d^{n_{\rm d}}x\; p_\delta\Big(\vec{d}-\vec{m}(\vec{\Theta},\vec{x})\Big)\,
  p(\vec{x})\;.
\end{equation}
We adopt a uniform prior $p(\vec{x})$ for $\vec{x}$ with the
additional condition that the variance of the aperture mass has to be
positive or zero.

The measurement noise $\delta\vec{d}$ in the aperture statistics
approximately obeys Gaussian statistics which is characterised by a
noise covariance \mbox{$\mat{N}=\ave{\delta\vec{d}\,\delta\vec{d}^{\rm
      T}}$}; the mean $\ave{\delta\vec{d}}$ vanishes by
definition. The exact covariance $\mat{N}$, however, is unknown so
that we estimate $\mat{N}$ from the data themselves by
$\widehat{\mat{N}}$, obtained with $n_{\rm jk}$ jackknife realisations
of the data (Appendix \ref{sect:estimators}). We include the
uncertainty of $\hat{\mat{N}}$ in the posterior of the spatial biasing
functions by analytically marginalising over its statistical error. As
shown in \citet{2016MNRAS.456L.132S}, this produces for Gaussian
$\delta\vec{d}$ a multivariate $t$-distribution for the noise model
$p_\delta(\delta\vec{d})$,
\begin{equation}
  \label{eq:likecond}
  -2\ln{p_\delta(\delta\vec{d})}
  =
  {\rm const}+
  n_{\rm jk}\,
  \ln{\left(1+\frac{\chi^2}{n_{\rm jk}-1}\right)}\;,
\end{equation}
where $\chi^2:=\delta\vec{d}^{\rm
  T}\,\widehat{\mat{N}}^{-1}\,\delta\vec{d}$.

To approximately evaluate \Ref{eq:likelihood}, we perform a numerical
Monte-Carlo integration
\begin{eqnarray}
  {\cal L}(\vec{d}|\vec{\Theta})&=&
  \int\d^{n_{\rm d}}\!x\,q(\vec{x})\,
 \frac{p_\delta\Big(\vec{d}-\vec{m}(\vec{\Theta},\vec{x})\Big)\,p(\vec{x})}
  {q(\vec{x})}
  \\
  \label{eq:likemargin}
  &\approx&
  \frac{1}{n_x}
  \sum_{i=1}^{n_x}
  \frac{p_\delta\Big(\vec{d}-\vec{m}(\vec{\Theta},\vec{x}_i)\Big)\,p(\vec{x}_i)}
  {q(\vec{x}_i)}\;,
\end{eqnarray}
for which
\begin{equation}
  \label{eq:importance}
  -2\ln{q(\vec{x})}=
  {\rm const}+
  \Big(\vec{x}-\vec{d}_{\rm map}\Big)^{\rm T}\mat{N}^{-1}_{\rm map}
  \Big(\vec{x}-\vec{d}_{\rm map}\Big)
\end{equation}
is a so-called importance function of the Monte-Carlo integral, and
$d_{{\rm map},j}=\ave{M^2_{\rm ap}(\theta_j)}$ are the measured
variances of the aperture mass at $\theta_j$; the vectors
$\vec{x}_i\sim q(\vec{x})$ are $n_{\rm x}$ random realisations of the
importance function; the matrix $\mat{N}_{\rm map}^{-1}$ denotes our
estimate for the inverse covariance of noise in $\vec{d}_{\rm map}$,
that is that of $\ave{M^2_{\rm ap}}$ alone, which we also obtain from
jackknife samples and the estimator in \cite{2007A&A...464..399H}. The
purpose of the importance function $q(\vec{x})$ is to improve the
convergence of the Monte-Carlo sum \Ref{eq:likemargin} by producing a
higher density of sampling points $\vec{x}_i$ where the most of the
probability mass of
\mbox{$p_\delta(\vec{d}-\vec{m}(\vec{\Theta},\vec{x}))$} is located
\citep[e.g.][]{2010MNRAS.405.2381K}. We note that for any $q(\vec{x})$
the sum always converges to the same ${\cal L}(\vec{d}|\vec{\Theta})$
as long as $q(\vec{x})$ is proper and \mbox{$q(\vec{x})>0$} for all
$\vec{x}$.  To save computation time, we initially prepare
\mbox{$n_x=10^3$} realisations $\vec{x}_i$ and reuse these for every
new estimation of the marginal likelihood in \Ref{eq:likemargin}.

We explore the posterior distribution of parameters $\vec{\Theta}$ in
the template, that is
\begin{equation}
  \label{eq:posterior}
  p(\vec{\Theta}|\vec{d})=
  E^{-1}(\vec{d})\,
  {\cal L}(\vec{d}|\vec{\Theta})\,p(\vec{\Theta})
  \propto{\cal L}(\vec{d}|\vec{\Theta})\,p(\vec{\Theta})\;,
\end{equation}
by applying sampling with the Multiple-Try Metropolis, where the
constant evidence $E(\vec{d})$ is not of interest here
\citep{2012arXiv1201.0646M}. We assume that the prior
$p(\vec{\Theta})$ is uniform on a linear scale for all parameters
within their defined boundaries, see Sect. \ref{sect:modimplement},
and \mbox{$0<b_{\rm ls}\le3$}. Different Monte-Carlo chains can be
combined by joining the different sets of sampling points from
independent Monte-Carlo runs. If the joint sample is too large to be
practical, a resampling can be applied. This means we randomly draw a
subset of points $\vec{\Theta}_i$ from the joint sample. Depending on
the details of the adopted MCMC algorithm, the probability of drawing
$\vec{\Theta}_i$ in the resampling has to be proportional to its
weight in case points are not equally weighted.

Finally to conclude the reconstruction, we map the Monte-Carlo
realisations of $\vec{\Theta}$ in the joint sample to a set of spatial
biasing functions. The final set then samples the posterior
distribution of $b(k)$ and $r(k)$.

\subsection{Marginalisation of errors in the galaxy-bias normalisation}

For our analysis, the fiducial cosmology and the intrinsic alignment
of sources is exactly known by the cosmological model in the mock
data. For future applications, however, it may be necessary to
additionally marginalise over an a priori uncertainty $p(\vec{\pi})$
of cosmological parameters $\vec{\pi}$ for the normalisation of the
galaxy bias, meaning that the $\vec{\Theta}$ posterior is
\begin{equation}
  \label{eq:post}
  p(\vec{\Theta}|\vec{d})\propto
  \int\d\pi\;p(\vec{\pi})\,
  {\cal L}(\vec{d}|\vec{\Theta},\vec{\pi})\,
  p(\vec{\Theta})
  \approx
  \sum_{i=1}^{n_\pi}\,
  \frac{{\cal L}(\vec{d}|\vec{\Theta},\vec{\pi}_i)\,
  p(\vec{\Theta})}{n_\pi}\;,
\end{equation}
where ${\cal L}(\vec{d}|\vec{\Theta},\vec{\pi})$ is the likelihood of
$\vec{d}$ for a given set $\vec{\Theta}$ and fiducial cosmology
$\vec{\pi}$. Numerically the marginalisation over $\vec{\pi}$ can be
achieved, as indicated by the right-hand side of \Ref{eq:post}, by (i)
randomly drawing a realisation $\vec{\pi}_i$ from the prior
$p(\vec{\pi})$, (ii) by performing the Monte-Carlo sampling of the
posterior in Eq. \Ref{eq:posterior} for the fixed fiducial cosmology
\mbox{$\vec{\pi}_i\sim p(\vec{\pi})$}, and (iii) by combining the
different chains with varying $\vec{\pi}$. Concretely, let us call the
resulting Monte-Carlo sample from step (ii) ${\cal M}_i$.  We repeat
this step $n_\pi$ times for different cosmologies. For joining the
chains in step (iii), we randomly draw one $\vec{\Theta}_i$ from each
sample ${\cal M}_i$ to produce $n_\pi$ new vectors $\vec{\Theta}$ that
go into the final sample. We repeat this random selection of
$n_\pi$-tupels until the final sample has the desired size. We may
apply the same technique to also marginalise over errors in the
redshift distributions of lenses and sources, or the uncertainties in
the II and GI models.

\subsection{Galaxy number density as prior}

The halo model provides a prediction of the mean galaxy density
$\bar{n}_{\rm g}(\vec{\Theta})$, Eq. \Ref{eq:nbar}, that can be
included in the template fit to improve the constraints on the
otherwise poorly constrained pivotal mass $m_{\rm piv}$. We may
achieve this by adding the log-normal likelihood
\begin{equation}
  \ln{{\cal L}(\bar{n}_{\rm g}|\vec{\Theta})}=
  -\frac{(\log_{10}\bar{n}_{\rm g}^{\rm est}-\log_{10}\bar{n}_{\rm
      g}(\vec{\Theta}))^2}{2\sigma_{\rm logn}^2}\;,
\end{equation}
to the logarithm of the marginal likelihood in
\Ref{eq:likemargin}. Here we denote by $\sigma_{\rm logn}^2$ the
root-mean-square (RMS) error of the logarithmic number density
$\bar{n}_{\rm g}^{\rm est}$ estimated from the data.

A reasonable prior on $\bar{n}_{\rm g}$ can also be found if
$\bar{n}_{\rm g}^{\rm est}$ is not available as it is assumed here.
The number density of galaxies is for redshifts $z\lesssim2$ typically
of the order of $10^{-2}$ to $10^{-1}\,h^3\,\rm Mpc^{-3}$, or smaller
for sub-samples \citep[e.g.][]{2016ApJ...830...83C}. Therefore in the
reconstruction of our biasing functions, we employ a weak Gaussian
prior of
\begin{equation}
  \log_{10}{(\bar{n}_{\rm g}^{\rm est}\,h^{-3}\rm Mpc^3)}\pm\sigma_{\rm
    logn}=-3\pm2
\end{equation}
for the galaxy number density, and we impose an upper limit of
$\bar{n}_{\rm g}(\vec{\Theta})\le1\,h^3\,\rm Mpc^{-3}$ to prevent an
nonphysically high number density of galaxies. We found that the upper
limit improves the convergence of the MCMCs as chains can get stuck at
low values of $m_{\rm piv}$ with unrealistically high values of
$\bar{n}_{\rm g}$.

\section{Results}
\label{sect:results}

In the following, we report our results for the reconstructed biasing
functions for the galaxy samples SM1 to SM6, RED, and BLUE inside the
two redshift bins low-$z$ (\mbox{$\bar{z}_{\rm d}\approx0.36$}) and
high-$z$ (\mbox{$\bar{z}_{\rm d}\approx0.52$}). We concentrate on the
reconstruction accuracy and precision although the template parameters
found in the reconstructions are also available in the Appendix
\ref{sect:physicaldetails}. If not stated otherwise, the results are
for mock sources with a shape-noise dispersion
\mbox{$\sigma_\epsilon=0.3$} and without reduced shear. As additional
test of the methodology, we use generic templates for a non-physical
model of the spatial biasing functions and compare the results to
those of our physical templates. Furthermore, we estimate the
systematic error in the bias normalisation originating from various
conceivable sources. The final sub-section is a demonstration of our
technique with data from the Garching-Bonn Deep Survey (\gabods).

\subsection{Reconstruction accuracy and precision}
\label{sect:mockanalysis}

\begin{figure*}[htb!]
  \begin{center}
    \epsfig{file=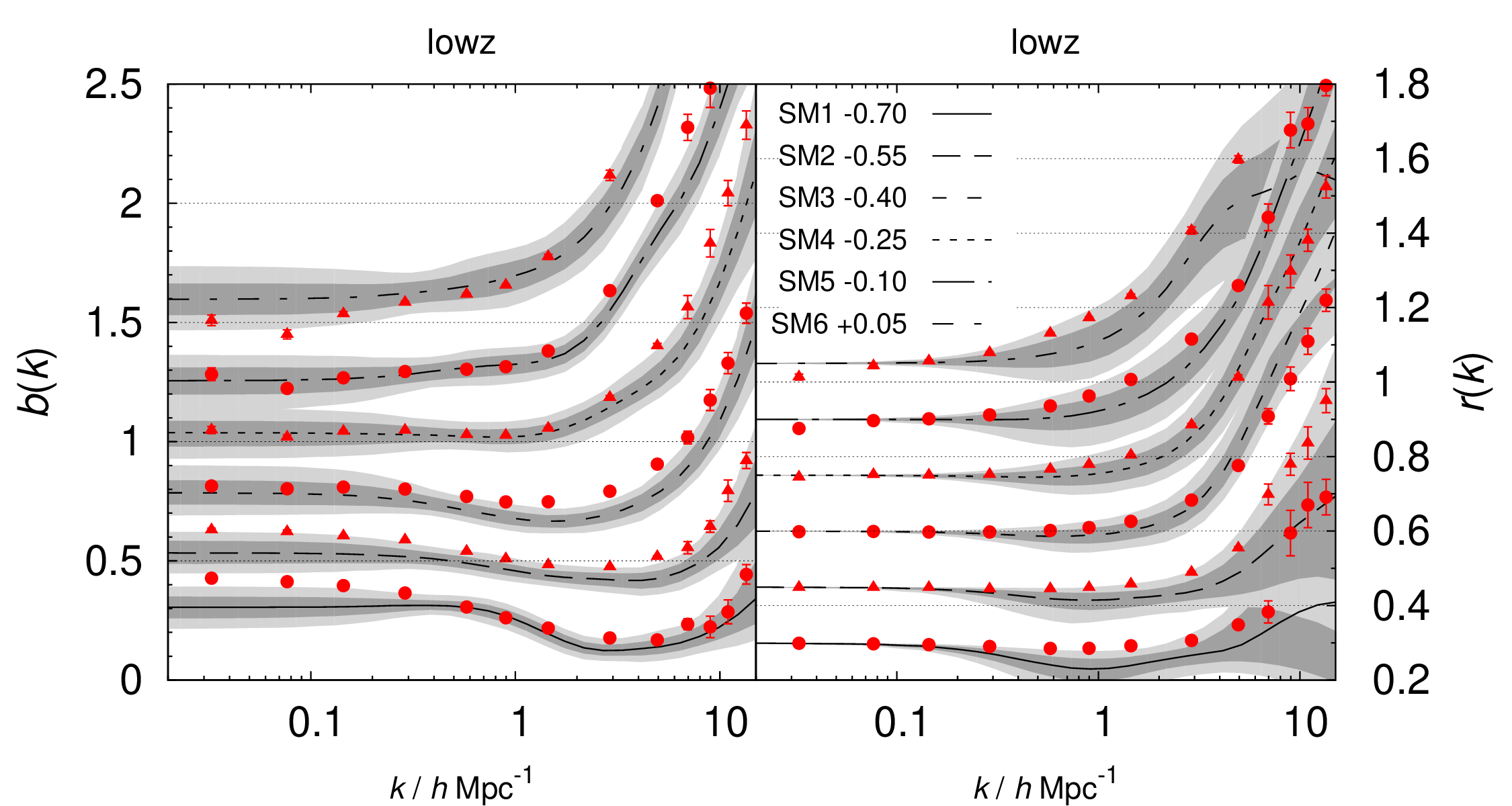,width=90mm,angle=-90}
    \vspace{0.5cm}
    \epsfig{file=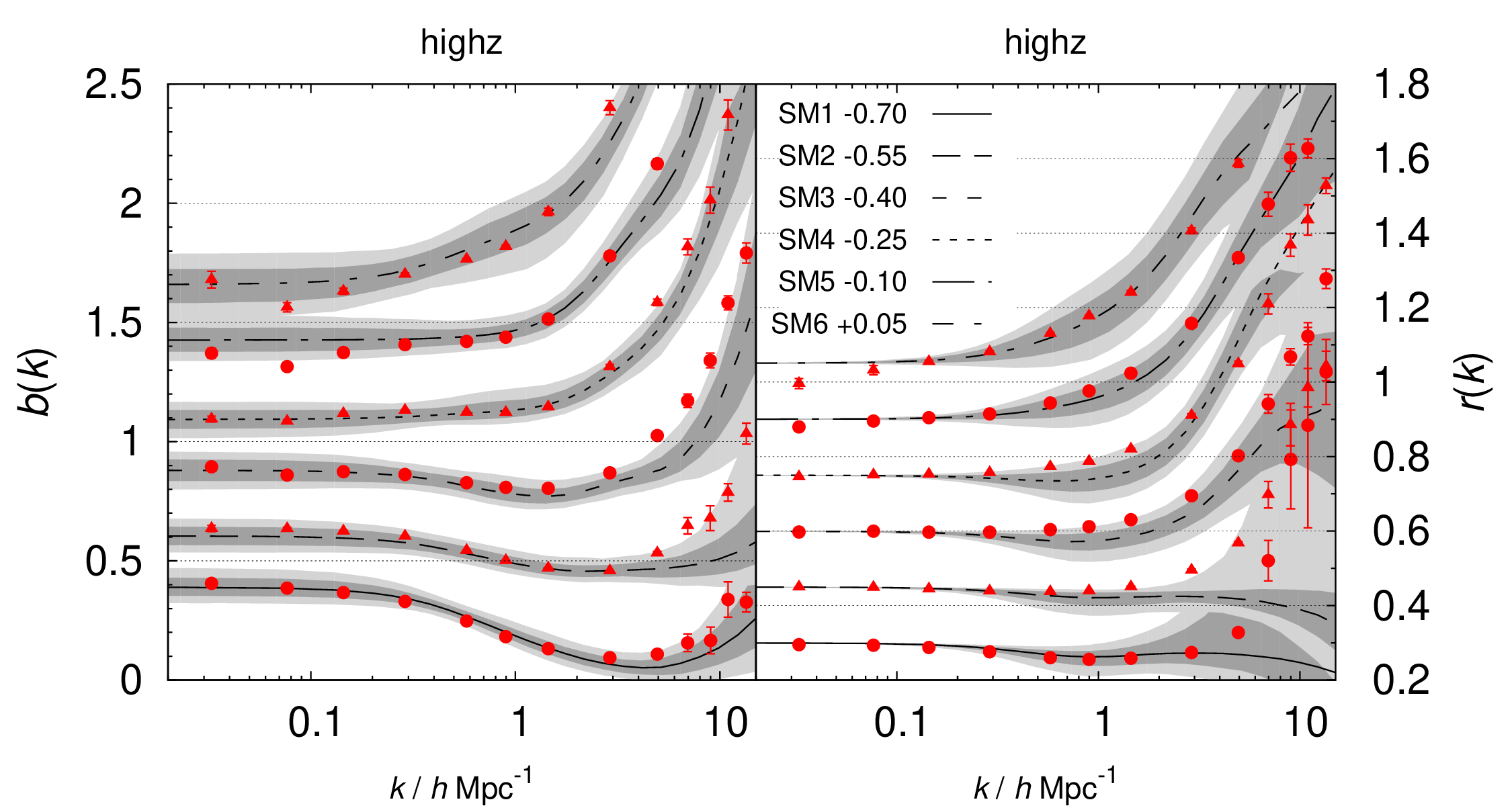,width=90mm,angle=-90}
    \vspace{-0.5cm}
  \end{center}
  \caption{\label{fig:brofksm} Biasing functions $b(k)$ (left panels)
    and $r(k)$ (right panels) for all mock galaxy samples SM1 to SM6
    and two redshift bins. The top figure is for the low-$z$ samples
    ($\bar{z}_{\rm d}\approx0.36$); the bottom figure for the high-$z$
    samples ($\bar{z}_{\rm d}\approx0.52$). The shaded regions
    indicate the $68\%$ and $95\%$ PI of the reconstructed biasing
    functions.  The red data points are the true basing functions for
    comparison. For more visibility, we shifted the biasing functions
    by the constant value in the figure key.}
\end{figure*}

\begin{figure*}
  \begin{center}
    \hspace{-0.5cm}
    \epsfig{file=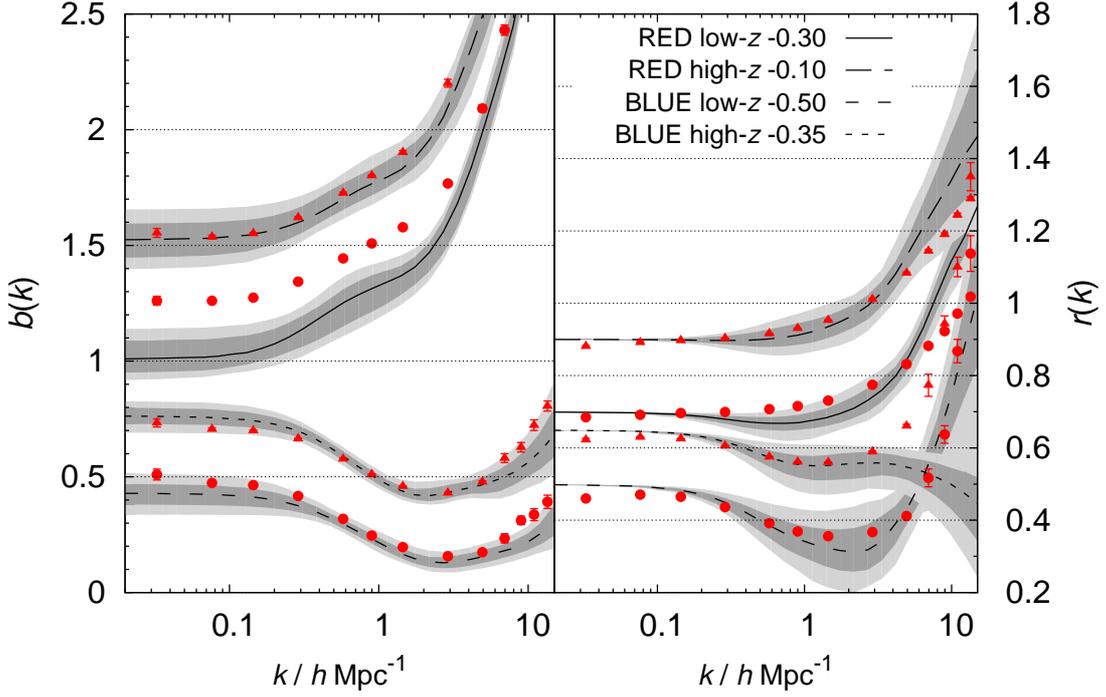,width=98mm,angle=-90}
  \end{center}
  \caption{\label{fig:brofkredblue} As in Fig. \ref{fig:brofksm} but
    now for the colour-selected samples RED and BLUE.}
\end{figure*}

\renewcommand{\arraystretch}{1.3}
\begin{table*}
  \caption{\label{tab:accuracy} Overview of the reconstruction accuracy by listing the mean fractional errors $\sigma_{\rm b,r}$ and extreme outliers $\Delta_{\rm b,r}$ of the inferred biasing functions $b(k)$ and $r(k)$, respectively, in per cent.}
  \begin{center}
    
 \begin{tabular}{lccccccccc}
   \hline\hline
           & \multicolumn{2}{c}{\emph{physical} low-$z$} & \multicolumn{2}{c}{\emph{physical} high-$z$} & & \multicolumn{2}{c}{\emph{generic} low-$z$} & \multicolumn{2}{c}{\emph{generic} high-$z$}\\
    Sample & $\sigma_{\rm b}~(\Delta_{\rm b})$ & $\sigma_{\rm r}~(\Delta_{\rm r})$ & $\sigma_{\rm b}~(\Delta_{\rm b})$ & $\sigma_{\rm r}~(\Delta_{\rm r})$ & & $\sigma_{\rm b}~(\Delta_{\rm b})$ & $\sigma_{\rm r}~(\Delta_{\rm r})$ & $\sigma_{\rm b}~(\Delta_{\rm b})$ & $\sigma_{\rm r}~(\Delta_{\rm r})$ \\
    \hline\\
    SM1 & 4.9~(3.0$\sigma$) & 3.9~(1.9$\sigma$) & 3.0~(2.1$\sigma$) & 1.4~(1.4$\sigma$) && 3.4~(1.8$\sigma$) & 4.9~(0.9$\sigma$) & 4.4~(\textbf{3.8}$\sigma$) & \textbf{6.0}~(\textbf{4.6}$\sigma$)  \\
    SM2 & \textbf{6.2}~(\textbf{3.1}$\sigma$) & 2.7~(1.9$\sigma$) & 4.4~(\textbf{3.9}$\sigma$) & 4.6~(\textbf{7.2}$\sigma$) && \textbf{6.9}~(\textbf{7.5}$\sigma$) & 4.0~(1.2$\sigma$) & 4.2~(\textbf{3.2}$\sigma$) & 1.9~(1.5$\sigma$)  \\
    SM3 & \textbf{6.5}~(2.8$\sigma$) & 1.7~(1.7$\sigma$) & 3.4~(1.8$\sigma$) & 2.9~(1.9$\sigma$) && \textbf{5.8}~(\textbf{3.4}$\sigma$) & 1.7~(1.6$\sigma$) & 2.4~(2.2$\sigma$) & 1.5~(1.5$\sigma$) 
\\
    SM4 & 3.2~(1.9$\sigma$) & 2.0~(1.0$\sigma$) & 2.4~(1.6$\sigma$) & 3.5~(1.7$\sigma$) && 2.8~(2.7$\sigma$) & 0.9~(1.1$\sigma$) & 4.1~(2.3$\sigma$) & 1.6~(1.3$\sigma$)  \\
    SM5 & 3.1~(1.4$\sigma$) & 3.2~(1.2$\sigma$) & 3.8~(1.5$\sigma$) & 1.4~(0.6$\sigma$) && 3.1~(1.6$\sigma$) & 0.9~(0.5$\sigma$) & \textbf{19.3}~(\textbf{3.5}$\sigma$) & \textbf{9.2}~(2.2$\sigma$)  \\
    SM6 & \textbf{5.2}~(1.4$\sigma$) & 4.5~(1.8$\sigma$) & \textbf{5.6}~(1.4$\sigma$) & 2.5~(1.0$\sigma$) && \textbf{5.8}~(1.7$\sigma$) & \textbf{5.7}~(1.5$\sigma$) & \textbf{7.6}~(2.5$\sigma$) & \textbf{7.9}~(1.3$\sigma$)  \\
    RED & \textbf{9.4}~(2.2$\sigma$) & 4.6~(\textbf{3.1}$\sigma$) & 3.0~(1.1$\sigma$) & 2.0~(1.2$\sigma$) && \textbf{11.1}~(\textbf{37.6}$\sigma$) & 3.7~(\textbf{6.9}$\sigma$) & \textbf{10.3}~(\textbf{19.8}$\sigma$) & 2.7~(\textbf{5.7}$\sigma$)  \\
    BLUE & 4.6~(\textbf{3.8}$\sigma$)  & 1.7~(1.4$\sigma$) & 3.2~(1.8$\sigma$) & \textbf{6.0}~(\textbf{5.6}$\sigma$) && \textbf{5.4}~(2.7$\sigma$)  & 3.1~(1.1$\sigma$) & 2.7~(1.9$\sigma$) & \textbf{8.9}~(2.0$\sigma$)  \\
   \hline\\

$\ave{\sigma_{\rm b,r}}$ & $5.4\pm2.9$ & $3.0\pm1.7$ & $3.6\pm1.7$ & $3.0\pm2.0$ && $5.5\pm3.4$ & $3.1\pm2.2$ & $6.9\pm6.2$ & $5.0\pm3.9$\\
$\ave{\Delta_{\rm b,r}}$ & $2.5\pm1.3\sigma$ & $1.8\pm0.9\sigma$ & $1.9\pm1.1\sigma$ & $2.6\pm2.6\sigma$ && $7.4\pm12.7\sigma$ & $1.9\pm2.2\sigma$ & $4.9\pm6.3\sigma$ & $2.5\pm1.9\sigma$\end{tabular}

  \end{center}
  \tablefoot{The columns `physical' refer to results with a physical
    model (Sect. \ref{sect:mockanalysis}), `generic' columns list the
    results with generic fitting functions
    (Sect. \ref{sect:generic}). Quoted values are for the errors in the
    domain $k\in[0.05,10]\,h\,\rm Mpc^{-1}$ for $b(k)$ and
    $k\in[0.3,10]\,h\,\rm Mpc^{-1}$ for $r(k)$. The values
    $\Delta_{\rm b}$ and $\Delta_{\rm r}$ inside the brackets are the
    most significant deviations between reconstructed and true biasing
    functions. Errors $\sigma_{\rm b,r}\ge5\%$ or outliers $\Delta_{\rm b,r}\ge3\sigma$ are
    quoted in boldface. Values in the last rows with $\ave{\sigma_{\rm b,r}}$ 
    (or $\ave{\Delta_{\rm b,r}}$) are  averages and dispersions for 
    $\sigma_{\rm b}$ and $\sigma_{\rm r}$ (or $\Delta_{\rm b}$ and $\Delta_{\rm r}$) of
    all samples in the same redshift bin.} 
\end{table*}
\renewcommand{\arraystretch}{1.0} 

The Figs. \ref{fig:brofksm} and \ref{fig:brofkredblue} are a direct
comparison of our reconstructed biasing functions for all samples
(shaded regions) to the true $b(k;\bar{z})$ and $r(k;\bar{z})$ in the
three-dimensional simulation cube of the MS shown as red data points;
we use the snapshot redshifts $\bar{z}=0.362,0.509$ for low-$z$ and
high-$z$ respectively. The shaded region indicate the 68\% and 95\%
posterior intervals (PI) of our posterior constraints.  In order to
accommodate many reconstructions, we have shifted the biasing function
along the $y$-axes by a constant value that is indicated in the legend
of each plot. We note that most functions are shifted downwards so
that relative errors might appear larger than in reality. The left
panels show $b(k)$, the right panels $r(k)$. Figure \ref{fig:brofksm}
displays only reconstructions for the stellar-mass samples where the
top row is for the low-$z$ samples and the bottom row for the high-$z$
samples. Similarly, Fig. \ref{fig:brofkredblue} shows the results for
the RED and BLUE samples, now low-$z$ and high-$z$ combined in one
figure.

Overall we find a good agreement between a reconstruction and the true
biasing functions although significant disagreements are also
visible. Most prominently, we find disagreements at large scales, this
means at small wave numbers \mbox{$k\approx0.05\,h\,\rm Mpc^{-1}$},
for the low-$z$ $b(k)$ of RED, SM1, and SM5; or at small scales,
\mbox{$k\approx10\,h^{-1}\,\rm Mpc$}, for the high-$z$ $r(z)$ of SM2
or BLUE; the function low-$z$ $b(k)$ of SM2 and SM3 is a few per cent
offset on all scales which may be an indication of a normalisation
error. The disagreement at high \mbox{$k\gtrsim10\,h^{-1}\,\rm Mpc$}
could be related to insufficient sampling by our MCMC because the
results improve significantly for samples without shape noise which
reduces the statistical error at $\theta_{\rm ap}\approx2^\prime$ (not
shown). It is also possible that the statistical model of the
likelihood in Eq. \Ref{eq:likecond} is inaccurate and, as a
consequence, underestimates the error distribution in the tail of the
posterior at large $k$.

To quantify the method accuracy we compare the reconstruction $b(k)$
or $r(k)$ to the true biasing function by the following metrics
$\sigma_{\rm f}^2$ and $\Delta_{\rm f}$; the subscript `f' is either
`b' for $b(k)$ or `r' for $r(k)$. The metrics compare the biasing
functions at a discrete set $\{k_i:i=1\ldots n_k\}$ of $n_k=10$ wave
numbers between \mbox{$0.05\le k\le10\,h\,\rm Mpc^{-1}$}, which we
equally space on a log-scale. In the equations, we denote by $f(k)$
the posterior median of either $b(k)$ or $r(k)$ in the reconstruction,
and $\sigma^2(k)$ is the variance of the posterior at a given $k$. In
addition, we denote by $f_{\rm true}(k)$ the true biasing function and
by $\sigma_{\rm true}^2(k)$ its standard error.  The variance
$\sigma^2_{\rm true}(k)$ is indicated by the error bars of the red
data points in the Figs. \ref{fig:brofksm} and \ref{fig:brofkredblue};
it is usually negligible compared to $\sigma^2(k)$. Our first metric
\begin{equation}
  \label{eq:metricfirst}
  \sigma_{\rm f}^2=
  \left(\sum_{i=1}^{n_k}\sigma_i^{-2}\right)^{-1}\,
  \sum_{i=1}^{n_k}\sigma_i^{-2}\,\left(\frac{f(k_i)}{f_{\rm true}(k_i)}-1\right)^2  
\end{equation}
then quantifies the average fractional error over the range of $k$,
weighted by the inverse statistical error
$\sigma_i^2=\sigma^2(k_i)+\sigma^2_{\rm true}(k_i)$. For $\sigma_{\rm
  r}^2$, we change the lower limit of $k$ to $0.3\,h\,\rm Mpc^{-1}$ to
avoid a seemingly too optimistic metric: by definition $r(k)$ is in
the reconstruction close to the true $r(k)=1$ of the MS data which
makes $\sigma_i$ relatively small and therefore assigns too much
weight to $k\lesssim0.3\,h\,\rm Mpc^{-1}$. The second metric
\begin{equation}
  \Delta_{\rm f}=
  \max{\left\{
      \sigma_i^{-1}\,\left|f(k_i)-f_{\rm true}(k_i)\right|:i=1\ldots n_k
    \right\}}
\end{equation}
yields the most significant deviation in units of $\sigma_i$; it is a
measure for the strongest outlier within the $k$-range. 

Table \ref{tab:accuracy} lists $\sigma_{\rm f}$ (in per cent) and
$\Delta_{\rm f}$ for all galaxy samples and redshift bins; the last
rows are averages and dispersions for each table column. The table
consists of two blocks of which we summarise the left-hand columns
`physical' here and the right-hand column `generic' in the
Sect. \ref{sect:generic} hereafter. The values for $\sigma_{\rm b}$
are typically in the range $5.4\pm2.9\%$ for low-$z$ samples and
slightly better with $3.6\pm1.7\%$ for the high-$z$ samples. The
accuracy of $\sigma_{\rm r}$ is consistently $3.0\pm2.0\%$ for both
redshift bins. For the outlier statistics, we find on average
$\Delta_{\rm b}=2.2\pm1.2\sigma$ and $\Delta_{\rm r}=2.2\pm1.8\sigma$
for all redshifts, which, however, can attain high values of
$6-7\sigma$ in a few cases; see high-$z$ BLUE and SM2 for instance. We
find these high values to be associated with mismatches of $r(k)$ at
\mbox{$k\approx10\,h\,\rm Mpc^{-1}$}. This corresponds to
\mbox{$\theta_{\rm ap}\approx1^\prime$}, thus to the lower limit of
the angular scales that we sample in the mock analysis (cf. bottom and
top $x$-axis in Fig. \ref{fig:models}).

Moreover, we quantify the statistical precision of our reconstruction
at wave number $k_i$ by the ratio of $\sigma(k_i)$ and the median of
the posterior of either $f(k_i)$. For an average over all galaxy
samples and the reconstruction within the range \mbox{$0.05\le
  k\le10\,h\,\rm Mpc^{-1}$}, we find a precision of $6.5\pm2.1\%$ for
$b(k)$ and $5.5\pm5.7\%$ for $r(k)$; we combine the low-$z$ and
high-$z$ samples because the precision is very similar for both
bins. The errors denote the RMS variance of the precision.

In summary, we find that a method accuracy of around $5\%$ with most
significant deviations at scales of \mbox{$k\gtrsim10\,h\,\rm
  Mpc^{-1}$} which, however, are not supported by the measurements and
have to be extrapolated by the templates. The statistical precision of
the reconstructions is typically between $5-10\%$ for our fiducial
survey and lens samples.

\subsection{Deprojection with generic templates}
\label{sect:generic}

We repeat the reconstruction of $b(k)$ and $r(k)$ for our mock data
with the Pad\'e approximants
\begin{equation}
  \label{eq:generic}
  b(k)=\frac{b_0+b_1\,k+b_2\,k^2}{1+b_3\,k+b_4\,k^2}
  ~;~
  r(k)=\frac{1+r_1\,k+r_2\,k^2}{1+r_3\,k+r_4\,k^2+r_0\,k^3}\;
\end{equation}
as generic templates of the biasing functions in
Sect. \ref{sect:statanalysis} (without a $\bar{n}_{\rm g}$
regularisation). The $b_i$ and $r_i$ denote ten coefficients which we
restrict to \mbox{$|b_i|,|r_i|\le100$} in the fit.  These generic
model templates are related to the fitting function for $b^2(k)$ in
\citet{2005MNRAS.362..505C}. We found that the Pad\'e approximants are
very good descriptions of the red data points in the
Figs. \ref{fig:brofksm} and \ref{fig:brofkredblue}. By fitting generic
templates we therefore investigate whether the foregoing inaccuracies
in the reconstruction with the physical templates might be related to
a model bias. If this is the case, we should obtain a better
reconstruction here. We note that the particular approximant of $r(k)$
asserts \mbox{$r\to1$} for \mbox{$k\to0$}, and that in the generic
templates, unlike the physical templates, $b(k)$ is independent from
$r(k)$.

Compared to the halo model, the generic templates produce a similar
(low-$z$) or somewhat worse (high-$z$) reconstruction but is prone to
more extreme deviations from the true biasing functions. The
right-hand block of values `generic' in Table \ref{tab:accuracy}
summarises the metrics of the reconstructions with the generic
templates and compares them to the metrics with the physical templates
`physical' on the left-hand side. We find an increased inaccuracy for
the high-$z$ samples, especially for SM5, SM6, and RED; in particular
the reconstruction of low-$z$ RED has not improved here. The worse
reconstruction for high-$z$ is because of the inability of the generic
templates to extrapolate to small spatial scales which is more
important for high-$z$ where the same angular range corresponds to
larger spatial scales. In a few cases, the generic templates produce
very significant deviations, mostly on small scales and indicated by
$\Delta_{\rm b,r}$, which are absent in the physical templates.

\subsection{Errors in the galaxy-bias normalisation}
\label{sect:calbias}

\begin{figure*}
  \begin{center}
    \epsfig{file=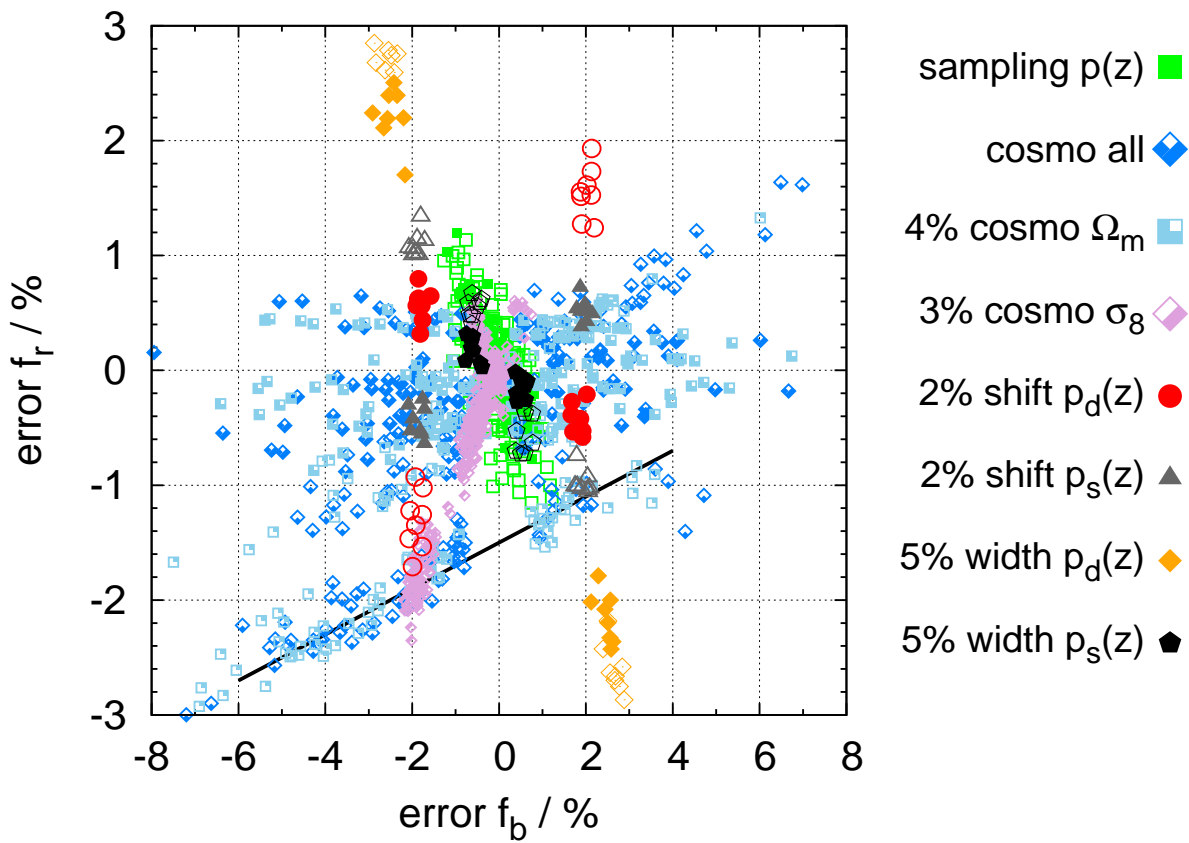,width=125mm,angle=-90}
    \vspace{-0.5cm}
  \end{center}
  \caption{\label{fig:calbias} Average error in the galaxy-bias
    normalisation $f_{\rm b}$ ($x$-axis) and $f_{\rm r}$
    ($y$-axis). The points show indistinguishable the errors of all
    galaxy samples SM1-4, BLUE, and RED together; the point styles
    indicate the redshift bin and what is varied. Symbols as in the
    figure key indicate the low-$z$ samples, inverted symbols indicate
    the high-$z$ samples (e.g. solid and open circles). The `cosmo'
    and `sampling p(z)' data points reuse the same galaxy samples many
    times with random normalisation errors. The solid line marks the
    estimated error for high-$z$ samples due to the baryon
    uncertainty. The fiducial cosmology is WMAP9. See text for more
    details.}
\end{figure*}

\renewcommand{\arraystretch}{0.8}
\begin{table}
  \caption{\label{tab:errors}Summary of possible systematic errors and
    their expected impact on the reconstruction of $b(k)$ or $r(k)$
    for a WMAP9 cosmology and our galaxy samples.}
  \begin{center}
    \begin{tabular}{lll}
      \hline\hline
      \\
      Origin & Error $b(k)$ & Error $r(k)$ \\
      \hline
      \\
      intr. align. $|A_{\rm ia}|\approx2$ & $\lesssim5.0\%$ & $\lesssim5.0\%$
      \\\\
      fiducial cosmology \\
      and model of $P_{\rm m}(k;\chi)$ & $2.8\%$ ($3.0\%$) & $0.4\%$ ($1.1\%$) 
      \\\\
      lens $p_{\rm d}(z)$; $\delta_\sigma=5\%$ & $2.5\%$ & $2.2\%$ ($2.7\%$) 
      \\\\
      lens $p_{\rm d}(z)$; $\delta_{z}=1\%$ & $1.9\%$ & $0.5\%$ ($1.4\%$) 
      \\\\
      source $p_{\rm s}(z)$; $\delta_{z}=1\%$ & $1.9\%$ & $0.5\%$ ($1.0\%$) 
      \\\\
      shear bias $m=1\%$ & $1.0\%$ & $0.0\%$
      \\\\
      shear bias $c\approx10^{-3}\%$ & $<1.0\%$ & $<1.0\%$
      \\\\
      source $p_{\rm s}(z)$; $\delta_\sigma=5\%$ & $0.8\%$ & $0.5\%$ ($0.3\%$)
      \\\\
      reduced shear & $\lesssim0.5\%$ & $\lesssim0.5\%$
      \\\\
      sampling noise of $p(z)$ & $0.4\%$ ($0.6\%$) & $0.4\%$ ($0.5\%$) 
      \\
      \hline
    \end{tabular}
  \end{center}
  \tablefoot{Values in brackets are for the high-$z$ samples ($\bar{z}_{\rm d}\approx0.52$) which are only shown if they differ from the low-$z$ values ($\bar{z}_{\rm d}\approx0.35$). Sources have a mean redshift of $\bar{z}_{\rm s}=0.93$. By $\delta_{z}$ and $\delta_\sigma$ we denote the relative error in the mean redshift and the redshift dispersion, respectively, which refer to either the lens redshift distribution, $p_{\rm d}(z)$, or that of the sources, $p_{\rm s}(z)$. We assume a constant residual shear bias $m$ here.}
\end{table}
\renewcommand{\arraystretch}{1.0}

The ratio statistics are normalised with respect to unbiased galaxies
in a fiducial model. Systematic errors in the normalisation affect the
amplitude of the deprojected biasing functions. Therefore, we explore
the robustness of the overall amplitude of $b(k)$ and $r(k)$ with
respect to changes in the fiducial cosmology and the adopted redshift
distributions in the normalisation; see the Eqs. \Ref{eq:calfb} and
\Ref{eq:calfr} that are evaluated for the unbiased galaxies. We note
that $f_{\rm b}$ and $f_{\rm r}$ normally show little dependence on
$\theta_{\rm ap}$ so that changes in the fiducial model mainly scale
the projected biasing functions up or down.

The functions $f_{\rm b}(\theta_{\rm ap})$ and $f_{\rm r}(\theta_{\rm
  ap})$ shall be the correct normalisation of the galaxy bias. For
Fig. \ref{fig:calbias}, we then compute $f_{\rm b}^\prime(\theta_{\rm
  ap})$ (and $f_{\rm r}^\prime(\theta_{\rm ap})$) for variations in
the normalisation parameters, and we compute the quadratic mean of
relative errors \mbox{$\delta_{\rm b}(\theta_{\rm ap})=f_{\rm
    b}^\prime(\theta_{\rm ap})/f_{\rm b}(\theta_{\rm ap})-1$} over the
angular range \mbox{$1^\prime\le\theta_{\rm ap}\le140^\prime$}. The
data points inside the figure indicate these means $\ave{\delta_{\rm
    b}^2(\theta_{\rm ap})}^{1/2}$ ($x$-axis) and $\ave{\delta_{\rm
    r}^2(\theta_{\rm ap})}^{1/2}$ ($y$-axis) for particular lens
samples. To have a good representation of the scatter between possible
lens-galaxy samples, we show results for all galaxy samples SM1-SM6,
RED, and BLUE in the same redshift bin together by the same point
style if they are subject to the same parameter variation.  We give
the normalisation errors a plus sign if the average of $\delta_{\rm
  b}(\theta_{\rm ap})$ is positive, and a negative sign
otherwise. This flags $b(k)$ (or $r(k)$) that are overall too high
(positive) or too low (negative). We apply variations relative to a
default model which has: WMAP9+eCMB+BAO+$H_0$ cosmological parameters
\citep{2013ApJS..208...19H}; redshifts distributions as shown in
Fig. \ref{fig:pofz}; a non-linear matter power-spectrum according to
\citet{2012ApJ...761..152T}. Inside the plot, data points have the
styles shown in the figure key for low-$z$ samples and an inverted
point style for high-$z$ samples, such as solid circles (low-$z$) and
open circles (high-$z$). We vary the following parameters in the
default model to quantify their impact on the normalisation.
\begin{itemize}
\item The data points `cosmo all' randomly draw combinations of
  cosmological parameters from an error distribution centred on the
  fiducial model
  \begin{multline}
    \label{eq:wmap9}
    \vec{\pi}=(\Omega_{\rm m},\Omega_\Lambda,\Omega_{\rm b},n_{\rm s},h,\sigma_8)
    \\
    =(0.288,0.712,0.0472,0.971,0.6933,0.83)\;.
  \end{multline}
  In this distribution, errors are uncorrelated and Gaussian with a
  dispersion of
  \begin{equation}
    \sigma_\pi=(4\%,{\rm n/a},2\%,1\%,2\%,3\%)
  \end{equation}
  relative to the fiducial $\vec{\pi}$. The exception is
  $\Omega_\Lambda$ which we set to $\Omega_\Lambda=1-\Omega_{\rm m}$
  in all realisations (a fixed \mbox{$K=0$} geometry). These errors
  are on the optimistic side but consistent with constraints from
  combined cosmological probes. In addition for each set of
  parameters, we plot data points for three different transfer
  functions of $P_{\rm m}(k;\chi)$: \citet{1986ApJ...304...15B}, and
  \citet{1998ApJ...496..605E} with and without BAOs. These are
  combined with two different \texttt{Halofit} models of the
  non-linear power spectrum: \citet{Smith03} and the more accurate
  \citet{2012ApJ...761..152T}. By these variations we mean to broadly
  account for model uncertainties in the non-linear power spectrum
  which produces extra scatter in the plot. In particular, the 10-20\%
  difference between the two versions of \texttt{Halofit} in the
  regime $k\gtrsim1\,h\,\rm Mpc^{-1}$ accounts to some extend for the
  theoretical uncertainty of baryons on the small-scale power spectrum
  \citep[e.g.][]{2017arXiv170703397J,2016MNRAS.463.3326F,2015MNRAS.450.1212H,2011MNRAS.417.2020S}. We
  find that errors in the cosmological parameters or the non-linear
  power spectrum mainly affect the normalisation of $b(k)$ which can
  be off by about $3.0\%$ (68\% confidence level, CL hereafter).  The
  error in $r(k)$ is within $1.1\%$ (68\% CL) for high-$z$ or smaller
  $0.4\%$ (68\% CL) for the low-$z$ samples (solid symbols). The
  straight line inside the figure indicates the locus of errors for
  the high-$z$ samples that are produced by the baryon uncertainty in
  the non-linear power spectrum.
\item For `cosmo $\Omega_{\rm m}$', we only vary $\Omega_{\rm m}$ in
  the cosmological parameters with the foregoing dispersion. This
  results in a distribution of data points that is very similar to
  `cosmo all'. For comparison, `cosmo $\sigma_8$' varies only
  $\sigma_8$. The scatter is now restricted to a small region.
  Therefore, the normalisation error owing to cosmological parameters
  is mainly explained by the variations in $\Omega_{\rm m}$.
\item For the data points `sampling $p(z)$', we add random shot noise
  to the redshift distributions. The idea here is that redshift
  distributions are estimated from a sub-sample of galaxies which
  gives rise to sampling noise in the estimated distributions used for
  normalisation; see e.g. \citet{2017MNRAS.465.1454H} which use a
  weighted sample of spectroscopic redshifts to model the redshift PDF
  of the full galaxy sample. To emulate the sampling shot-noise, we
  randomly draw $n$ redshifts \mbox{$z\sim p(z)$} from the true $p(z)$
  to build a finely binned histogram of a noisy redshift distribution
  ($\Delta z=0.015$). We then employ this histogram for $f^\prime_{\rm
    b}$ and $f^\prime_{\rm r}$.  As fiducial values for our $1024\,\rm
  deg^2$ survey, we adopt $n=10^4$ for the lenses and $n=10^5$ for the
  sources. These fiducial values imply that we estimate $p(z)$ from
  spectroscopic redshifts of \mbox{$\sim0.5\%$} of the sources and
  roughly $1\%, 2\%, 20\%, 1\%$ of the lenses in the samples SM1, SM4,
  SM6, RED/BLUE, respectively.  The result is a similar scatter for
  the low-$z$ and the high-$z$ samples in Fig.\ref{fig:pofz}. The
  error is typically within $0.5\%$ for $b(k)$ and $r(k)$ (68\% CL).
\item The data points `shift $p_{\rm d}(z)$' vary the mean in the lens
  redshift distribution. For this, we systematically shift $z\mapsto
  z\,(1+\delta_z)$ by $\delta_z=\pm2\%$, which is twice as large as
  the typical error on the mean redshift reported in
  \citet{2017MNRAS.465.1454H}. The impact differs for the low-$z$
  (solid circles) and the high-$z$ samples (open circles). For
  systematically higher redshifts in low-$z$, this means $\delta_z>0$,
  $b(k)$ is too large and $r(k)$ is too low. For high-$z$ and
  \mbox{$\delta_z>0$}, we find that both $b(k)$ and $r(k)$ are too
  high in amplitude. For \mbox{$\delta_z<0$}, the effects are exactly
  reversed. The overall systematic normalisation error is nevertheless
  not greater than typically $2\%$ for $b(k)$ and $1-2\%$ for $r(k)$.
\item The data points `width $p_{\rm d}(z)$' vary the width of the
  lens redshift distribution. This we emulate by mapping $p_{\rm
    d}(z)\mapsto p_{\rm d}(z)^{1/(1-\delta_\sigma)^2}$ to a new PDF
  that is then used for the normalisation. For a Gaussian density
  $p_{\rm d}(z)$, this maps the dispersion to
  $\sigma\mapsto\sigma\,(1-\delta_\sigma)$ while leaving the mean and
  Gaussian shape in the new PDF unchanged. For skewed distributions,
  \mbox{$\delta_\sigma\ne0$} also moves the mean of the PDF. To
  account for this unwanted (small) side effect, we shift every PDF to
  assure that it retains its original mean redshift. We consider
  \mbox{$\delta_\sigma=\pm5\%$} here. The effect of squeezing, this
  means \mbox{$\delta_\sigma>0$}, is similar for low-$z$ and high-$z$:
  $b(k)$ is too low, $r(k)$ is too high, with errors of around $2-3\%$
  for $b(k)$ and $r(k)$. A stretching, $\delta_\sigma<0$, has the
  reverse effect on both redshift bins.
\item The data points `shift $p_{\rm s}(z)$' and `width $p_{\rm
    s}(z)$' explore the effect of errors in the mean or width of the
  source $p_{\rm s}(z)$.  Shifting by \mbox{$\delta_z=+2\%$} produces
  too high $b(k)$ for low-$z$ and high-$z$ ($1.9\%$), too high $r(k)$
  for low-$z$ ($0.5\%$), and a too low $r(k)$ for high-$z$
  ($1.0\%$). The reverse behaviour is present for systematically lower
  redshifts with \mbox{$\delta_z=-2\%$}. Changes in the width of the
  distribution with \mbox{$\delta_\sigma=\pm5\%$} have a $0.5\%$
  effect for $b(k)$ and $r(k)$, with low-$z$ samples being slightly
  less affected: a systematically wider distribution gives a too low
  $b(k)$ and a too high $r(k)$; the reverse effects apply for
  systematically narrower distributions, that is for
  \mbox{$\delta_\sigma>0$}.
\item The intrinsic alignment of sources contributes to both
  $\ave{M^2_{\rm ap}}$ and $\ave{{\cal N}M_{\rm ap}}$ and thereby can
  have an impact on $b_{\rm 2D}$ and $r_{\rm 2D}$. We account for this
  in the normalisation by II and GI models; see
  Sect. \ref{sect:IIandGI}. If unaccounted for, as assumed here, we
  bias $b_{\rm 2D}$ by the error in $\ave{M_{\rm ap}^2}^{1/2}_{\rm
    th}$ that is used in the normalisation $f_{\rm b}$,
  Eq. \Ref{eq:calfb}. This error is plotted in
  Fig. \ref{fig:fbrGIandII} for varying values of $A_{\rm ia}$ and
  angular scales $\theta_{\rm ap}$. The normalisation error in $r_{\rm
    2D}$ is determined by error in $\ave{{\cal N}M_{\rm ap}}_{\rm
    th}^{-1}\,\ave{M_{\rm ap}^2}^{1/2}_{\rm th}$ used for $f_{\rm r}$,
  Eq. \Ref{eq:calfr}, which is plotted in Fig. \ref{fig:frGIhighz} for
  SM4 high-$z$ as example; the errors of other high-$z$ samples are
  comparable. For the low-$z$ samples, the overlap of lens and source
  redshifts is small so that the error in $\ave{{\cal N}M_{\rm
      ap}}_{\rm th}^{-1}$ is negligible compared to the error in
  $\ave{M_{\rm ap}^2}^{1/2}_{\rm th}$. Therefore, the normalisation
  error for $r_{\rm 2D}$ in low-$z$ samples is approximately that of
  $b_{\rm 2D}$ in Fig. \ref{fig:fbrGIandII}. For \mbox{$|A_{\rm
      ia}|\lesssim2$}, the normalisation error of $b_{\rm 2D}$ and
  $r_{\rm 2D}$ is typically within $\pm5\%$ at scales
  \mbox{$\theta_{\rm ap}\gtrsim1^\prime$}.
\end{itemize}

A summary of normalisation errors and their estimated magnitude is
listed in Table \ref{tab:errors}.  We find that the response to errors
in the redshift distributions is approximately linear for $\delta_{z}$
and $\delta_\sigma$ that are within several per cent so that the
quoted values could be scaled.

\begin{figure}
  \begin{center}
    \epsfig{file=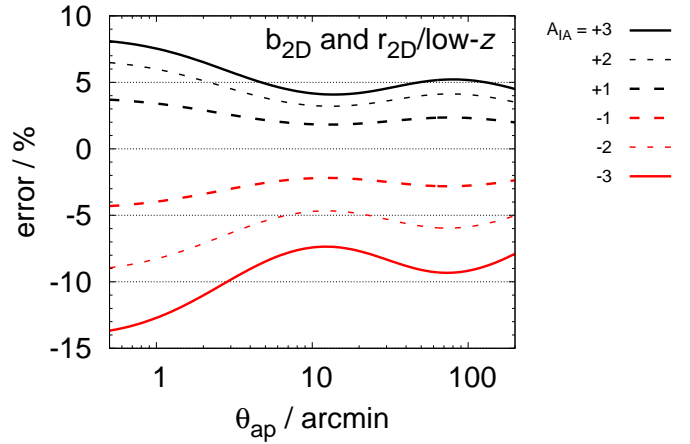,width=65mm,angle=-90}
  \end{center}
  \caption{\label{fig:fbrGIandII} Systematic relative errors in
    $b_{\rm 2D}(\theta_{\rm ap})$ (low-$z$ and high-$z$) and $r_{\rm
      2D}(\theta_{\rm ap})$ (only low-$z$) when II and GI terms are
    ignored in the normalisation of the galaxy bias.  Different lines
    show predictions for different values of $A_{\rm ia}$ with sources
    as in our mock survey. The fiducial cosmology is WMAP9.}
\end{figure}

\begin{figure}
  \begin{center}
    \epsfig{file=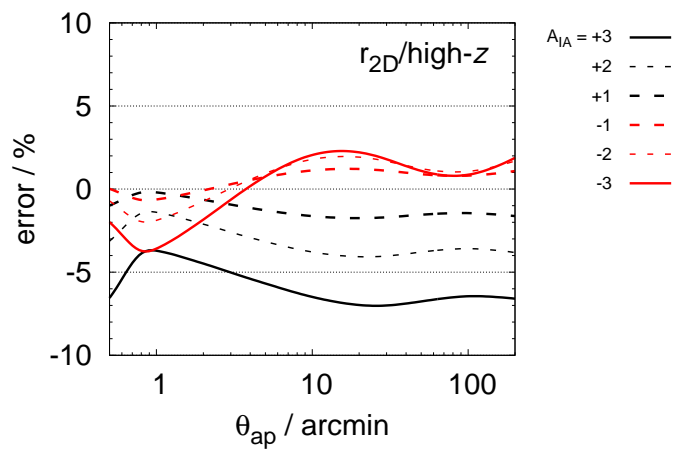,width=65mm,angle=-90}
  \end{center}
  \caption{\label{fig:frGIhighz} As in Fig. \ref{fig:fbrGIandII} but
    now for $r_{\rm 2D}(\theta_{\rm ap})$ high-$z$. Shown are results
    form SM4 but the values are similar for the other samples.}
\end{figure}

\subsection{Shear bias and reduced shear}
\label{sect:othersys}

As another source of systematic error, we consider a residual bias in
the shear estimators that has not properly corrected for in the
lensing pipeline. Following \citet{2012MNRAS.423.3163K} (K+12
hereafter), we quantify a shear bias by $\ave{\gamma}=(1+m)\,\gamma+c$
for average estimated shear $\ave{\gamma}$ in an ensemble of sources
that are subject to the same $\gamma$: $m$ is the so-called
multiplicative bias and $c$ is the additive bias. For a crude estimate
of the impact of $m$ on the measurement of $b_{\rm 2D}(\theta_{\rm
  ap})$ and $r_{\rm 2D}(\theta_{\rm ap})$, we assume a constant and
real-valued $m$. A value of \mbox{$m\ne0$} produces a bias of $1+m$ in
the measured aperture statistics $\ave{M_{\rm ap}^2}^{1/2}$ and
$\ave{{\cal N}M_{\rm ap}}$. Therefore, applying our methodology while
ignoring $m$ will scale the amplitude of $b(k)$, Eq. \Ref{eq:b2dobs},
by $(1+m)^{-1}=1-m+{\cal O}(m^2)$ but it will not change $r(k)$ in
Eq. \Ref{eq:r2dobs}. Contemporary lensing techniques reach a typical
accuracy of \mbox{$|m|\approx1\%$}, therefore we expect a similarly
small systematic error for $b(k)$ (K+12).

A residual additive bias $c$ does not affect the aperture statistics
if it is constant.  If, on the other hand, $c$ varies at a scale
within the sensitive $\ell$-range of the aperture filter, we could
have significant contributions to the measured $\ave{M_{\rm ap}^2}$,
depending on the power of the $c$-fluctuations. Our polynomial filter
in Eq. \Ref{eq:apfilter} has its maximum sensitivity for the angular
wave number \mbox{$\ell_{\rm c}\approx4.25/\theta_{\rm
    ap}\approx1.5\times10^4\,(\theta_{\rm ap}/1^\prime)^{-1}$} or
angular scale $\theta_{\rm c}=2\pi/\ell_{\rm
  c}\approx1.44\,\theta_{\rm ap}$ \citep{1998A&A...334....1V}.  The
typical residual amplitudes of $c$ \emph{after} a calibration
correction of $\xi_\pm$ are of the order of $10^{-5}$ (K+12; Appendix
D4 in \citealt{2017MNRAS.465.1454H}) so that systematic errors owing
to $c$ fluctuations are probably below a per cent for
\mbox{$\ave{M^2_{\rm ap}}^{1/2}\gtrsim10^{-3}$}, which is the case for
\mbox{$\theta_{\rm ap}\lesssim2\,\rm deg$} and typical sources with
\mbox{$z_{\rm s}\approx1$}; see the data points in
Fig. \ref{fig:GIandII}. The statistic $\ave{{\cal N}M_{\rm ap}}$ is
not affected by the additive shear bias in the likely absence of
correlations between lens positions and fluctuations of $c$, or is
presumably corrected for by subtracting the correlation between random
lens positions and shear in the data; see the estimator in
Eq. \Ref{eq:estggl}.

With regard to reduced shear, our analysis assumes that the
$\epsilon_i$ are estimates of shear $\gamma(\vec{\theta}_i)$, whereas
they are in reality estimates of the reduced shear
\mbox{$g_i=\gamma_i/(1-\kappa_i)$}. While $\ave{\epsilon_i}=\gamma_i$
is a good approximation for weak gravitational-lensing and
substantially simplifies the formalism in
Sect. \ref{sect:projectedbias}, we will have some systematic error. To
quantify this error, we redo the reconstruction of the biasing
functions for a new shear catalogue where the intrinsic source
ellipticities are sheared by $g_i$ rather than $\gamma_i$; source
positions and intrinsic shapes do not match between the old and new
catalogues.  For the new catalogues, we obtain a set of values
$\sigma_{\rm f}^{\rm red}$, Eq. \Ref{eq:metricfirst}, which we
statistically compare to the previous values $\sigma_{\rm f}$ in Table
\ref{tab:accuracy} by fitting an average parameter $\delta_{\rm red}$
for the relative difference, defined by \mbox{$\sigma_{\rm f}^{\rm
    red}=(1+\delta_{\rm red})\,\sigma_{\rm f}$}, to all samples and
redshift bins. For all values of $\sigma_{\rm b}$ and $\sigma_{\rm r}$
combined, we find no significant differences between the new and old
shear catalogues, this means $\delta_{\rm red}$ is consistent with
zero; the upper limit is $\delta_{\rm red}\lesssim13\%$ ($68\%$
CL). For an average of $\ave{\sigma_{\rm f}}=3.8\%$, the additional
inaccuracy due to reduced shear is therefore less than
\mbox{$13\%\times\ave{\sigma_{\rm f}}\approx0.5\%$}.

\subsection{Garching-Bonn Deep Survey}
\label{ap:gabods}

Finally, we apply our procedure in a first demonstration to data in
the \gabods~ (\citealt{2007A&A...461..861S}, SHS07 hereafter;
\citealt{2007A&A...468..859H}).  Because of its comparatively small
effective survey area of roughly 15 square degree, the statistical
power of \gabods~ is no longer competitive to measurements in
contemporary surveys. Nevertheless, the results presented here shed
some new light on the nature of the lens galaxies in SHS07 and round
off the past \gabods~ analysis. We plan to apply our methodology to
more recent lensing data in an upcoming paper.

\begin{figure}[htb]
  \begin{center}
  \epsfig{file=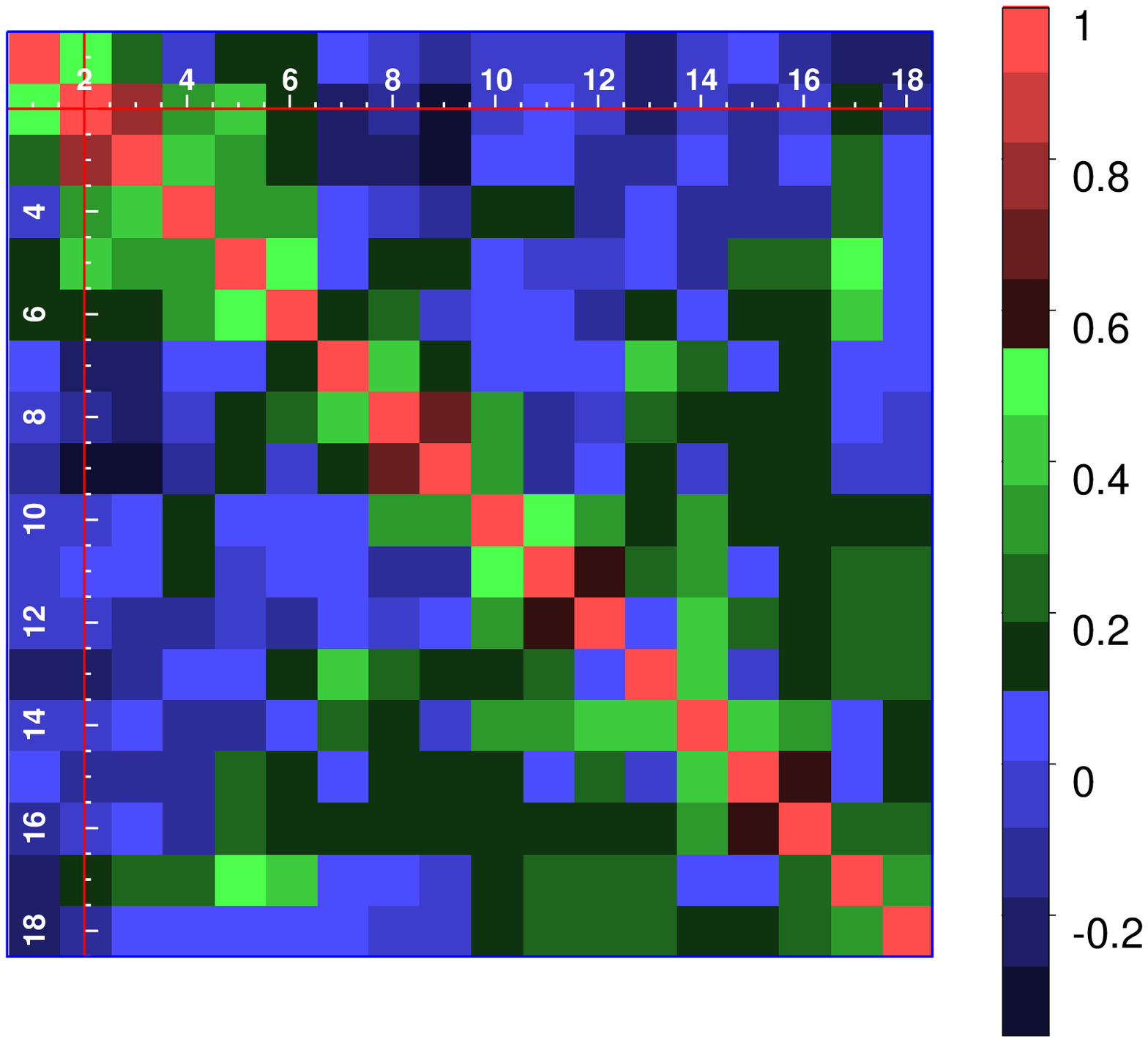,width=100mm,angle=0}
  \end{center}
  \caption{\label{fig:gabodscorr} Correlation matrix $C_{ij}$ of
    measurements errors for three kinds of aperture statistics of
    FORE-I lenses in the \gabods~ analysis. The integers on the two
    axes inside the matrix refer to either the $i$ or $j$
    index. Values $1\le k\le6$ for $k$ being either $i$ or $j$ refer
    to errors of $\ave{{\cal N}^2(\theta_k)}$, values of $7\le k\le12$
    to $\ave{{\cal N}M_{\rm ap}(\theta_{k-6})}$, and values $13\le
    k\le18$ to $\ave{M_{\rm ap}^2(\theta_{k-12})}$. The aperture
    scales $\{\theta_k/{\rm arcmin}\}$ are
    $\{2,3.3,5.3,8.7,14.1,23\}$. The matrix is estimated from 52
    jackknife samples.}
\end{figure}

As lens sample in \gabods~ we choose FORE-I galaxies, which comprise
\mbox{$R\le21.0$} flux-limited galaxies with mean redshift
$\bar{z}_{\rm d}=0.35$; the RMS dispersion of the lens redshifts is
0.16. The source galaxies are flux-selected between $21.5\le R\le24.0$
and have $\bar{z}_{\rm s}=0.68$; see Figure 3 in SHS07 for the
redshift distributions of lenses and sources in these samples. For the
estimators, we bin the two-point correlation functions
\Ref{eq:estxipm}-\Ref{eq:estomega} between 7 arcsec and 46 arcmin
using 4100 linear bins and merge the catalogues of the $n_{\rm
  patch}=52$ \gabods~ fields also used in SHS07. In contrast to SHS07,
we only use six aperture scales between 2 and 23 arcmin, equidistant
on a logarithmic scale, because of the strong correlation of errors
between similar aperture scales. The correlation matrix of statistical
(jackknife) errors can be found in
Fig. \ref{fig:gabodscorr}. Furthermore, we normalise the new
measurements by a WMAP9 cosmology, Eq. \Ref{eq:wmap9}. In contrast to
the foregoing analyses with our mock MS data, for which we measure the
aperture statistics up to degree scales, we here have to use
Eq. \Ref{eq:lsbias} to extrapolate the large-scale bias $b_{\rm ls}$,
which is then no longer a free parameter. For the halo bias-factor
$b_{\rm h}(m)$, needed in this extrapolation, we employ the fitting
formula in \citet{2005ApJ...631...41T}.  Owing to lacking information
on an IA of \gabods~ sources, we do assume \mbox{$A_{\rm ia}=0$}. A
value of \mbox{$|A_{\rm ia}|\lesssim2$} could therefore shift the
amplitude of $b_{\rm 2D}$ by up to 10 to 15 per cent, mainly because
of the GI term, and that of $r_{\rm 2D}$ by up to 2 per cent.

\begin{figure*}
  \epsfig{file=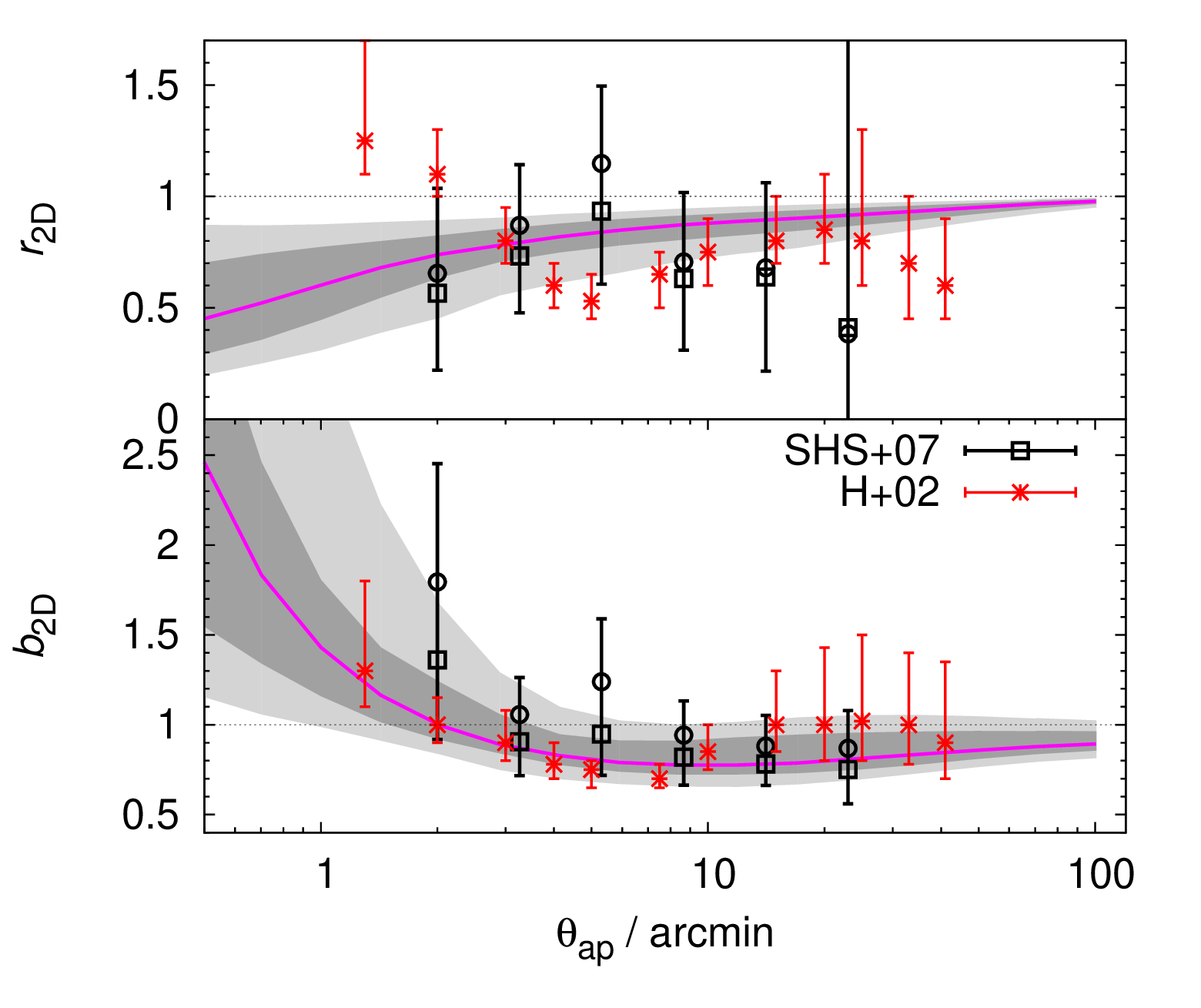,width=75mm,angle=-90}
  \epsfig{file=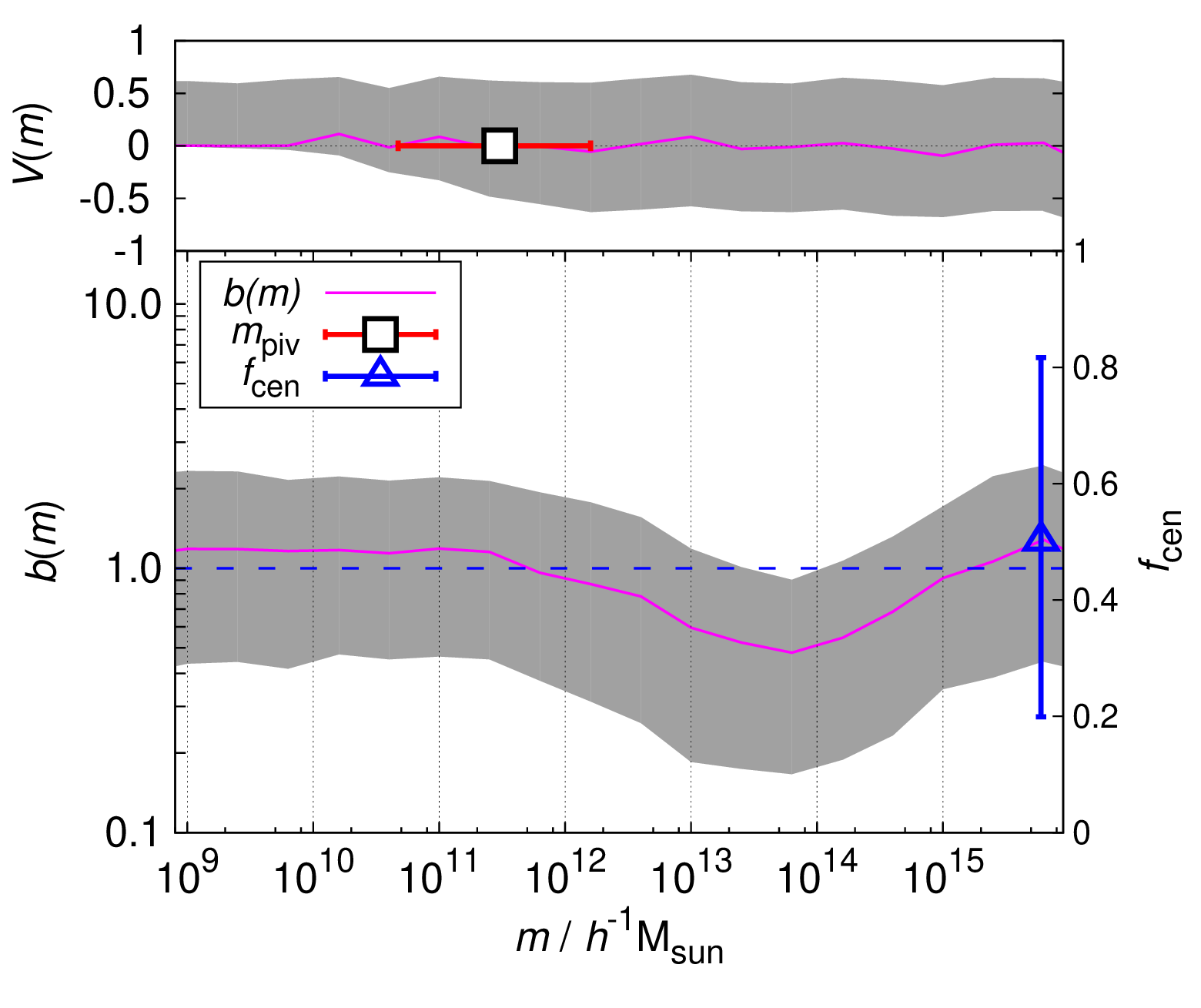,width=75mm,angle=-90}
  \caption{\label{fig:gabods} \emph{Left}: Posterior model of $r_{\rm
      2D}(\theta_{\rm ap})$ (top) and $b_{\rm 2D}(\theta_{\rm ap})$
    (bottom) based on the \gabods~ measurements \mbox{FORE-I} (shaded
    regions with 68\% and 95\% PI). Shown as black open squares are
    the median values and a 68\% interval around the median for the
    measured $b_{\rm 2D}$ and $r_{\rm 2D}$; the open circles indicate
    the mean. The red-star data points H+02 show the measurements by
    \citet{2002ApJ...577..604H} for comparison. \emph{Right}: 68\% PI
    posterior of the excess HOD variance $V(m)$ with open box for the
    mass scale of the pivotal mass $m_{\rm piv}$ (top); 68\% PI
    posterior of the mean biasing function $b(m)$ and $f_{\rm cen}$ as
    open triangle (bottom). The fiducial model has WMAP9 parameters.}
\end{figure*}

Our updated measurements are shown in the left panel of
Fig. \ref{fig:gabods} as $b_{\rm 2D}$ and $r_{\rm 2D}$ by the black
data points designated SHS+07. To obtain these points from the
observed aperture-moment statistics, we randomly draw realisations of
the aperture statistics from a Gaussian likelihood based on our
jackknife data covariance.  The open squares show the median and 68
percentiles of the normalised bias parameters from this Monte-Carlo
process, computed with Eqs. \Ref{eq:b2dobs} and \Ref{eq:r2dobs} for
each realisation; the open circles are the mean of the realisations
which are different to the median owing to the skewness in the error
distribution.  The shaded regions indicate the $68\%$ and $95\%$ PI of
the posterior (projected) biasing functions. The red stars are
measurements in \mbox{VIRMOS/DESCART}, broadly consistent with ours,
for flux-limited galaxies with a similar selection function
\citep{2002ApJ...577..604H}.

The right panel of Fig. \ref{fig:gabods} depicts the posterior of the
template parameters that provide a physical interpretation of the
galaxy bias. We take from here that the scale-dependence of the galaxy
bias mainly originates in a scale-dependence of $b(m)$: between halo
masses of $10^{13}$ to $10^{14}\,h^{-1}\msol$ there is a relative
scarcity of galaxies, which is qualitatively comparable to the BLUE
low-$z$ sample (see Fig. \ref{fig:deltag}).  The HOD variance is
consistent with a Poisson model, that means \mbox{$V(m)=0$}, albeit
only weakly constrained. The 68\% PI of the pivotal halo mass is
\mbox{$m_{\rm piv}=10^{11.48+0.72-0.81}\,h^{-1}\msol$}, and the
fraction \mbox{$f_{\rm cen}=0.50\pm0.31$} of halos open for central
galaxies is essentially the uniform prior which has the variance
$1/\sqrt{12}$ and the mean $0.5$. The posterior galaxy number density
is \mbox{$\bar{n}_{\rm g}=0.19^{+0.33}_{-0.13}\,h^3\rm Mpc^{-3}$}.

\begin{figure}
  \begin{center}
  \epsfig{file=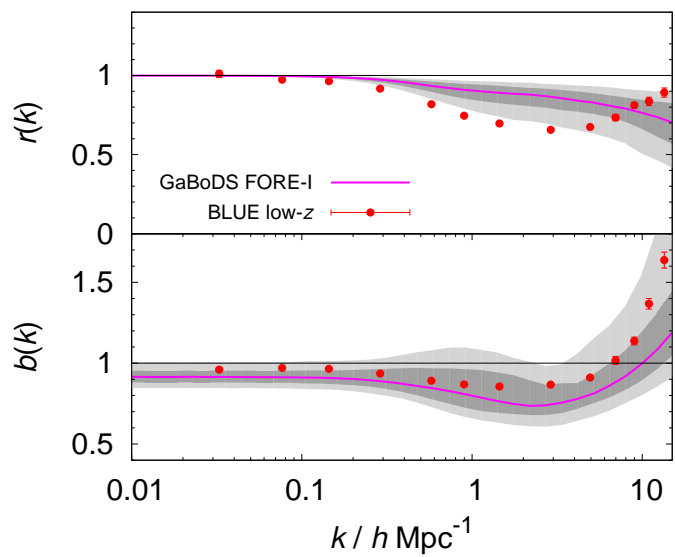,width=80mm,angle=-90}
  \end{center}
  \caption{\label{fig:gabodsbrofk} Reconstructed biasing functions of
    FORE-I galaxies in \gabods.  Shown are the 68\% and 95\% PI of
    $b(k)$ in the bottom panel and that of $r(k)$ in the top
    panel. The biasing function are an average over the redshift range
    $0.34\pm0.16$ for a WMAP9 cosmology. The red data points show the
    biasing function of BLUE low-$z$, which have a similar $b(m)$.}
\end{figure}

Fig. \ref{fig:gabodsbrofk} displays the posterior distribution of the
deprojected biasing functions and the 68\% PI for FORE-I galaxies. The
biasing functions are an average for the redshift range covered by the
lens galaxies. The \gabods~ data probe primarily the one-halo regime
\mbox{$\theta_{\rm ap}\lesssim20\,\rm arcmin$}; the large-scale bias
of \mbox{$b_{\rm ls}=0.92^{+0.04}_{-0.03}$} visible at $k\ll1\,h\,\rm
Mpc^{-1}$ is extrapolated. The red data points show the biasing
functions of BLUE low-$z$ for a qualitative comparison.

\section{Discussion}
\label{sect:discussion}

In this study, we have outlined and successfully tested a refined
technique to measure in contemporary lensing surveys the
scale-dependent galaxy bias down to non-linear scales of
\mbox{$k\sim10\,h^{-1}\,\rm Mpc$} for lens galaxies at
\mbox{$z\lesssim0.6$}.  To test our reconstruction technique, we
employ a fiducial survey with a sky coverage of $\sim1000\,\rm deg^2$,
and a photometry and a survey depth as in CFHTLenS. To construct
realistic samples of lenses and sources, we have prepared mock
catalogues that are consistent with those used in SES13 and
\cite{2016arXiv160808629S}. Despite some variations in survey depth
and area, these survey parameters are similar to the ongoing
Kilo-Degree Survey (KiDS), Dark Energy Survey (DES), or the survey
with the Hyper Suprime-Cam
\citep{2015MNRAS.454.3500K,2016PhRvD..94b2002B,2017arXiv170405858A}. If
the galaxy-bias normalisation is perfect, our technique applied to
these data can achieve a statistical precision within the range of
$5-10\%$ ($68\%$ CL), if similar lens and source samples are targeted,
and a slightly better accuracy of $3-7\%$ ($68\%$ CL), see Table
\ref{tab:accuracy}. For the high-$z$ samples, the accuracy will be
somewhat higher with $3-5\%$. On the other hand, it is clear from our
overview Table \ref{tab:errors} that the accuracy of the galaxy-bias
normalisation is in fact limited, mainly by our knowledge of the
intrinsic alignment of sources, cosmological parameters, and the
galaxy redshift-distributions. With a broad knowledge of
\mbox{$|A_{\rm ia}|\lesssim2$} and the specifications for the
normalisation errors in Table \ref{tab:errors}, we conclude that
systematic errors would potentially degrade the overall accuracy to
approximately $15\%$ for $b(k)$ and $10\%$ for $r(k)$. For fully
controlled intrinsic alignment of sources, these errors could be
reduced by $5\%$. An additional reduction by $3\%$ may be possible by
controlling the redshift distributions (their mean and variance) in
the normalisation to $1\%$ accuracy. For the fiducial cosmology, the
knowledge of $\Omega_{\rm m}$ is of most importance while the
normalisation of the ratio statistics is less affected by $\sigma_8$.

For a future method improvement, various problems could be of
interest: (i) approximations in the formalism or estimators of
Sect. \ref{sect:projectedbias}; (ii) an inaccurate statistical model
for the likelihood function; (iii) a model bias in the templates.  We
discuss a few problems in the following. With regard to our
statistical model, we find indeed evidence for deviations from a
Gaussian model of the joint aperture statistics which is explicitly
assumed in Eq. \Ref{eq:likelihood} (see Appendix
\ref{sect:nongauss}). However, the magnitude of a bias owing to a
Gaussian model is not clear and requires more research. For example,
deviations from a Gaussian distribution in broadly related
cosmological analyses with the aperture mass $M_{\rm ap}$ are reported
in \cite{2015MNRAS.449.1505S} and \cite{2009A&A...504..689H} where
non-Gaussian corrections to the likelihood produce insignificant
changes in one case but not in the other. Interestingly for our data,
the most inaccurate reconstruction (for small $k$) is that of RED
low-$z$ which shows a strong indication of a non-Gaussian
error-distribution for $\ave{{\cal N}^2}$ at large angular scales; see
Table \ref{tab:gausstest}. Moreover, our likelihood model employs an
error covariance that we estimate by the jackknife technique. The
jackknife technique is known to underestimate cosmic variance, in
particular for angular scales comparable to the size of sub-fields
used for the jackknife sample \citep{2016MNRAS.456.2662F}. However,
this problem is partly addressed in our analysis by using ratio
statistics which is less affected by cosmic variance
\citep{2011MNRAS.416.3009B}. While this may not be sufficient for
future surveys, it seems to be so for contemporary surveys because
cosmic variance is included in our assessment of the reconstruction
accuracy.  Finally, a model bias in our templates for $b(k)$ and
$r(k)$ is arguably unlikely, at least for our simulated galaxy
samples, because the purely generic models in Eq. \Ref{eq:generic} do
not produce a more accurate reconstruction of the biasing functions
although they are excellent fits to the true biasing functions (see
Table \ref{tab:accuracy}).  Nevertheless, a relevant model bias could
arise through our assumption of a non-evolving galaxy bias for galaxy
samples with a distance distribution that is broad compared to the
galaxy-bias evolution.

Our physical templates for the biasing functions $b(k)$ and $r(k)$ are
also insightful for a basic physical interpretation of the
scale-dependence of galaxy bias. On the one hand, the physical
parameters in the physical templates describe the HOD of the actual
galaxy population. On the other hand, these HOD parameters have only a
moderate accuracy because our relatively simple halo model lacks the
implementation of recently identified effects such as halo exclusion,
non-linear or stochastic halo clustering, assembly bias, galaxy
conformity, or a scale-dependent halo-bias function
\citep{2013PhRvD..88h3507B,2007MNRAS.377L...5G,2013MNRAS.430.1447K,2005ApJ...631...41T}. And
our model has a comparably simplistic treatment of central
galaxies. According to \cite{2012MNRAS.426..566C}, by taking ratios of
the aperture statistics we are, however, probably less sensitive to
these shortfalls in the halo model. We therefore expect the HOD
parameters in our templates not to be more accurate than $10-20\%$
compared to the true HOD in the lens sample, based on the reported
biases in the cited literature.  We stress that this does not
necessarily pose a problem for the deprojection as long as the
templates are good fits to the true biasing functions. With regard to
a basic interpretation of galaxy bias, we nevertheless take from the
discussion in Sect. \ref{sect:modeldiscussion} that central galaxies
and a non-Poisson HOD variance produce a scale-dependent bias most
prominently towards small scales, namely in the regime that is
dominated by low-occupancy halos with $m\lesssim m_{\rm piv}$. A
strong scale-dependence over a wider range of spatial or angular
scales and a non-monotonic behaviour may be produced by a mean biasing
function $b(m)$ that varies with halo mass $m$; in particular only
$b(m)$ affects the large-scale bias.  Interestingly here, the effect
of central galaxies is different from that of a non-Poisson variance:
central galaxies increase both $b(k)$ and $r(k)$ for larger $k$,
whereas a non-Poisson variance induces opposite trends for $b(k)$ and
$r(k)$. Therefore, the measurement of biasing functions can in
principle constrain both $b(m)$ and the excess variance $V(m)$ to test
galaxy models, although predictably with limited accuracy in
contemporary surveys (see Fig. \ref{fig:Vm}).

A demonstration of our reconstruction technique to data in the
\gabods~ suggests that the \mbox{$R\le21$} flux-limited sample of lens
galaxies FORE-I consists mainly of blue galaxies in the
field. Fig. \ref{fig:gabodsbrofk} reports our reconstruction of the
biasing functions for the FORE-I sample in
\cite{2007A&A...461..861S}. The physical parameters in the right panel
of Fig. \ref{fig:gabods} show that these galaxies tend to avoid halos
in the broad mass-range $10^{13}-10^{14}\,h^{-1}\,\msol$ and thereby
produce the relatively low (mean) values of \mbox{$b_{\rm
    2D}\approx0.8$} and \mbox{$r_{\rm 2D}\approx0.6$} and their
scale-dependence between a few and 20 arcmin (left panel); see also
the measurements by H+02 for similar lens galaxies with comparable
results. Consequently, they are presumably in majority field and group
galaxies. The reconstructed biasing functions also broadly match those
of BLUE low-$z$ which supports this interpretation. Clearly, the BLUE
low-$z$ sample does not have the same selection function as FORE-I so
that this comparison is certainly only qualitative. For a quantitative
test of galaxy models with more recent galaxy surveys, simulated and
observed galaxies have to be carefully selected to obtain consistent
samples. If this succeeds, both our little demonstration with the
$15\,\rm deg^2$ \gabods~ data and the multiplicity of biasing
functions visible in the Figs. \ref{fig:brofksm} and
\ref{fig:brofkredblue} promise useful constraints for galaxy models.

\section*{Acknowledgements}

We thank Hananeh Saghiha for preparing the RED and BLUE galaxy
samples. We also thank Catherine Heymans and Indiarose Friswell for
comments on the shear bias, and Peter Schneider for general comments
on the paper. This work has been supported by Collaborative Research
Center TR33 `The Dark Universe' and by the Deutsche
Forschungsgemeinschaft through the project SI 1769/1-1.  Patrick Simon
also acknowledges support from the German Federal Ministry for
Economic Affairs and Energy (BMWi) provided via DLR under project
no. 50QE1103. Stefan Hilbert acknowledges support by the DFG cluster
of excellence ‘Origin and Structure of the Universe’
(\url{www.universe-cluster.de}).

\bibliographystyle{aa}
\bibliography{galaxybias}

\appendix
\counterwithin{figure}{section}

\section{Impact of shot-noise subtraction on real-space biasing
  functions}
\label{ap:realspacecorr}

Let the functions $\xi_{\rm g}(x)=\ave{\delta_{\rm g}(0)\delta_{\rm
    g}(x)}$, $\xi_{\rm mg}(x)=\ave{\delta_{\rm g}(0)\delta_{\rm
    m}(x)}$, and $\xi_{\rm m}(x)=\ave{\delta_{\rm m}(0)\delta_{\rm
    m}(x)}$ be the correlation between density contrasts of galaxies
and matter at lag $x$, and
\begin{equation}
  b(x)=\sqrt{\frac{\xi_{\rm g}(x)}{\xi_{\rm m}(x)}}~;~
  r(x)=\frac{\xi_{\rm mg}(x)}{\sqrt{\xi_{\rm g}(x)\,\xi_{\rm m}(x)}}
\end{equation}
the biasing functions in real space.  We show for two specific
scenarios of the toy model in Sect. \ref{sect:toymodel} that the
subtraction of Poisson-shot noise can produce \mbox{$r(x)>1$} for
\mbox{$x>0$}. To this end, we first work out the real-space biasing
functions $b(x)$ and $r(x)$ for the toy model. The correlation
function $\xi(x)$ for a given power spectrum $P(k)$ is
\begin{equation}
  \xi(x)=[P](x):=
  \frac{1}{2\pi^2x}\int_0^\infty\d k\;k\,P(k)\,\sin{(k\,x)}\;,
\end{equation}
where we have defined the integral operator $[P](x)$ on the function
$P(k)$. For our toy model, we hence find $\xi_{\rm g}(x)=[P_{\rm
  g}](x)$, $\xi_{\rm gm}(x)=[P_{\rm gm}](x)$, and $\xi_{\rm
  m}(x)=[P_{\rm m}](x)$ with the one-halo terms
Eqs. \Ref{eq:pg}--\Ref{eq:pgm} and $n(m)\propto\delta_{\rm D}(m-m_0)$.
We assume $x>0$ in the following. After some algebra, we find
\begin{eqnarray}
  b(x)&=&
  \frac{[\tilde{u}_{\rm m}\cdot\tilde{u}^q_{\rm g}](x)}
  {[\tilde{u}_{\rm m}^2](x)}\,\frac{1}{r(x)}\;;
  \\
  r(x)&=&
  \frac{[\tilde{u}_{\rm m}\cdot\tilde{u}^q_{\rm g}](x)}
  {\sqrt{[\tilde{u}_{\rm g}^{2p}](x)\,[\tilde{u}_{\rm m}^2](x)}}\,
  \left(1+\frac{\Delta\sigma_N^2(m_0)}{\ave{N|m_0}}\right)^{-1/2}\;;
\end{eqnarray}
where $[\tilde{f}\cdot\tilde{g}](x)$ denotes the Fourier
back-transform of the product $\tilde{f}(k)\times\tilde{g}(k)$; it is
hence the convolution of $f(x)$ and $g(x)$. These equation assume
$u_{\rm m}(x)\ge0$ for all lags $x$ such that the convolution
$[\tilde{u}_{\rm m}^2](x)=[\tilde{u}_{\rm m}\cdot\tilde{u}_{\rm
  m}](x)$ is positive definite as well. Now, for faithful galaxies we
have $\tilde{u}_{\rm m}(k,m_0)=\tilde{u}_{\rm g}(k,m_0)$ and
\mbox{$p=q=1$}, and therefore analogous to $(b(k),r(k))$
\begin{equation}
  b(x)\times r(x)=1
  ~;~
  r(x)=\left(1+\frac{\Delta\sigma_N^2(m_0)}{\ave{N|m_0}}\right)^{-1/2}\;.
\end{equation}
Clearly, we find \mbox{$r(x)>1$} for a sub-Poisson HOD variance also
for real-space biasing functions. Moreover, for galaxies with
$\tilde{u}_{\rm m}(k,m_0)=\tilde{u}_{\rm g}(k,m_0)$, Poisson HOD
variance ($\Delta\sigma_N^2=0$), and central galaxies in low-occupancy
halos ($p=1/2,q=0$), we arrive at
\begin{equation}
  \frac{b(x)}{r(x)}=1
  ~;~
  r(x)=
  \sqrt{\frac{[\tilde{u}_{\rm m}](x)}{[\tilde{u}_{\rm m}^2](x)}}\;.
\end{equation}
Therefore, also here we find \mbox{$r(x)>1$} for at least some lags
$x$ because the convolution of $u_{\rm m}(x)$ with itself has to be
\mbox{$[\tilde{u}_{\rm m}^2](x)<u_{\rm m}(x)$} for some
$x$. 

\section{Estimators of aperture statistics}
\label{sect:estimators}

The aperture statistics can be obtained from three kinds of two-point
correlation functions based on the positions $\vec{\theta}^{\rm d}_i$
of $n_{\rm d}$ lens galaxies on the sky and the positions
$\vec{\theta}^{\rm s}_i$, shear estimators $\epsilon_i$, and
statistical weights $w_i$ of $n_{\rm s}$ source galaxies. We estimate
these correlation functions as follows.

First, we estimate the shear-shear correlation functions
$\xi_\pm(\vartheta)=\ave{\gamma_{\rm
    t}(\vec{\theta}+\vec{\vartheta})\gamma_{\rm t}(\vec{\theta})}\pm
\ave{\gamma_\times(\vec{\theta}+\vec{\vartheta})\gamma_\times(\vec{\theta})}$
for a separation $\vartheta$ of two sources, where the tangential,
$\gamma_{\rm t}$, and cross components, $\gamma_\times$, of $\gamma$
at the source positions are defined relative to the vector
$\vec{\vartheta}$ connecting a source pair through $\gamma_{\rm
  t}+\i\,\gamma_\times=-\vec{\vartheta}^\ast/\vec{\vartheta}\,\gamma$;
position or separation vectors use the usual complex-valued notation
in a local Cartesian frame on the sky \citep{2001PhR...340..291B}. We
define for all estimators a galaxy pair $ij$ with positions
$\vec{\theta}_i^{\rm x}$ and $\vec{\theta}_j^{\rm y}$ to be within the
separation bin $(\overline{\vartheta},\Delta\vartheta)$ if
$\Delta_{ij}^{\rm xy}(\overline{\vartheta},\Delta\vartheta)\ne0$ for
\begin{equation}
  \Delta_{ij}^{\rm xy}(\overline{\vartheta},\Delta\vartheta):=
  \left\{
    \begin{array}{ll}
      1 & {\rm if~}\overline{\vartheta}-\Delta\vartheta/2\le|\vec{\theta}_i^{\rm x}-
      \vec{\theta}_j^{\rm y}|<\overline{\vartheta}+\Delta\vartheta/2\\
      0 & {\rm otherwise}
    \end{array}
  \right.\;.
\end{equation}
Then we estimate the average $\xi_\pm(\vartheta)$ for source pairs
within the bin $(\overline{\vartheta},\Delta\vartheta)$ by
\begin{equation}
  \label{eq:estxipm}
  \widehat{\xi_\pm}(\overline{\vartheta},\Delta\vartheta)=
  \frac{\sum_{i,j=1}^{n_{\rm s}}w_i\,w_j\,\Delta_{ij}^{\rm ss}(\overline{\vartheta},\Delta\vartheta)\,
    \left(\epsilon_{{\rm t},i}\epsilon_{{\rm t},j}\pm\epsilon_{\times,i}\epsilon_{\times,j}\right)}
  {\sum_{i,j=1}^{n_{\rm s}}w_i\,w_j\,\Delta_{ij}^{\rm ss}(\overline{\vartheta},\Delta\vartheta)}\;,
\end{equation}
where $\epsilon_{{\rm t},i}$ and $\epsilon_{{\rm t},j}$ refer to the
tangential components of the shear estimator of the $i$th or $j$th
source in the pair $ij$ relative to $\vec{\vartheta}=\vec{\theta}^{\rm
  s}_i-\vec{\theta}^{\rm s}_j$, and likewise for $\epsilon_{\times,i}$
and $\epsilon_{\times,j}$ \citep{2002A&A...396....1S}.

Second, we estimate the mean tangential shear $\overline{\gamma}_{\rm
  t}(\vartheta)=\ave{\gamma_{\rm t}(\vec{\theta}^{\rm
    d}+\vec{\vartheta})|\vec{\theta}^{\rm d}}$ of sources at
separation $\vartheta$ from lenses located at $\vec{\theta}^{\rm d}$
by
\begin{equation}
  \label{eq:estggl}
  \widehat{\gamma_{\rm
      t}}(\overline{\vartheta},\Delta\vartheta)=
  \frac{\sum_{i=1}^{n_{\rm d}}\sum_{j=1}^{n_{\rm s}}w_j\,
    \Delta_{ij}^{\rm ds}(\overline{\vartheta},\Delta\vartheta)\,\epsilon_{{\rm t},j}}
  {\sum_{i=1}^{n_{\rm d}}\sum_{j=1}^{n_{\rm s}}w_j\,
    \Delta_{ij}^{\rm ds}(\overline{\vartheta},\Delta\vartheta)}-
  \widehat{\gamma_{\rm
      t}^{\rm rnd}}(\overline{\vartheta},\Delta\vartheta)\;,
\end{equation}
where now $\epsilon_{{\rm t},j}$ is the tangential component of
$\epsilon_j$ relative to $\vec{\vartheta}=\vec{\theta}^{\rm
  d}_i-\vec{\theta}^{\rm s}_j$, and $\widehat{\gamma_{\rm t}^{\rm
    rnd}}(\overline{\vartheta},\Delta\vartheta)$ is the first term on
the right-hand side of \Ref{eq:estggl} for a large sample of random
lens positions \citep{2016arXiv161100752S}.

Third, for the correlation function
$\omega(\vartheta)=\ave{\kappa_{\rm
    g}(\vec{\theta}+\vec{\vartheta})\kappa_{\rm g}(\vec{\theta})}$ of
the lens clustering on the sky, we employ the estimator in
\cite{1993ApJ...412...64L},
\begin{equation}
  \label{eq:estomega}
  \widehat{\omega}(\overline{\vartheta},\Delta\vartheta)=
  \frac{dd(\overline{\vartheta},\Delta\vartheta)}{rr(\overline{\vartheta},\Delta\vartheta)}-
  2\times\frac{dr(\overline{\vartheta},\Delta\vartheta)}{rr(\overline{\vartheta},\Delta\vartheta)}
  +1\;,
\end{equation}
where $dd$ is the normalised number of lens pairs in the separation
bin, $rr$ the normalised number of pairs with random positions
$\vec{\theta}^{\rm r}_i$ out of $n_{\rm r}\gg n_{\rm d}$ in total, and
$dr$ is the normalised number of lens-random pairs:
\begin{multline}
  dd(\overline{\vartheta},\Delta\vartheta)=
  \sum_{i,j=1}^{n_{\rm d}}\frac{\Delta_{ij}^{\rm dd}(\overline{\vartheta},\Delta\vartheta)}{n_{\rm d}^2}\;;
  \\
  rr(\overline{\vartheta},\Delta\vartheta)=
  \sum_{i,j=1}^{n_{\rm r}}\frac{\Delta_{ij}^{\rm rr}(\overline{\vartheta},\Delta\vartheta)}{n_{\rm r}^2}~;~
  dr(\overline{\vartheta},\Delta\vartheta)=
  \sum_{i,j=1}^{n_{\rm d},n_{\rm d}}\frac{\Delta_{ij}^{\rm dr}(\overline{\vartheta},\Delta\vartheta)}
  {n_{\rm d}\,n_{\rm r}}\;.
\end{multline}

To combine the estimates from $n_{\rm patch}$ different patches, we
merge their lens and source catalogues with constant position offsets
for each patch such that we never have pairs of galaxies from
different patches inside a separation bin. The probability to find a
random-lens position inside a particular patch in the merged catalogue
is proportional to the effective, unmasked area of the patch, which is
always $16\,\rm deg^2$ for the mock data. For the analysis of the 64
patches mock-data, we use as angular binning 5000 linear bins between
$1.4\,\rm arcsec$ and $5.7\,\rm deg$.

We transform the estimates of the three two-point correlation
functions into estimates of the aperture statistics for several
\mbox{$\theta_{\rm ap}\in[1^\prime,140^\prime]$} by a numerical
integration of the following equations
\begin{multline}
  \left.
    \begin{array}{ll}
      \ave{M_{\rm ap}^2}(\theta_{\rm ap})
      \\
      \ave{{\cal N}M_{\rm ap}}(\theta_{\rm ap})
      \\
      \ave{{\cal N}^2}(\theta_{\rm ap})
    \end{array}
  \right\}=
  \\
  \int_0^\infty\!\!\!\!\d x\;x\,
  \times\left\{
    \begin{array}{ll}
      \frac{1}{2}
      \left(\xi_+(x\theta_{\rm ap})\,T_+(x)+\xi_-(x\theta_{\rm ap})\,T_-(x)\right)
      \\
      \overline{\gamma}_{\rm t}(x\theta_{\rm ap})\,F(x)
      \\
      \omega(x\theta_{\rm ap})\,T_+(x)
   \end{array}
  \right.
\end{multline}
based on the auxiliary functions
\begin{equation}
  \left.
    \begin{array}{ll}
      T_{+,-}(x)\\
      F(x)
    \end{array}
  \right\}=
  (2\pi)^2\,\int_0^\infty\d s\;s\,I^2(s)
  \times\left\{
    \begin{array}{ll}
      {\rm J}_{0,4}(s\,x)\\
      {\rm J}_2(s\,x)
    \end{array}
  \right.
\end{equation}
with analytic expressions for $T_\pm(x)$ and $F(x)$ as in
\cite{2007A&A...461..861S}. Because of the lower cutoff at 1.4 arcsec
in the correlation functions (we set them zero here), we cannot use
values of the aperture statistics below around two arcmin where the
transformation bias grows to about 10 per cent
\citep{2006A&A...457...15K}.  To estimate the statistical errors or
covariances between the three aperture statistics and angular scales
$\theta_{\rm ap}$, we employ the jackknife technique with $n_{\rm
  patch}$ sub-samples that we obtain by removing one patch from the
merged catalogue at a time \citep{2016MNRAS.456.2662F}.

In Fig. \ref{fig:GIandII}, we plot estimates of $\ave{M^2_{\rm
    ap}}(\theta_{\rm ap})$ in our mock data as points with statistical
errors obtained with the jackknife technique (inflated by a factor of
five). The three different styles of the data points are for: (i)
shear with shape noise of sources (solid squares); (ii) reduced shear
with shape noise (open circles); and (iii) shear without shape noise
(solid triangles). The data points are a very good match to a
theoretical model GG for the MS cosmology (blue solid line), and they
are consistent with each other at the same $\theta_{\rm ap}$. The
statistical errors of data without shape noise are similar to those
for sources with shape noise for \mbox{$\theta_{\rm
    ap}\gtrsim10^\prime$} which indicates that cosmic variance
dominates in this regime for our fiducial survey.

\renewcommand{\arraystretch}{1.3}
\begin{table*}
  \begin{center}
     
 \caption{\label{tab:gausstest} Results of KS-test for Gaussianity for
   three kinds of the aperture statistic at scales
   $\theta=2,20,60,120$ arcmin (in this order). Quoted are the
   $p$-values of the test statistic. Values $p\le0.05$ indicate a
   tension with a Gaussian distribution at 95\% confidence or higher
   (bold face). The tests use the empirical distribution of the
   aperture statistics in 64 fields of our mock data with
   $4\times4\,\rm deg^2$ area each. Realisations of the source
   shape-noise and positions are different for each lens sample.}
 \begin{tabular}{lllllcllllcllll}
   \hline\hline
   Sample\tablefootmark{a} & 
   \multicolumn{4}{c}{$\ave{{\cal N}^2(\theta)}$\tablefootmark{b}} & &
   \multicolumn{4}{c}{$\ave{{\cal N}M_{\rm ap}(\theta)}\tablefootmark{c}$} & &
   \multicolumn{4}{c}{$\ave{M^2_{\rm ap}(\theta)}\tablefootmark{d}$}\\
   \hline\\
   SM1 low-$z$  & 0.54 & 0.08 & 0.35 & 0.16 &  & 0.44 & 0.28 & 0.06 & 0.05 &  & \textbf{0.01} & \textbf{0.02} & \textbf{0.00} & \textbf{0.01} \\
   SM2 low-$z$  & 0.52 & 0.23 & 0.97 & 0.35 &  & 0.71 & 0.90 & 0.25 & 0.06 &  & 0.31 & \textbf{0.04} & 0.06 & \textbf{0.01} \\
   SM3 low-$z$  & \textbf{0.04} & 0.18 & 0.35 & \textbf{0.00} &  & 0.44 & 0.74 & 0.61 & 0.84 &  & 0.76 & 0.28 & \textbf{0.03} & \textbf{0.00} \\
   SM4 low-$z$  & 0.08 & 0.16 & 0.20 & 0.76 &  & 0.86 & 0.94 & 0.23 & \textbf{0.01} &  & 0.41 & 0.05 & 0.07 & 0.16 \\
   SM5 low-$z$  & 0.52 & 0.05 & 0.95 & 0.16 &  & 0.16 & \textbf{0.04} & 0.31 & \textbf{0.01} &  & 0.16 & 0.20 & 0.20 & 0.16 \\
   SM6 low-$z$  & 0.22 & 0.44 & 0.65 & 0.82 &  & 0.08 & 0.18 & 0.84 & 0.08 &  & 0.12 & 0.22 & \textbf{0.02} & \textbf{0.00} \\
   RED low-$z$  & 0.31 & \textbf{0.00} & \textbf{0.00} & \textbf{0.00} &  & 0.67 & \textbf{0.02} & 0.56 & \textbf{0.00} &  & 0.06 & \textbf{0.03} & \textbf{0.02} & 0.06 \\
   BLUE low-$z$  & 0.47 & 0.96 & 0.47 & 0.71 &  & 0.77 & 0.89 & 0.12 & 0.77 &  & 0.31 & 0.07 & 0.20 & 0.05 \\
   \\
   SM1 high-$z$  & 0.08 & 0.07 & 0.93 & 0.28 &  & 0.33 & 0.44 & \textbf{0.02} & 0.24 &  & 0.24 & 0.08 & \textbf{0.00} & \textbf{0.00} \\
   SM2 high-$z$  & 0.57 & 0.37 & 0.50 & 0.44 &  & 0.79 & 0.64 & 0.44 & 0.41 &  & 0.31 & 0.12 & \textbf{0.00} & \textbf{0.01} \\
   SM3 high-$z$  & 1.00 & 0.28 & 0.12 & 0.89 &  & 0.84 & 0.23 & 0.06 & 0.87 &  & 0.44 & \textbf{0.04} & \textbf{0.04} & \textbf{0.00} \\
   SM4 high-$z$  & 0.50 & 0.44 & 0.50 & 0.06 &  & 0.94 & 0.50 & 0.96 & 0.64 &  & 0.08 & 0.22 & \textbf{0.00} & \textbf{0.00} \\
   SM5 high-$z$  & 0.71 & 0.13 & 0.61 & \textbf{0.01} &  & 0.47 & 0.24 & 0.71 & \textbf{0.00} &  & 0.86 & 0.06 & \textbf{0.00} & 0.44 \\
   SM6 high-$z$  & 0.64 & 0.08 & 0.64 & 0.18 &  & 0.52 & 0.86 & 0.61 & 0.23 &  & 0.52 & \textbf{0.01} & 0.06 & 0.18 \\
   RED high-$z$  & 0.07 & 0.41 & 0.20 & \textbf{0.00} &  & 0.41 & 0.05 & 0.41 & 0.54 &  & 0.31 & 0.20 & 0.25 & \textbf{0.00} \\
   BLUE high-$z$  & 0.08 & 0.67 & 0.08 & 0.55 &  & 0.46 & 0.22 & 0.63 & \textbf{0.02} &  & 0.86 & 0.17 & 0.55 & 0.08 
  \end{tabular}
 \tablefoot{
   \tablefoottext{a}{mock sample used in analysis (lenses);}
   \tablefoottext{b}{variances of aperture number count for sequence of aperture scales;}
   \tablefoottext{c}{sequence of covariances between aperture number count and aperture mass;}
   \tablefoottext{d}{sequence of variances of aperture mass}
  }

  \end{center}
\end{table*}
\renewcommand{\arraystretch}{1.0}

\section{Non-Gaussianity of measurement errors}
\label{sect:nongauss}

In this appendix, we test for a non-Gaussian distribution of
statistical errors in the aperture statistics $\ave{{\cal
    N}^2}(\theta_{\rm ap})$, $\ave{{\cal N}M_{\rm ap}}(\theta_{\rm
  ap})$, and $\ave{M_{\rm ap}^2}(\theta_{\rm ap})$ with a
one-dimensional Kolmogorov-Smirnov (KS) test. We perform the KS tests
separately for each aperture statistics, denoted by $x$ in the
following, and the aperture scales $\theta_{\rm
  ap}\in\{2^\prime,20^\prime,60^\prime,120^\prime\}$. Since we have
only one simulated lensing survey with $1024\,\rm deg^2$ of data, we
split the data into independent patches and test the empirical
distributions of measurements $x_i$ in $n=64$ patches of area
$4\times4\,\rm deg^2$ each. We standardise the measurements by
computing $z_i=(x_i-\bar{x})/\sigma_{\rm x}$, where $\bar{x}$ and
$\sigma_{\rm x}$ are the mean and standard deviation, respectively, of
the sample $\{x_i:i=1\ldots n\}$. For the KS test, we then compare the
distribution \mbox{$F_n(z)=n^{-1}\,\sum_{i=1}^n{\rm I}(z_i-z)$} in the
mock data to the average distribution $F(z)$ of $n$ normally
distributed measurements using the test statistic
$D_n=\sup_z\{|F_n(z)-F(z)|\}$; here we define the function
\mbox{$I(z)=1$} for \mbox{$z\le0$} and \mbox{$I(z)=0$} otherwise.

The resulting $p$-values of our $D_n$ are listed in Table
\ref{tab:gausstest}; $p$-values in boldface are smaller than $0.05$
and indicate a conflict with a Gaussian distribution ($95\%$ CL). We
perform the test for each galaxy sample and redshift bin. For every
galaxy sample, we make new realisations of the source catalogues by
randomly changing the source positions in the patch and the intrinsic
shapes. This explains the differences in the test results for
$\ave{M_{\rm ap}^2}$ at identical angular scales. While we expect some
failures of the KS test by chance, tensions are clearly visible for
$\ave{M^2_{\rm ap}}$ at \mbox{$\theta_{\rm ap}\gtrsim60^\prime$} and
for $\ave{{\cal N}^2}$ of the strongly clustered sample \mbox{RED
  low-$z$}.

In summary, for $\ave{{\cal N}^2}$ and $\ave{{\cal N}M_{\rm ap}}$ a
Gaussian likelihood of errors is a fair approximation, whereas for
$\ave{M_{\rm ap}^2}$ and around degree scales or more, non-Gaussian
features in the error distribution, mainly cosmic variance, are
detectable. We note that evidence for non-Gaussian distributions does
not necessarily mean that a Gaussian likelihood is an insufficient
approximation for a shear analysis.

\section{Template parameters of reconstructed biasing functions}
\label{sect:physicaldetails}

\renewcommand{\arraystretch}{1.3}
\begin{table*}
  \caption{\label{tab:results} Summary of estimated model parameters in our simulated lensing analysis for the various galaxy 
    samples and redshift bins. The quoted values are medians and 68\% PIs of the marginal posterior distributions. The statistical errors assume our fiducial survey for sources including shape noise.}
  \begin{center}
      
 \begin{tabular}{lcccccc}
   \hline\hline
   Sample\tablefootmark{a} & $\log_{10}{(\bar{n}_{\rm g}/h^3{\rm Mpc}^{-3})}$\tablefootmark{b} & $b_{\rm ls}$\tablefootmark{c} & $\log_{10}{(m_{\rm piv}/h^{-1}{\rm M_\odot})}$\tablefootmark{d} & $f_{\rm cen}$\tablefootmark{e} \\
    \hline\\
   SM1 low-$z$ & $-1.69^{+1.38}_{-0.51}$ & $1.00^{+0.05}_{-0.05}$ & $12.44^{+0.74}_{-1.29}$ & $0.59^{+0.27}_{-0.32}$ \\
   SM2 low-$z$ & $-2.08^{+0.42}_{-0.26}$ & $1.09^{+0.05}_{-0.05}$ & $12.92^{+0.39}_{-0.60}$ & $0.75^{+0.17}_{-0.29}$ \\
   SM3 low-$z$ & $-2.69^{+0.17}_{-0.16}$ & $1.19^{+0.05}_{-0.05}$ & $13.52^{+0.24}_{-0.27}$ & $0.86^{+0.09}_{-0.15}$ \\
   SM4 low-$z$ & $-3.07^{+0.23}_{-0.32}$ & $1.28^{+0.05}_{-0.06}$ & $13.80^{+0.49}_{-0.32}$ & $0.82^{+0.12}_{-0.18}$ \\
   SM5 low-$z$ & $-3.48^{+0.18}_{-0.26}$ & $1.35^{+0.06}_{-0.06}$ & $14.26^{+0.31}_{-0.30}$ & $0.80^{+0.13}_{-0.17}$ \\
   SM6 low-$z$ & $-3.66^{+0.20}_{-0.18}$ & $1.55^{+0.06}_{-0.07}$ & $14.31^{+0.34}_{-0.32}$ & $0.82^{+0.11}_{-0.15}$ \\
   BLUE low-$z$ & $-1.97^{+0.32}_{-0.24}$ & $0.93^{+0.04}_{-0.05}$ & $13.06^{+0.61}_{-0.53}$ & $0.69^{+0.21}_{-0.29}$ \\
   RED low-$z$ & $-3.37^{+0.21}_{-0.14}$ & $1.30^{+0.08}_{-0.06}$ & $14.00^{+0.31}_{-0.21}$ & $0.90^{+0.07}_{-0.10}$ \\
   \\
   SM1 high-$z$ & $-0.86^{+0.65}_{-1.31}$ & $1.09^{+0.04}_{-0.04}$ & $11.89^{+1.30}_{-0.87}$ & $0.56^{+0.28}_{-0.32}$ \\
   SM2 high-$z$ & $-0.51^{+0.35}_{-0.86}$ & $1.15^{+0.04}_{-0.04}$ & $11.43^{+0.93}_{-0.52}$ & $0.52^{+0.31}_{-0.33}$ \\
   SM3 high-$z$ & $-2.57^{+0.29}_{-0.31}$ & $1.28^{+0.04}_{-0.04}$ & $13.31^{+0.41}_{-0.32}$ & $0.70^{+0.20}_{-0.27}$ \\
   SM4 high-$z$ & $-3.13^{+0.24}_{-0.32}$ & $1.34^{+0.04}_{-0.04}$ & $13.85^{+0.44}_{-0.28}$ & $0.85^{+0.10}_{-0.18}$ \\
   SM5 high-$z$ & $-3.44^{+0.17}_{-0.30}$ & $1.53^{+0.05}_{-0.05}$ & $14.09^{+0.46}_{-0.34}$ & $0.79^{+0.14}_{-0.18}$ \\
   SM6 high-$z$ & $-3.82^{+0.21}_{-0.22}$ & $1.61^{+0.07}_{-0.07}$ & $14.41^{+0.34}_{-0.27}$ & $0.90^{+0.07}_{-0.11}$ \\
   BLUE high-$z$ & $-0.42^{+0.28}_{-0.41}$ & $1.11^{+0.04}_{-0.03}$ & $11.34^{+0.57}_{-0.48}$ & $0.51^{+0.32}_{-0.32}$ \\
   RED high-$z$ & $-3.55^{+0.32}_{-0.81}$ & $1.62^{+0.07}_{-0.08}$ & $14.13^{+0.82}_{-0.34}$ & $0.88^{+0.08}_{-0.14}$ 
    \end{tabular}
 \tablefoot{
  \tablefoottext{a}{sample used for the emulated lensing analysis; SM samples are selected in stellar mass, RED and BLUE are selected by colour (see Table \ref{tab:samples}); low-$z$ samples have a mean redshift of $\bar{z}\approx0.36$, high-$z$ samples have $\bar{z}\approx0.52$;}
  \tablefoottext{b}{mean galaxy number-density (comoving);}
  \tablefoottext{c}{large-scale bias factor;}
  \tablefoottext{d}{pivital halo mass;}
  \tablefoottext{e}{fraction of halos with central galaxies (\emph{not} the fraction of central galaxies)}
  }

  \end{center}
\end{table*}
\renewcommand{\arraystretch}{1.0}

\begin{figure*}[htb]
  \begin{center}
    \epsfig{file=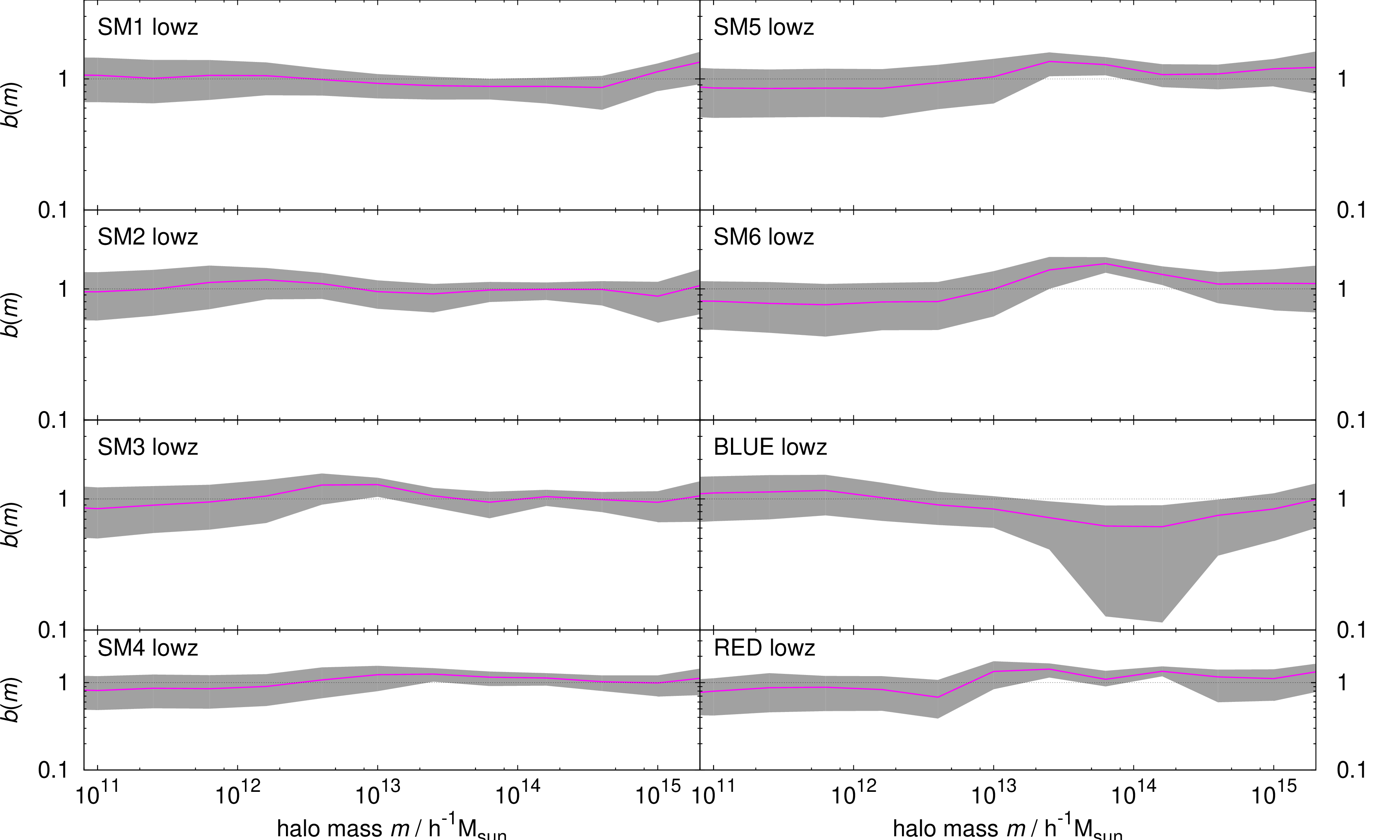,width=110mm,angle=-90}

    \vspace{0.5cm}
    \epsfig{file=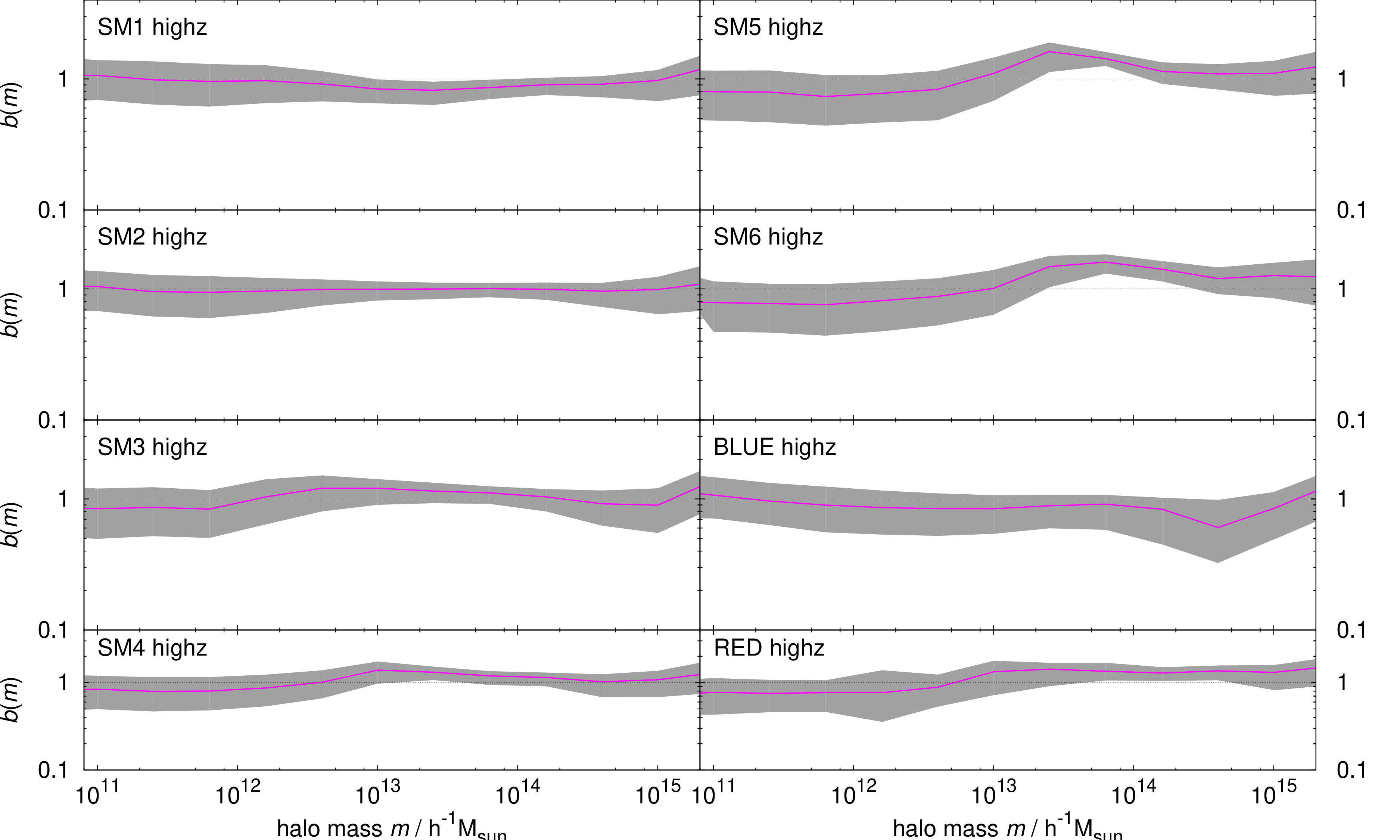,width=110mm,angle=-90}
  \end{center}
  \caption{\label{fig:deltag} Mean biasing function $b(m)$ for the
    mock galaxy samples SM1 to SM6 and the colour-selected samples
    BLUE and RED in the low-$z$ (top) and high-$z$ (bottom) redshift
    bin. The shaded regions indicate the 68\% PI about the median for
    our fiducial mock survey.}
\end{figure*}

\begin{figure*}[htb]
  \begin{center}
    \epsfig{file=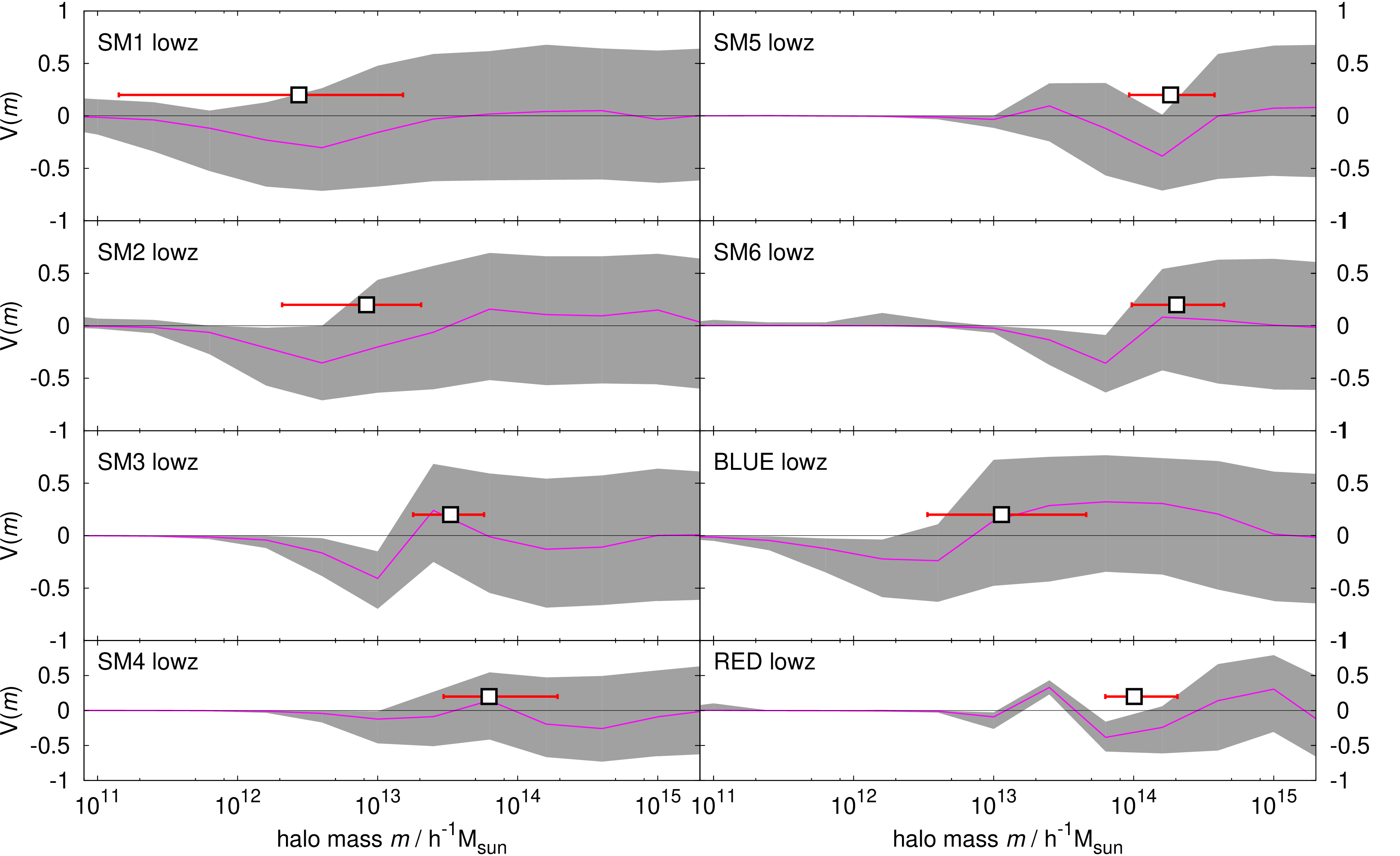,width=110mm,angle=-90}

    \vspace{0.5cm}
    \epsfig{file=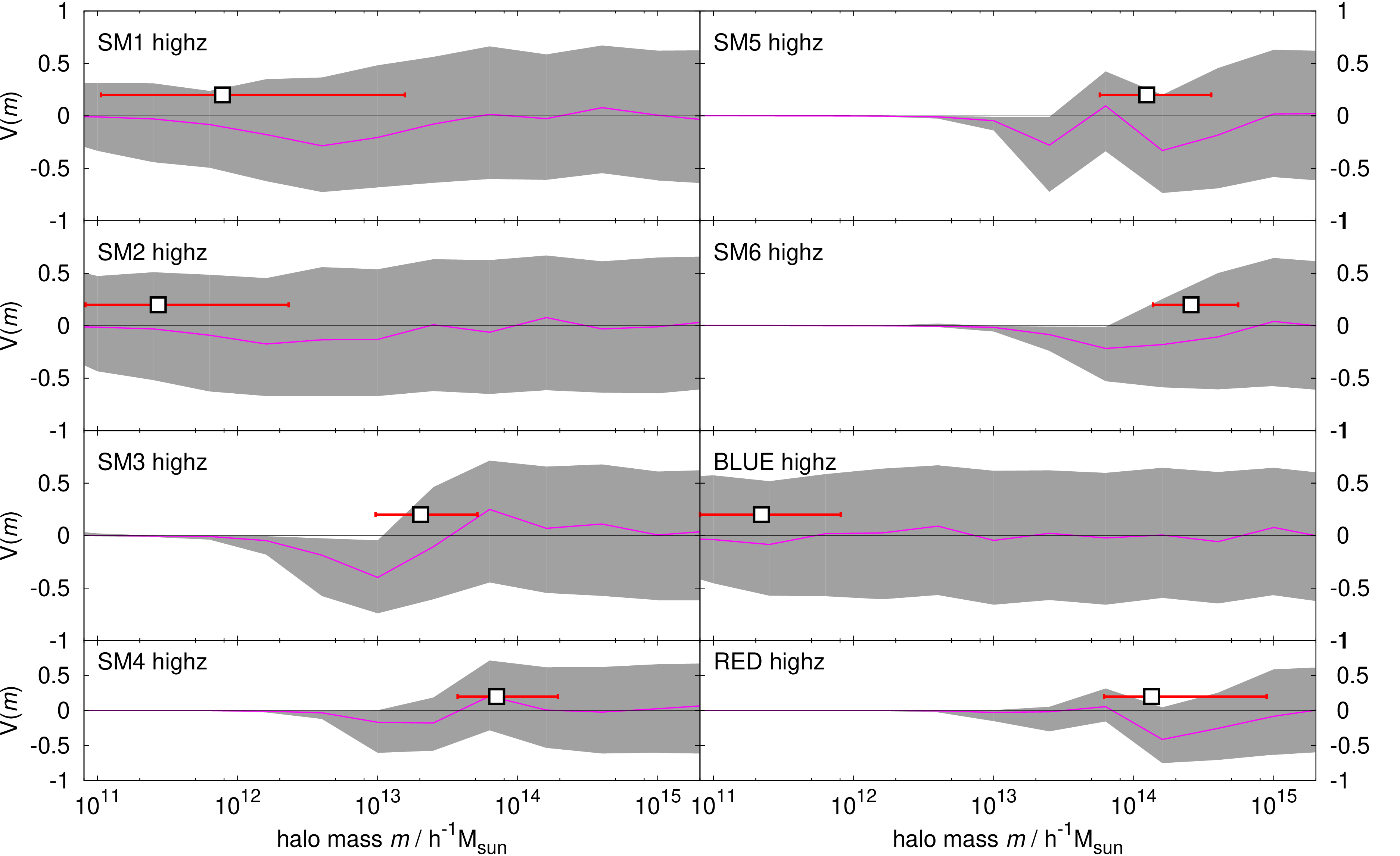,width=110mm,angle=-90}
  \end{center}
  \caption{\label{fig:Vm} Excess variance $V(m)$ for the mock galaxy
    samples SM1 to SM6 and the colour-selected samples BLUE and RED in
    the low-$z$ (top) and high-$z$ (bottom) redshift bin. The shaded
    regions indicate the 68\% PI about the median for our fiducial
    mock survey. For each panel, the open square shows the median
    mass-scale of the pivotal mass $m_{\rm piv}$ and its errorbars the
    68\% PI.}
\end{figure*}

Table \ref{tab:results} and the Figs. \ref{fig:Vm} and
\ref{fig:deltag} summarise the posterior distribution of template
parameters that are the basis for the inferred biasing function shown
in the Figs. \ref{fig:brofksm} and \ref{fig:brofkredblue}. The high
uncertainties of most parameters therein reflect the high degeneracy
of the template model.  We see weak trends for the mean biasing
function $b(m)$ in Fig. \ref{fig:deltag}: galaxies with low stellar
masses or blue galaxies prefer a relatively high number of galaxies
inside halos below \mbox{$\sim10^{13}\,h^{-1}\,{\rm M}_\odot$} while
higher stellar masses or red galaxies are underrepresented in this
regime; see in particular the RED high-$z$ sample. The excess variance
$V(m)$, shown in Fig. \ref{fig:Vm}, is almost always consistent with a
Poisson variance although a very tentative sub-Poisson variance may be
visible just below $m_{\rm piv}$ in some cases (SM3 or SM4, for
instance), but usually gets smeared out by the uncertainty of $m_{\rm
  piv}$.
  
\end{document}